\documentclass{article}
\pdfoutput=1

\pdfoutput=1

\usepackage{xcolor}

\usepackage{mathtools}
\usepackage{bbm}
\usepackage{enumitem}
\usepackage{dsfont}
\usepackage{appendix}
\usepackage{multirow}
\usepackage{subcaption}

\usepackage{pdfpages}
\usepackage{arxiv-jcgs}
\usepackage{natbib}
\usepackage{latexsym}
\usepackage{amsmath}
\usepackage{amsfonts}
\usepackage{amsthm}
\usepackage{amssymb}
\usepackage{psfrag}
\usepackage{graphicx}
\usepackage{url}
\usepackage{enumitem}
\usepackage{algorithm}
\usepackage{algpseudocode}
\usepackage{lineno}
\usepackage{appendix}

\usepackage{graphicx}
\DeclareGraphicsExtensions{.pdf,.png,.jpg}
\graphicspath{ {fig/} }
\pdfoutput=1

\DeclareMathOperator*{\argmax}{arg\,max}
\DeclareMathOperator*{\argmin}{arg\,min}

\DeclareMathOperator*{\logit}{logit}

\def\bfalpha{\boldsymbol{\alpha}}
\def\bfbeta{\boldsymbol{\beta}}
\def\bfdelta{\boldsymbol{\delta}}
\def\bfeta{\boldsymbol{\eta}}

\def\bfphi{\boldsymbol{\phi}}
\def\bfPhi{\boldsymbol{\Phi}}
\def\bfxi{\boldsymbol{\xi}}
\def\bftheta{\boldsymbol{\theta}}
\def\bfGamma{\boldsymbol{\Gamma}}
\def\bfgamma{\boldsymbol{\gamma}}
\def\bfSigma{\boldsymbol{\Sigma}}

\def\bfQ{\mathbf{Q}}

\def\bfX{\mathbf{X}}
\def\bfY{\mathbf{Y}}

\def\bfy{\mathbf{y}}
\def\bfz{\mathbf{z}}

\def\bbE{\mathbb{E}}

\def\bbN{\mathbb{N}}

\def\bbP{\mathbb{P}}

\def\bbR{\mathbb{R}}

\def\calN{\mathcal{N}}
\def\calO{\mathcal{O}}

\def\calS{\mathcal{S}}

\DeclarePairedDelimiterX\bigCond[2]{[}{]}{#1 \;\delimsize\vert\; #2}

\def\onevec{\mathbf{1}}

\def\part{\pi} 

\newtheorem{theorem}{Theorem}

\newtheorem{lemma}{Lemma}

\date{}
\begin{document}

\title{Variance-Reduced Stochastic Optimization for Efficient Inference of Hidden Markov Models}

\author{
  \textbf{Evan Sidrow} \\
  Department of Statistics \\
  University of British Columbia\\
  Vancouver, Canada \\
  \texttt{evan.sidrow@stat.ubc.ca} \\
  \and
  \textbf{Nancy Heckman} \\
  Department of Statistics \\
  University of British Columbia \\
  Vancouver, Canada \\
  \and
  \textbf{Alexandre Bouchard-C\^ot\'e} \\
  Department of Statistics \\
  University of British Columbia \\
  Vancouver, Canada \\
  \and
  \textbf{Sarah M. E. Fortune} \\
  Department of Oceanography \\
  Dalhousie University \\
  Halifax, Canada \\
  \and
  \textbf{Andrew W. Trites} \\
  Department of Zoology \\
  Institute for the Oceans and Fisheries \\
  University of British Columbia \\
  Vancouver, Canada \\
  \and
  \textbf{Marie Auger-M\'eth\'e} \\
  Department of Statistics \\
  Institute for the Oceans and Fisheries \\
  University of British Columbia \\
  Vancouver, Canada \\
}
\maketitle

\bigskip
\begin{abstract}
    Hidden Markov models (HMMs) are popular models to identify a finite number of latent states from sequential data. However, fitting them to large data sets can be computationally demanding because most likelihood maximization techniques require iterating through the entire underlying data set for every parameter update.  We propose a novel optimization algorithm that updates the parameters of an HMM without iterating through the entire data set. Namely, we combine a partial E step with variance-reduced stochastic optimization within the M step. We prove the algorithm converges under certain regularity conditions. We test our algorithm empirically using a simulation study as well as a case study of kinematic data collected using suction-cup attached biologgers from eight northern resident killer whales ({\em{Orcinus orca}}) off the western coast of Canada. In both, our algorithm converges in fewer epochs and to regions of higher likelihood compared to standard numerical optimization techniques. Our algorithm allows practitioners to fit complicated HMMs to large time-series data sets more efficiently than existing baselines.
\end{abstract}

\vfill 

\section{Introduction}

Hidden Markov models (HMMs) are statistical models for sequential data that are widely used to model time series in fields such as speech recognition \citep{Gales:2008}, geology \citep{Bebbington:2007}, neuroscience \citep{Kottaram:2019}, finance \citep{Mamon:2007}, and ecology \citep{McClintock:2020}. Such models are often used to predict a latent process of interest (e.g. a spoken phrase or an animal's behavioral state) from an observed time series (e.g. raw audio or time-depth data). Many practitioners estimate the parameters of an HMM by maximizing the likelihood function using either gradient-based numerical maximization or the Baum-Welch algorithm \citep{Baum:1970}. The latter is a special case of the expectation-maximization (EM) algorithm \citep{Dempster:1977}. 

One serious concern for both numerical maximization and the Baum-Welch algorithm is that every parameter update requires iterating though the full set of observations to calculate either the likelihood or its gradient. This concern is likely to only worsen in the future, as time-series data sets are increasingly collected at high frequencies and contain large numbers of observations \citep{Patterson:2017,Li:2020}. These data sets require progressively complex HMMs which can be computationally expensive to fit \citep{Adam:2019,Sidrow:2021}. In addition, many model validation techniques such as cross-validation require repeated parameter estimation which can be prohibitive even for relatively simple HMMs \citep{Pohle:2017}.

Many inference techniques for independent data sets do not require iterating through the entire data set to update a model's parameters. We henceforth refer to these techniques as \textit{sub-linear} methods. One example of a sub-linear method is stochastic gradient descent, which evaluates the gradient of the likelihood over a random subset of the data \citep{Robbins:1951}. It is ubiquitous in the optimization literature and has inspired several extensions \citep{Johnson:2013, Defazio:2014, Kingma:2014}. Similarly, the incremental EM algorithm is a generalization of the EM algorithm that updates only a subset of hidden variables at each E step \citep{Neal:1998, Thiesson:2001, Karimi:2019}.  However, these methods assume that the underlying data set is comprised of independent subsets. This assumption is violated for HMMs since the underlying data set exhibits sequential dependence. 

Some work has been done to apply sub-linear inference methods to HMMs. For example, \citet{Gotoh:1998} divide the sequence of observations into subsequences and use the incremental EM algorithm to perform inference. This approach assumes that subsequences are independent of one another, which is not true in general. Alternatively, \citet{Ye:2017} define a sufficiently large ``buffer" before and after subsequences of data to minimize the effect of serial dependence. However, the appropriate size of the buffer can be difficult to calculate. More examples of sub-linear inference techniques are given by \citet{Khreich:2012}, who review on-line and incremental methods for HMM inference. However, most of these methods assume either that the M step of the EM algorithm is tractable (see section 3.2.1 of \citet{Khreich:2012}), or that the emissions of the HMM are discrete \citep{Baldi:1993}. 

In this paper, we introduce a new inference method for HMMs based on variance-reduced stochastic optimization. This inference method updates the HMM parameters without iterating through the entire data set. Critically, it does not require a closed-form solution for its M step, does not require any buffer tuning, and does not introduce error into the HMM likelihood.

We begin with a formal definition of HMMs, a brief review of standard inference techniques for HMMs, and a description of stochastic optimization algorithms before introducing our algorithm. We then prove that our algorithm converges to a local maximum of the likelihood (under standard regularity assumptions) and note several practical considerations regarding its implementation. Finally, we compare the efficiency of our new algorithm to that of standard optimization techniques using several simulation studies and a kinematic case study of eight northern resident killer whales ({\em{Orcinus orca}}) off the western coast of Canada. 

\section{Background}
\subsection{Hidden Markov Models}

HMMs are common statistical models used to describe time series that exhibit state-switching behavior. An HMM models an observed sequence of length $T$, $\bfY = \{Y_t\}_{t=1}^T$, together with an unobserved (or  ``hidden") sequence $\bfX = \{X_t\}_{t=1}^T$. The hidden sequence $\bfX$ is a Markov chain, and each observation $Y_t$ is a random variable, where $Y_t$ given all other observations $(\bfY \setminus \{Y_t\})$ and hidden states $(\bfX)$ depends only on $X_t$. We assume $X_t \in \{1,\ldots,N\}$ for some finite $N$. The unconditional distribution of $X_1$ is denoted by the row-vector
$\bfdelta = \begin{pmatrix} \delta^{(1)} & \cdots & \delta^{(N)} \end{pmatrix}$,
where $\delta^{(i)} = \bbP(X_1 = i)$. Further, the distribution of $X_t$ for $t > 1$ conditioned on $X_{t-1}$ is denoted by an $N$-by-$N$ transition probability matrix 
\begin{equation}
    \bfGamma_t = \begin{pmatrix} 
    \Gamma_t^{(1,1)} & \cdots & \Gamma_t^{(1,N)} \\
    \vdots & \ddots & \vdots \\
    \Gamma_t^{(N,1)} & \cdots & \Gamma_t^{(N,N)} \\
    \end{pmatrix},
\end{equation}
where $\Gamma_t^{(i,j)} = \bbP(X_t = j \mid X_{t-1} = i)$. For simplicity, we assume that $\bfGamma_t$ does not change over time (i.e. $\bfGamma_t = \bfGamma$ for all $t$) unless stated otherwise. 

To ensure that all entries are positive and all rows sum to one, it is convenient to reparameterize the transition probability matrix $\bfGamma \in \bbR^{N \times N}$ and initial distribution $\bfdelta \in \bbR^N$ in terms of an auxiliary variable $\bfeta$:
\begin{equation}
    \Gamma^{(i,j)}(\bfeta) = \frac{\exp(\eta^{(i,j)})}{\sum_{k=1}^N \exp(\eta^{(i,k)})}, \qquad \delta^{(i)}(\bfeta) = \frac{\exp(\eta^{(i)})}{\sum_{k=1}^N \exp(\eta^{(k)})},
    \label{eqn:reparam}
\end{equation}
where $i,j = 1,\ldots,N$ and $\eta^{(i,i)}$ and $\eta^{(1)}$ are set to zero for identifiability. This formulation simplifies likelihood maximization by removing constraints in the optimization problem. One may also incorporate covariates into $\bfGamma$ by setting $\eta_t^{(i,j)}(\bfbeta) = \left(\bfbeta^{(i,j)}\right)^{\top} \bfz_t$, where $\bfz_t$ is a column vector of known covariates at time index $t$ and $\bfbeta^{(i,j)}$ is a column vector of unknown regression coefficients. While $\bfGamma$ and $\bfdelta$ are functions of $\bfeta$, we abuse notation in future sections and treat $\bfGamma$ and $\bfdelta$ as variables since the mapping is a bijection.

\begin{figure}
    \centering
    \includegraphics[width=5in]{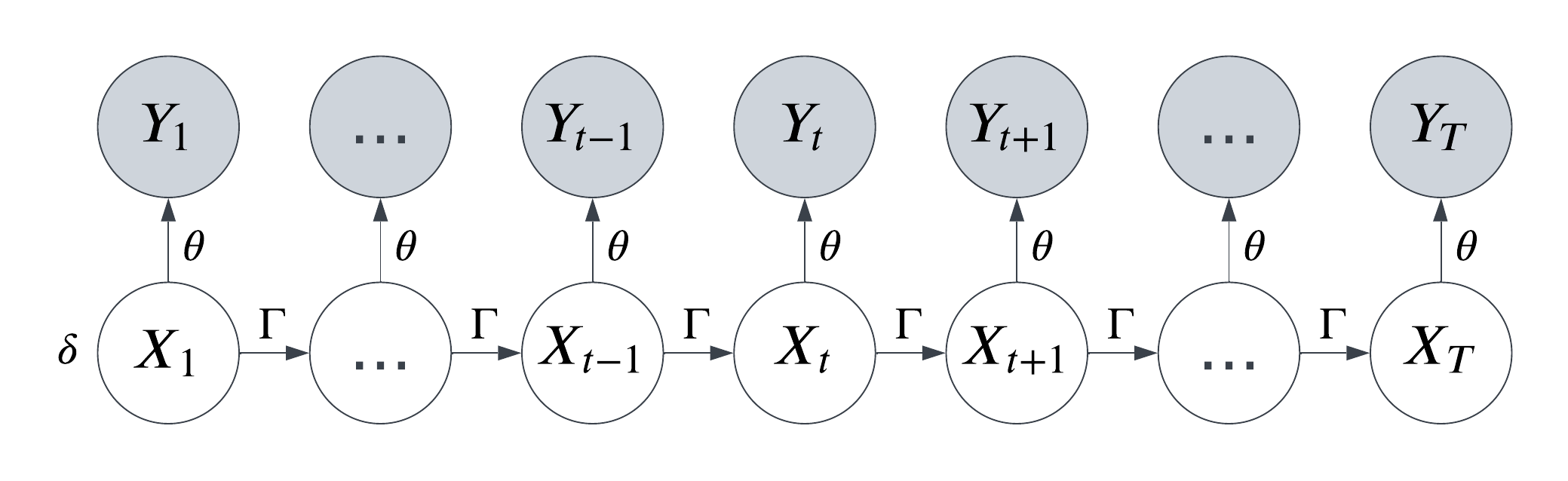}
    \caption{Graphical representation of an HMM. $X_t$ corresponds to an unobserved latent state at time $t$ whose distribution is described by a Markov chain. $Y_t$ corresponds to an observation at time $t$, where $Y_t$ given all other observations $\bfY \setminus \{Y_t\}$ and hidden states $\bfX$ depends only on $X_t$.}
    \label{fig:HMM}
\end{figure}

If $X_t=i$, then we denote the conditional density or probability mass function of $Y_t$ as $f^{(i)}(\cdot ; \theta^{(i)})$, where $\theta^{(i)}$ are the parameters describing the state-dependent distribution of $Y_t$. The collection of all state-dependent parameters is $\bftheta = \{\theta^{(i)}\}_{i=1}^N$. For brevity, we denote the full set of parameters as $\bfphi = \{\bftheta,\bfeta\}$. Figure (\ref{fig:HMM}) shows an HMM as a graphical model.

The joint likelihood of an HMM given observations $\bfY$ and latent states $\bfX$ is
\begin{equation}
    p(\bfX,\bfY;\bfphi) = \delta^{(X_1)} f^{(X_1)}(Y_1; \theta^{(X_1)}) \prod_{t=2}^T \Gamma^{(X_{t-1},X_t)} f^{(X_t)}(Y_t; \theta^{(X_t)}).
    \label{eqn:like}
\end{equation}
Alternatively, the marginal likelihood of the observed data $\bfY$ alone is 
\begin{equation}
    p(\bfY;\bfphi) = \bfdelta P(Y_1;\bftheta) \prod_{t=2}^T \bfGamma P(Y_t;\bftheta) \mathbf{1}^\top_N,
    \label{eqn:like_marginal}
\end{equation}
where $\mathbf{1}_N$ is an $N$-dimensional row vector of ones and $P(Y_t;\bftheta)$ is an $N \times N$ diagonal matrix with entry $(i,i)$ equal to $f^{(i)}(Y_t; \theta^{(i)})$. For a more complete introduction to HMMs, see \citet{Zucchini:2016}.
\subsection{State Decoding}

One appealing feature of HMMs is that it is simple to determine the distribution of a given hidden state ($X_t$) conditioned on the set of observations $\bfY$. Assuming that the HMM parameters $\bfphi$ are fixed, define the probability density of the observations between times $s$ and $t$ as $p(Y_{s:t};\bfphi)$. Likewise, define \textit{forward probabilities} $\alpha^{(i)}_t = p(Y_{1:t},X_t = i;\bfphi)$ (for $i = 1,\ldots,N$ and $t = 1,\ldots,T$) and \textit{backward probabilities} $\beta^{(i)}_t = p(Y_{(t+1):T} \mid X_t = i;\bfphi)$ (for $i = 1,\ldots,N$ and $t = 1,\ldots,T-1$). By convention, $\beta^{(i)}_T = 1$ for $i = 1,\ldots,N$. We thus define the row vectors $\bfalpha_t = \begin{pmatrix} \alpha_t^{(1)} & \cdots & \alpha_t^{(N)} \end{pmatrix}$ and $\bfbeta_t = \begin{pmatrix} \beta_t^{(1)} & \cdots & \beta_t^{(N)} \end{pmatrix}$ . Both $\bfalpha_t$ and $\bfbeta_t$ can be calculated using the following recursions: 

\begin{gather}
    \bfalpha_1 = \bfdelta ~ P(Y_1;\bftheta), \qquad 
    \bfalpha_t = \bfalpha_{t-1} ~ \bfGamma ~ P(Y_t;\bftheta), \quad t = 2,\ldots,T, \label{eqn:alpha} \\
    \bfbeta^\top_T = \mathbf{1}_N^\top, \qquad
    \bfbeta^\top_t = \bfGamma ~ P(Y_{t+1};\bftheta) ~ \bfbeta^\top_{t+1}, \quad t = 1,\ldots,T-1. \label{eqn:beta}
\end{gather}
We define conditional probabilities $\gamma_t^{(i)} = \bbP(X_t = i \mid \bfY ~;~ \bfphi)$ and $\xi_t^{(i,j)} = \bbP(X_{t-1} = i, X_t = j \mid \bfY ~;~ \bfphi)$. We also define the row vector $\bfgamma_t = \begin{pmatrix} \gamma_t^{(1)} & \cdots & \gamma_t^{(N)} \end{pmatrix}$ and the matrix 

\begin{equation*}
    \bfxi_t = \begin{pmatrix} 
    \xi_t^{(1,1)} & \cdots & \xi_t^{(1,N)} \\
    \vdots & \ddots & \vdots \\
    \xi_t^{(N,1)} & \cdots & \xi_t^{(N,N)} \\
    \end{pmatrix}, \qquad t = 2,\ldots,T. 
\end{equation*}
Both $\bfgamma_t$ and $\bfxi_t$ can be calculated from $\bfalpha_{t-1}$, $\bfalpha_t$, $\bfbeta_t$, $\bfGamma$, and $\bftheta$. Let $\text{diag}(\cdot)$ map a row vector to the diagonal matrix with that row vector as its diagonal. Then,

\begin{gather}
    \gamma^{(i)}_t = \frac{\alpha^{(i)}_{t} ~ \beta^{(i)}_{t}}{\bfalpha_{t} ~ \bfbeta_t^\top}, \qquad \bfgamma_t = \frac{\bfalpha_t ~ \text{diag}(\bfbeta_t)}{\bfalpha_t ~ \bfbeta_t^\top} \label{eqn:gamma}, \\
    \xi_{t}^{(i,j)} = \frac{\alpha_{t-1}^{(i)} ~ \Gamma^{(i,j)} ~ f^{(j)}(y_{t};\theta^{(j)}) ~ \beta_{t}^{(j)}}{\bfalpha_{t-1} ~ \bfGamma ~ P(y_{t};\bftheta) ~ \bfbeta_{t}^\top}, \qquad \bfxi_t = \frac{\text{diag}(\bfalpha_{t-1}) ~ \bfGamma ~ P(y_t;\bftheta) ~ \text{diag}(\bfbeta_t)}{\bfalpha_{t-1} ~ \bfGamma ~ P(y_{t};\bftheta) ~ \bfbeta_{t}^\top} \label{eqn:xi}.
\end{gather}
For shorthand, we define the sets $\{\bfalpha, \bfbeta, \bfgamma, \bfxi\} = \{\bfalpha_t, \bfbeta_t, \bfgamma_t, \bfxi_t\}_{t=1}^T$ to summarize the conditional probabilities for all $t$. In future sections, when $\bfphi$ is not fixed (e.g. during the parameter estimation procedures), we add an argument to the conditional probabilities and write $\{\bfalpha(\bfphi), \bfbeta(\bfphi), \bfgamma(\bfphi), \bfxi(\bfphi)\} = \{\bfalpha_t(\bfphi), \bfbeta_t(\bfphi), \bfgamma_t(\bfphi), \bfxi_t(\bfphi)\}_{t=1}^T$ to highlight the dependence on $\bfphi$. 

\subsection{The Baum-Welch Algorithm}

The Baum-Welch algorithm is a specific instance of the EM algorithm used to estimate the parameters of an HMM. At iteration $k$ of the EM algorithm, denote the current parameter estimates as $\bfphi_{k}$. One iteration of the EM algorithm consists of an expectation (or E) step, followed by a maximization (or M) step. For the E step, the function value $Q(\bfphi \mid \bfphi_{k})$ is defined as the expected value of the joint log-likelihood $\log p(\bfX,\bfY; \bfphi)$ taken with respect to $\bfX$, where $\bfX$ has conditional probability mass function $p(\bfX \mid \bfY ; \bfphi_{k})$. For the M step, the next parameter estimate $\bfphi_{k+1}$ is found by maximizing $Q(\bfphi \mid \bfphi_{k})$ with respect to $\bfphi$:

\begin{gather}
    Q(\bfphi \mid \bfphi_{k}) = \bbE_{\bfphi_{k}}\left[\log p(\bfX,\bfY;\bfphi) \mid \bfY \right] \label{eqn:Q}, \\
    \bfphi_{k+1} = \argmax_{\bfphi} Q(\bfphi \mid \bfphi_{k}). \label{eqn:BW_update}
\end{gather}
For notational convenience, we occasionally denote the set of conditional probabilities $\{\bfalpha_t(\bfphi_k), \bfbeta_t(\bfphi_k), \bfgamma_t(\bfphi_k), \bfxi_t(\bfphi_k)\}$ as $\{\bfalpha_{k,t},\bfbeta_{k,t},\bfgamma_{k,t},\bfxi_{k,t}\}$. Substituting Equation (\ref{eqn:like}) into Equation (\ref{eqn:Q}) and performing some algebra yields a closed form expression for $Q$: 
\begin{equation}
    Q(\bfphi \mid \bfphi_{k})
    = \sum_{i=1}^N \gamma^{(i)}_{k,1} \log \delta^{(i)}(\bfeta) + \sum_{t = 1}^T \sum_{i=1}^N \gamma^{(i)}_{k,t} \log f^{(i)}(y_t;\theta^{(i)}) + \sum_{t=2}^{T} \sum_{i=1}^N \sum_{j=1}^N \xi^{(i,j)}_{k,t} \log \Gamma^{(i,j)}(\bfeta).
    \label{eqn:Q_sum}
\end{equation}
The conditional probabilities $\gamma_{k,t}^{(i)}$ and $\xi_{k,t}^{(i,j)}$ thus act as weights for $\log \delta^{(i)}(\bfeta)$, $\log f^{(i)}(y_t;\theta^{(i)})$, and $\log \Gamma^{(i,j)}(\bfeta)$ for $i,j = 1,\ldots,N$. We thus refer to $\bfgamma$ and $\bfxi$ as weights in future sections. Detailed pseudocode for the E and the M step are given in Algorithms (\ref{alg:E}) and (\ref{alg:EM}) below.
\begin{algorithm}
\caption{\texttt{E-step}($\bfphi$)}\label{alg:E}
\begin{algorithmic}[1]
\Require Parameter value $\bfphi = \{\bftheta,\bfeta\}$.
\State $\bfdelta = \bfdelta(\bfeta), \quad \bfGamma = \bfGamma(\bfeta)$
\State $\bfalpha_1 = \bfdelta ~ P(y_1;\bftheta)$
\State $\bfbeta^\top_T = \mathbf{1}_N^\top$
\For{$t = 2,\ldots,T$}
    \State $$\bfalpha_t = \bfalpha_{t-1} ~ \bfGamma ~ P(y_t;\bftheta), \qquad \bfbeta^\top_{T-t+1} = \bfGamma ~ P(y_{T-t+2};\bftheta) ~ \bfbeta^\top_{T-t+2}$$
\EndFor
\State $\bfgamma_1 = \frac{\bfalpha_1 ~ \text{diag}(\bfbeta_1)}{\bfalpha_1 ~ \bfbeta_1^\top}$
\For{$t = 2,\ldots,T$}
    \State $$\bfgamma_t = \frac{\bfalpha_t ~ \text{diag}(\bfbeta_t)}{\bfalpha_t ~ \bfbeta_t^\top}, \qquad \bfxi_t = \frac{\text{diag}(\bfalpha_{t-1}) ~ \bfGamma ~ P(y_t;\bftheta) ~ \text{diag}(\bfbeta_t)}{\bfalpha_{t-1} ~ \bfGamma ~ P(y_{t};\bftheta) ~ \bfbeta_{t}^\top}$$
\EndFor
\State \Return $\{\bfalpha_t,\bfbeta_t,\bfgamma_t,\bfxi_t\}_{t=1}^T$
\end{algorithmic}
\end{algorithm}
\begin{algorithm}
\caption{\texttt{Baum-Welch}$(\bfphi_0,K)$}\label{alg:EM}
\begin{algorithmic}[1]
\Require Initial parameter values $\bfphi_0$, number of iterations $K$
\For{$k = 0,\ldots,K-1$}
    \State $\{\bfalpha_{k,t},\bfbeta_{k,t},\bfgamma_{k,t},\bfxi_{k,t}\}_{t=1}^T = \texttt{E-step}(\bfphi_{k})$ \Comment{E step} 
    \State \Comment{M step} \small $$\bfphi_{k+1} = \argmax_{\{\bftheta,\bfeta\}} \sum_{i=1}^N \gamma^{(i)}_{k,1} \log \delta^{(i)}(\bfeta) + \sum_{t = 1}^T \sum_{i=1}^N \gamma^{(i)}_{k,t} \log f^{(i)}(y_t;\theta^{(i)}) + \sum_{t=2}^{T} \sum_{i=1}^N \sum_{j=1}^N \xi_{k,t}^{(i,j)} \log \Gamma^{(i,j)}(\bfeta)$$ \normalsize
\EndFor
\State \Return $\bfphi_K$
\end{algorithmic}
\end{algorithm}
In some simple scenarios, the maximization problem in Equation (\ref{eqn:BW_update}) above has a closed-form solution. However, this maximization problem is not always straightforward and often requires numerical maximization techniques. We thus review different methods for numerical maximization via stochastic optimization. 

\subsection{Stochastic Optimization}
\label{subsec:stoch_optim}

Stochastic optimization involves a class of optimization methods that use random variables to maximize or minimize an objective function. We focus on optimization methods to minimize objective functions that can be written as a sum of many terms \citep{Robbins:1951}, namely to solve
\begin{equation}
    \bfphi^* = \argmin_{\bfphi} F(\bfphi), \quad \text{where} \quad F(\bfphi) = \frac{1}{T}\sum_{t = 1}^T F_t(\bfphi).
    \label{eqn:stoch_opt}
\end{equation}

We denote the parameter values at step $m$ of an optimization scheme as $\bfphi^{(m)} = \{\bftheta^{(m)},\bfeta^{(m)}\}$ to distinguish between the parameter values $\bfphi_k$ at iteration $k$ of the Baum-Welch algorithm. The bold parameters $\bftheta^{(m)}$ and $\bfeta^{(m)}$ correspond to optimization step $m$ and are not to be confused with $\theta^{(i)}$ and $\eta^{(i)}$, which correspond to hidden state $i$ of the HMM. Standard gradient descent at a given step $m$ with step size $\lambda$ updates the parameter value $\bfphi^{(m)}$ by moving in the direction of the negative gradient of $F$. Formally, the update step is
$\bfphi^{(m+1)} = \bfphi^{(m)} - \lambda \nabla F(\bfphi^{(m)})$, or $\bfphi^{(m+1)} = \bfphi^{(m)} - (\lambda/T) \sum_{t=1}^T \nabla F_t(\bfphi^{(m)})$ for our problem.
The step size $\lambda$ is user-defined. It should be large enough so $\bfphi^{(m+1)}$ moves quickly towards a minimum of $F$, but not so large that $\bfphi^{(m+1)}$ ``overshoots" the minimum. This update requires evaluating a gradient for all $t = 1,\ldots,T$, which can be prohibitively expensive if $T$ is large.

In contrast, stochastic gradient descent (SGD) updates $\bfphi$ using
$\bfphi^{(m+1)} = \bfphi^{(m)} - \lambda_m \nabla F_{t_m}(\bfphi^{(m)}),$
where $t_m \in \{1,\ldots,T\}$ is selected uniformly at random at step $m$ of the algorithm \citep{Robbins:1951}. Stochastic gradient descent reduces the amount of time between updates by using an unbiased estimate of the gradient to update $\bfphi^{(m)}$. However, the gradient estimates can have very high variance, so stochastic gradient descent requires that the step size $\lambda_m$ is smaller than that of full gradient descent. The step size must also decay to zero as $m \to \infty$ to ensure convergence. Further, SGD has slower convergence rates than full gradient descent \citep{Schmidt:2017}.

Variance-reduced stochastic optimization techniques such as stochastic average gradient descent (SAG) \citep{Schmidt:2017}, stochastic variance reduced gradient descent (SVRG) \citep{Johnson:2013}, and SAGA \citep{Defazio:2014} enjoy the speed of stochastic gradient descent as well as the convergence rates of full gradient descent. These algorithms involve storing gradient approximations at each gradient step $m$, $\widehat \nabla F_{t}^{(m)}$ for $t = 1,\ldots,T$, whose average approximates the full gradient $\nabla F(\bfphi^{(m)})$. The gradient approximations are updated at various stages in the optimization algorithm and are used to reduce the variance of the full gradient estimate. For example, SVRG and SAGA update $\bfphi^{(m)}$ via
$\bfphi^{(m+1)} = \bfphi^{(m)} - \lambda_m \left[\nabla F_{t_m}(\bfphi^{(m)}) - \widehat \nabla F_{t_m}^{(m)} + \widehat \nabla F^{(m)} \right],$
where $t_m \in \{1,\ldots,T\}$ is chosen uniformly at random and
$\widehat \nabla F^{(m)} = (1/T) \sum_{t=1}^T \widehat \nabla F_{t}^{(m)}$.
Like SGD, this stochastic update has the same expectation as that of standard gradient descent, but it empirically exhibits lower variance than SGD and it is guaranteed to converge without decaying the step size $\lambda_m$ (under certain regularity conditions). After updating the parameters at step $m$, SAGA updates the gradient approximation $\widehat \nabla F_{t_m}^{(m+1)}$ and recalculates the resulting gradient average $\widehat \nabla F^{(m+1)}$.
Algorithm (\ref{alg:VRSO}) outlines SVRG and SAGA in pseudocode, and we denote it as \textit{variance-reduced stochastic optimization}, or $\texttt{VRSO}$. 

\begin{algorithm}
\caption{\texttt{VRSO}$(F,\bfphi^{(0)},\lambda,A,M)$}\label{alg:VRSO}
\begin{algorithmic}[1]
\Require Loss function $F = \frac{1}{T}\sum_{t=1}^T F_t$, initial value $\bfphi^{(0)}$, step size $\lambda$, algorithm $A \in \{\text{SVRG, SAGA}\}$, and number of iterations $M$.
\vspace{5pt}
\For{$t = 1,\ldots,T$} \Comment{initialize gradients}
\State $\widehat \nabla F_{t}^{(0)} = \nabla F_t (\bfphi^{(0)})$
\EndFor
\State $\widehat \nabla F^{(0)} = (1/T) \sum_{t=1}^T \widehat \nabla F_{t}^{(0)}$
\For{$m = 0,\ldots,M-1$}:
    \State Pick $t_m \in \{1,\ldots,T\}$ uniformly at random.
    \State \Comment{update parameters}
    \begin{gather}
        \bfphi^{(m+1)} = \bfphi^{(m)} - \lambda \left[\nabla F_{t_m}(\bfphi^{(m)}) - \widehat \nabla F_{t_m}^{(m)} + \widehat \nabla F^{(m)} \right]
        \label{eqn:SAGA_update0}
    \end{gather}
    \State $\widehat \nabla F_{t}^{(m+1)} = \widehat \nabla F_{t}^{(m)} \enspace$ for $t = 1,\ldots,T$ 
    \Comment{update gradient approximations and average}
    \If{$A$ = SAGA}:
        \begin{gather}
            \widehat \nabla F_{t_m}^{(m+1)} = \nabla F_{t_m}(\bfphi^{(m)}) \\
            \widehat \nabla F^{(m+1)} = \widehat \nabla F^{(m)} + \frac{1}{T} \left(\widehat \nabla F_{t_m}^{(m+1)} - \widehat \nabla F_{t_m}^{(m)}\right)
        \end{gather}
    \EndIf
\EndFor
\State \Return $\bfphi^{(M)}$
\end{algorithmic}
\end{algorithm}

SAGA involves running Algorithm (\ref{alg:VRSO}) until convergence. However, the gradient approximations $\widehat \nabla F_{t}^{(m)}$ are never updated when using SVRG within Algorithm (\ref{alg:VRSO}). As such, SVRG requires repeatedly running Algorithm (\ref{alg:VRSO}) with $M$ scaling approximately with $T$ so the set of gradient approximations remains up-to-date.

\section{Stochastic Optimization for HMM Inference}
Both the E step and the M step of the Baum-Welch algorithm are expensive when the length of the observation sequence ($T$) is large. The E step is expensive because $\bfgamma_t(\bfphi_{k})$ and $\bfxi_t(\bfphi_{k})$ must be calculated for $t = 1,\ldots,T$ to define $Q(\bfphi \mid \bfphi_k)$. If closed-form solutions to (\ref{eqn:BW_update}) are not readily available, then the M step is also expensive because evaluating full gradients of $Q(\bfphi \mid \bfphi_{k})$ takes $\calO(T)$ time. In this section, we introduce an original algorithm that speeds up both the expensive M step as well as the expensive E step of the Baum-Welch algorithm. 

\subsection{Variance-Reduced Stochastic M Step}
\label{subsec:stoch_M}

To speed up the expensive M step, we notice from Equation (\ref{eqn:Q_sum}) that $Q$ is a large sum and thus implement variance-reduced stochastic optimization. It is straightforward to re-frame the M step of iteration $k$ of the Baum Welch algorithm from Equation (\ref{eqn:BW_update}) so it looks like the minimization problem from Equation (\ref{eqn:stoch_opt}). To do so, we define $\bfxi_1 = \emptyset$ and the loss function $F(\cdot \mid \bfgamma,\bfxi)$ as follows:

\begin{gather}
    F(\bfphi \mid \bfgamma, \bfxi) = \frac{1}{T}\sum_{t=1}^T F_t(\bfphi \mid \bfgamma_t , \bfxi_t), \qquad \text{where} \\
    F_1(\bfphi \mid \bfgamma_1,\bfxi_1) = - \sum_{i=1}^N \gamma^{(i)}_1 \log f^{(i)}(y_t;\theta^{(i)}) - \sum_{i=1}^N \gamma^{(i)}_1 \log \delta^{(i)}(\bfeta), \\
    F_t(\bfphi \mid \bfgamma_t , \bfxi_t) = - \sum_{i=1}^N \gamma^{(i)}_t \log f^{(i)}(y_t;\theta^{(i)}) - \sum_{i=1}^N \sum_{j=1}^N \xi^{(i,j)}_t \log \Gamma^{(i,j)}(\bfeta), \quad t = 2, \ldots, T.
\end{gather}
The two functions $F$ and $Q$ are closely related to one another, as $F(\bfphi \mid \bfgamma(\bfphi_k), \bfxi(\bfphi_k)) = - Q(\bfphi \mid \bfphi_k) / T$. However, we make a distinction between the two to bridge the gap between existing EM literature (which uses $Q$) and stochastic optimization literature (which uses $F$). At any iteration $k$ of the EM algorithm, the loss function $F(\cdot \mid \bfgamma(\bfphi_k),\bfxi(\bfphi_k))$ can be minimized using Algorithm (\ref{alg:VRSO}). 

There are additional reasons to use SAGA and SVRG within the Baum-Welch algorithm beyond the standard benefits of variance-reduced stochastic optimization. Traditionally, SAGA is more memory intensive than SVRG because the gradient at every index must be stored. However, the Baum-Welch algorithm involves storing weights for each time index $t$ to define $F(\cdot \mid \bfgamma(\bfphi_k), \bfxi(\bfphi_k))$, so storing each gradient for SAGA is not considerably more memory intensive than the Baum-Welch algorithm itself. Alternatively, SVRG can be more computationally expensive than SAGA partially because it requires periodically re-calculating the full gradient approximation $\widehat \nabla F^{(0)}$, and this involves a full pass of the underlying data set. However, the E step of the Baum-Welch algorithm also involves a full pass of the data set, so using SVRG is not considerably more computationally expensive than the Baum-Welch algorithm itself. In this way, using either SAGA or SVRG in the M step adds minimal computational and memory complexity to the Baum-Welch algorithm.

\subsection{Partial E Step within the M step}
\label{subsec:stoch_E}

Variance-reduced stochastic optimization reduces the computational cost of the M step, but the E step itself still has a time complexity of $\calO(T)$, which can be prohibitive for large $T$. To decrease this computational burden, \citet{Neal:1998} justify a partial E step within the EM algorithm for general latent variable models. However, they assume that the optimization of the M step has a closed-form solution. We use their method as inspiration and add a partial E step to the stochastic M step of the Baum-Welch algorithm. 

Consider running one iteration of our version of the Baum-Welch algorithm with an initial parameter estimate $\bfphi^{(0)}$. The E step involves calculating the conditional probabilities $\bfgamma(\bfphi^{(0)})$ and $\bfxi(\bfphi^{(0)})$, and the M step involves running variance-reduced stochastic optimization with loss function $F(\cdot \mid \bfgamma(\bfphi^{(0)}),\bfxi(\bfphi^{(0)}))$ and initial parameter value $\bfphi^{(0)}$. Now, suppose $\bfphi^{(m)}$ is to be updated using a gradient estimate using a random observation index $t_m$. The function $F_{t_m}(\cdot \mid \bfgamma_{t_m}(\bfphi^{(0)}),\bfxi_{t_m}(\bfphi^{(0)}))$ depends on $\bfgamma_{t_m}(\bfphi^{(0)})$ and $\bfxi_{t_m}(\bfphi^{(0)})$, each of which are vectors of conditional probabilities given $\bfphi^{(0)}$. However, $\bfphi^{(0)}$ is an out-of-date parameter estimate since the current parameter estimate is $\bfphi^{(m)}$. Therefore, it is natural to update $\bfgamma_{t_m}$ and $\bfxi_{t_m}$ and redefine $F_{t_m}(\cdot \mid \bfgamma_{t_m}, \bfxi_{t_m})$ before calculating $\bfphi^{(m+1)}$. 

A naive method would be to calculate the new conditional probabilities $\bfgamma_{t_m}(\bfphi^{(m)})$ and $\bfxi_{t_m}(\bfphi^{(m)})$ and then update $F_{t_m}$ as $F_{t_m}(\cdot \mid \bfgamma_{t_m}(\bfphi^{(m)}), \bfxi_{t_m}(\bfphi^{(m)}))$. This would ensure that $\bfgamma_{t_m}$ and $\bfxi_{t_m}$ are completely up-to-date, but evaluating $\bfgamma_{t_m}(\bfphi^{(m)})$ and $\bfxi_{t_m}(\bfphi^{(m)})$ takes $\calO(TN^2)$ time and requires a full E step. Instead, our goal is to update $\bfgamma_{t_m}$ and $\bfxi_{t_m}$ in a way that does not scale with $T$.

To this end, we define the mappings $\widetilde \bfalpha_t$, $\widetilde \bfbeta_t$, $\widetilde \bfgamma_t$, and $\widetilde \bfxi_t$ for $t = 1,\ldots,T$ similarly to Equations (\ref{eqn:alpha}--\ref{eqn:xi}):

\begin{gather}
    \widetilde \bfalpha_1(\mathbf{a},\bfphi) = \bfdelta ~ P(y_1;\bftheta), \qquad \widetilde \bfalpha_t(\mathbf{a},\bfphi) = \mathbf{a} ~ \bfGamma ~ P(y_t;\bftheta), \quad t = 2,\ldots,T, \label{eqn:tilde_alpha} \\
    \widetilde \bfbeta^\top_T(\mathbf{b},\bfphi) = \mathbf{1}_N^\top, \qquad \widetilde \bfbeta^\top_t(\mathbf{b},\bfphi) = \bfGamma ~ P(y_{t+1};\bftheta) ~ \mathbf{b}^\top, \quad t = 1,\ldots,T-1, \label{eqn:tilde_beta} \\ \nonumber \\
    \widetilde \bfgamma_t(\mathbf{a},\mathbf{b}) = \frac{\mathbf{a} ~ \text{diag}(\mathbf{b})}{\mathbf{a} ~ \mathbf{b}^T}, \quad t = 1,\ldots,T, \label{eqn:tilde_gamma} \\ \nonumber \\
    \widetilde \bfxi_{t}(\mathbf{a},\mathbf{b},\bfphi) = \frac{\text{diag}(\mathbf{a}) ~ \bfGamma ~ P(y_t;\bftheta) ~ \text{diag}(\mathbf{b})}{\mathbf{a} ~ \bfGamma ~ P(y_{t};\bftheta) ~ \mathbf{b}^\top}, \quad t = 2,\ldots,T \label{eqn:tilde_xi},
\end{gather}
all of which take $\calO(N^2)$ time to compute. Further, at the beginning of the M step we define conditional probability \textit{approximations} $\widehat \bfalpha_{t}^{(0)} = \bfalpha_t(\bfphi^{(0)})$, $\widehat \bfbeta_{t}^{(0)} = \bfbeta_t(\bfphi^{(0)})$, $\widehat \bfgamma_{t}^{(0)} = \bfgamma_t(\bfphi^{(0)})$, and $\widehat \bfxi_{t}^{(0)} = \bfxi_t(\bfphi^{(0)})$ for $t = 1,\ldots,T$. This is simply the E step of the Baum-Welch algorithm and takes $\calO(TN^2)$ time to compute. Then, at any given step $m$ of the stochastic M step, we update $F_{t_m}$ by first updating $\widehat \bfalpha_{t_m}^{(m+1)} = \widetilde \bfalpha_{t_m}\left(\widehat \bfalpha_{t_m-1}^{(m)} ~,~ \bfphi^{(m)}\right)$ and $\widehat \bfbeta_{t_m}^{(m+1)} = \widetilde \bfbeta_{t_m}\left(\widehat \bfbeta_{t_m+1}^{(m)} ~,~ \bfphi^{(m)}\right)$, followed by $\widehat \bfgamma_{t_m}^{(m+1)} = \widetilde \bfgamma_{t_m}\left(\widehat \bfalpha_{t_m}^{(m+1)} ~,~ \widehat \bfbeta_{t_m}^{(m+1)}\right)$ and $\widehat \bfxi_{t_m}^{(m+1)} = \widetilde \bfxi_{t_m}\left(\widehat \bfalpha_{t_m-1}^{(m+1)} ~,~ \widehat \bfbeta_{t_m}^{(m+1)} ~,~ \bfphi^{(m)}\right)$. Finally, the loss function at index $t_m$ can be defined as $F_{t_m}\left(\cdot ~ \Big | ~ \widehat \bfgamma_{t_m}^{(m+1)},\widehat \bfxi_{t_m}^{(m+1)}\right)$. Updating $F_{t_m}$ in this way take a total of $\calO(N^2)$ time, which accomplishes a parameter update step that does not scale with $T$. Algorithm (\ref{alg:VRSO-PE}) outlines the M step of the Baum-Welch algorithm with a partial E step integrated in. 

The partial E step detailed above is closely related to belief propagation, an algorithm that calculates conditional probabilities of variables within graphical models \citep{Pearl:1982}. In fact, the E step of the Baum-Welch algorithm is a specific instance of belief propagation, where Equations (\ref{eqn:tilde_alpha} -- \ref{eqn:tilde_xi}) above correspond to ``passing messages" within the graphical model. Belief propagation can only perform exact inference on acyclic graphical models (including HMMs), but a generalization called loopy belief propagation can perform approximate inference on general graphical models \citep{Pearl:1988}. Practitioners balance approximation error and computational complexity to decide how long to run loopy belief propagation. Likewise, we consider computational complexity when running belief propagation and evaluate Equations (\ref{eqn:tilde_alpha} -- \ref{eqn:tilde_xi}) only once to approximate $\bfgamma_{t_m}(\bfphi^{(m)})$ and $\bfxi_{t_m}(\bfphi^{(m)})$. This forms the basis of the partial E step within the stochastic M step of our modified Baum-Welch algorithm.

\begin{algorithm}
\caption{\texttt{VRSO-PE}$(\{\widehat \bfalpha_t^{(0)}, \widehat \bfbeta_t^{(0)}, \widehat \bfgamma_t^{(0)}, \widehat \bfxi_t^{(0)}\}_{t=1}^T,\bfphi^{(0)},\lambda,A,P,M)$}\label{alg:VRSO-PE}
\begin{algorithmic}[1]
\Require Initial conditional probability approximations $\{\widehat \bfalpha^{(0)}, \widehat \bfbeta^{(0)}, \widehat \bfgamma^{(0)}, \widehat \bfxi^{(0)}\}$, initial parameters $\bfphi^{(0)}$, step size $\lambda$, algorithm $A \in \{\text{SVRG, SAGA}\}$, whether to do a partial-E step $P \in \{\texttt{True,False}\}$, and number of iterations $M$.
\For{$t=1,\ldots,T$} \Comment{initialize gradients}
    \State $\widehat \nabla F^{(0)}_{t} = \nabla F_t\left(\bfphi^{(0)} ~ \Big | ~ \widehat \bfgamma^{(0)}_{t}, \widehat \bfxi^{(0)}_{t}\right)$ 
\EndFor
\State $\widehat \nabla F^{(0)} = (1/T) \sum_{t=1}^T \widehat \nabla F^{(0)}_{t}$
\For{$m = 0,\ldots,M-1$}:
    \State Pick $t_m \in \{1,\ldots,T\}$ uniformly at random.
    \State $\left\{\widehat \bfalpha^{(m+1)}_t, \widehat \bfbeta^{(m+1)}_t, \widehat \bfgamma^{(m+1)}_t, \widehat \bfxi^{(m+1)}_t\right\} = \left\{\widehat \bfalpha^{(m)}_{t}, \widehat \bfbeta^{(m)}_{t}, \widehat \bfgamma^{(m)}_{t}, \widehat \bfxi^{(m)}_{t}\right\} \enspace$ for $t = 1,\ldots,T$.
    \If{$P = \texttt{True}$} \Comment{partial E step}
    \State $\widehat \bfalpha_{t_m}^{(m+1)} = \widetilde \bfalpha_{t_m}\left(\widehat \bfalpha_{t_m-1}^{(m)},\bfphi^{(m)}\right), \quad \widehat \bfbeta_{t_m}^{(m+1)} = \widetilde \bfbeta_{t_m}\left(\widehat \bfbeta_{t_m+1}^{(m)},\bfphi^{(m)}\right)$ 
    \State $\widehat \bfgamma_{t_m}^{(m+1)} = \widetilde \bfgamma_{t_m}\left(\widehat \bfalpha_{t_m}^{(m+1)},\widehat \bfbeta_{t_m}^{(m+1)}\right), 
    \quad \widehat \bfxi_{t_m}^{(m+1)} = \widetilde \bfxi_{t_m}\left(\widehat \bfalpha_{t_m-1}^{(m+1)},\widehat \bfbeta_{t_m}^{(m+1)},\bfphi^{(m)}\right)$
    \EndIf
    \State \Comment{update parameters}
    \begin{gather}
        \bfphi^{(m+1)} = \bfphi^{(m)} - \lambda \left[\nabla F_{t_m}\left(\bfphi^{(m)} ~ \Big | ~ \widehat \bfgamma_{t_m}^{(m+1)}, \widehat \bfxi_{t_m}^{(m+1)}\right) - \widehat \nabla F^{(m)}_{t_m} + \widehat \nabla F^{(m)} \right]
    \end{gather}
    \State $\widehat \nabla F_{t}^{(m+1)} = \widehat \nabla F_{t}^{(m)} \enspace$ for $t = 1,\ldots,T$ \Comment{update gradients}
    \If{$A$ = SAGA}:
        \begin{gather}
            \widehat \nabla F_{t_m}^{(m+1)} = \nabla F_{t_m}\left(\bfphi^{(m)} ~ \Big | ~ \widehat \bfgamma_{t_m}^{(m+1)}, \widehat \bfxi_{t_m}^{(m+1)}\right), \\
            \widehat \nabla F^{(m+1)} = \widehat \nabla F^{(m)} + \frac{1}{T} \left( \widehat \nabla F_{t_m}^{(m+1)} - \widehat \nabla F_{t_m}^{(m)}\right).
        \end{gather}
    \EndIf
\EndFor
\State \Return $\bfphi^{(M)}$
\end{algorithmic}
\end{algorithm}

\subsection{Full Algorithm}

In principal, it is possible to run Algorithm (\ref{alg:VRSO-PE}) alone without ever performing a full E step. However, if no partial E step is used (i.e. $P = \texttt{False}$) or if SVRG is used as the optimization algorithm, then either the conditional probability approximations $\left\{\widehat \bfalpha_t^{(m)}, \widehat \bfbeta_t^{(m)}, \widehat \bfgamma_t^{(m)}, \widehat \bfxi_t^{(m)} \right\}_{t=1}^T$ or the gradient approximations $\left\{\widehat \nabla F_{t}^{(m)} \right\}_{t=1}^T$ will not be updated and become out-of-date. To avoid this issue, Algorithm (\ref{alg:EM-VRSO}) combines the M step defined in Algorithm (\ref{alg:VRSO-PE}) with a full E step to complete our new Baum-Welch algorithm for HMMs.

\begin{algorithm}
\caption{\texttt{EM-VRSO}$(\bfphi_0,\lambda, A, P, M, K)$ (Version 1)}\label{alg:EM-VRSO}
\begin{algorithmic}[1]
\Require Initial parameters ($\bfphi_{0}$), step size ($\lambda$), algorithm $A \in \{\text{SVRG, SAGA}\}$, whether to do a partial E step $P \in \{\texttt{True,False}\}$, iterations per update ($M$), and number of updates ($K$).
\For{$k = 0,\ldots,K-1$}
\State $\{\bfalpha_{k,t}, \bfbeta_{k,t}, \bfgamma_{k,t}, \bfxi_{k,t}\}_{t=1}^T = \texttt{E-step}(\bfphi_{k})$ \Comment{E step}
\State $\ell \gets 0$ \Comment{M step}
\While{$\ell = 0$ or $\log p(\bfy; \bfphi_{k,\ell}) < \log p(\bfy;\bfphi_{k})$} 
\State $\ell \gets \ell+1$
\State $\bfphi_{k,\ell} = \texttt{VRSO-PE}(\{\bfalpha_{k,t}, \bfbeta_{k,t}, \bfgamma_{k,t}, \bfxi_{k,t}\}_{t=1}^T,\bfphi_k,\lambda,A,P,M)$
\EndWhile
\State $\bfphi_{k+1} = \bfphi_{k,\ell}$
\EndFor
\State \Return $\bfphi_K$
\end{algorithmic}
\end{algorithm}

There are two versions of \texttt{EM-VRSO}. Version 1 (Algorithm \ref{alg:EM-VRSO}), requires the likelihood to not decrease (i.e. $\log p(\bfy;\bfphi_{k,\ell}) \geq \log p(\bfy;\bfphi_{k})$) in order to exit the while loop of the M step. Version 2 (Algorithm 6) requires the likelihood to \textit{strictly} increase by some threshold to exit the while loop of the M step. We use version 2 to prove theoretical results, but the strict threshold relies on values that are usually not known in practice. Therefore, we use version 1 in our simulation and case studies and defer version 2 to the online appendix. Our simulation and case studies show that version 1 of \texttt{EM-VRSO} converges to local maxima of the log-likelihood function in practice. 

At first, it seems troubling to require $\log p(\bfy;\bfphi_{k,\ell}) \geq \log p(\bfy;\bfphi_{k})$ to exit the while loop of \texttt{EM-VRSO}, since this requirement may cause an infinite loop if it cannot be met. Denote $\ell^*(k)$ as the (random) number of runs through the inner loop of \texttt{EM-VRSO} for iteration $k$ (i.e. the maximum value obtained by $\ell$ for a given value of $k$). We prove in Theorem 1 below that $\bbP(\ell^*(k) < \infty) = 1$. 
One final concern is whether the sequence $\{\bfphi_{k}\}_{k=0}^\infty$ generated by \texttt{EM-VRSO} converges to a local maximum of the likelihood as $K \to \infty$. Theorem 1 below guarantees such convergence under standard regularity conditions. We prove Theorem 1 in the appendix.

    
\begin{theorem}

    Suppose the following conditions are met in Algorithm (6) with $P = \texttt{False}$ and $A = \text{SVRG}$:
    
    \begin{enumerate}
        \item The parameters $\bfphi$ lie in $\bfPhi = \bbR^r$ for some dimension $r$.
        \item $\bfPhi_{\bfphi_0} = \{\bfphi \in \bfPhi: \log p(\bfy;\bfphi) \geq \log p(\bfy;\bfphi_0)\}$ is compact for all $\bfphi_0$ if $\log p(\bfy;\bfphi_0) > -\infty$.
        \item $\log p(\bfy;\bfphi)$ is differentiable in $\bfphi$ for all $\bfphi \in \bfPhi$.
        \item $F_t(\bfphi \mid \bfgamma_t(\bfphi'), \bfxi_t(\bfphi'))$ is convex with respect to $\bfphi$ and $F(\bfphi \mid \bfgamma(\bfphi'), \bfxi(\bfphi'))$ is strongly convex with respect to $\bfphi$ for all $\bfphi'$ with constant $C > 0$. Namely, for all $\bfphi$, $\bfphi_0$ and $\bfphi'$:
        \small
        \begin{equation}
            F_t(\bfphi \mid \bfgamma_t(\bfphi'), \bfxi_t(\bfphi')) \geq F_t(\bfphi_0 \mid \bfgamma_t(\bfphi'), \bfxi_t(\bfphi')) + \nabla F_t(\bfphi_0 \mid \bfgamma_t(\bfphi'), \bfxi_t(\bfphi'))^T(\bfphi-\bfphi_0),
        \end{equation}
        \begin{equation}
            F(\bfphi \mid \bfgamma(\bfphi'), \bfxi(\bfphi')) \geq F(\bfphi_0 \mid \bfgamma(\bfphi'), \bfxi(\bfphi')) + \nabla F(\bfphi_0 \mid \bfgamma(\bfphi'), \bfxi(\bfphi'))^T(\bfphi-\bfphi_0) + \frac{C}{2} \|\bfphi - \bfphi_0\|_2^2.
        \end{equation}
        \normalsize
        \item $F_t(\bfphi \mid \bfgamma_t(\bfphi'), \bfxi_t(\bfphi'))$ is uniformly Lipschitz-smooth with respect to $\bfphi$ for all $t$ and $\bfphi'$ with constant $L \geq C > 0$. Namely, for all $t$, $\bfphi$, $\bfphi_0$ and $\bfphi'$:
        \small
        \begin{equation}
            F_t(\bfphi \mid \bfgamma_t(\bfphi'), \bfxi_t(\bfphi')) \leq F_t(\bfphi_0 \mid \bfgamma_t(\bfphi'), \bfxi_t(\bfphi')) + \nabla F_t(\bfphi_0 \mid \bfgamma_t(\bfphi'), \bfxi_t(\bfphi'))^T(\bfphi - \bfphi_0) + \frac{L}{2} \| \bfphi - \bfphi_0 \|_2^2.
        \end{equation}
        \normalsize
        \item The step size $\lambda$ is sufficiently small and $M$ is sufficiently large such that
        \begin{equation}
            \zeta = \frac{1}{C \lambda(1-2L\lambda)M} + \frac{2L\lambda}{1-2L\lambda} < 1.
        \end{equation}
        \item $\nabla F_t(\bfphi \mid \bfgamma_t(\bfphi'), \bfxi_t(\bfphi'))$ is uniformly continuous in $(\bfphi,\bfphi')$ for all $t$.
    \end{enumerate}
    
    Further, let $\calS = \{\ell^*(k) < \infty \text{ for all } k \}$. Then, $\bbP\{\calS\} = 1$, and the following holds on $\calS$: all limit points of $\{\bfphi_{k}\}_{k=0}^\infty$  are stationary points of $\log p(\bfy;\cdot)$, and $\log p(\bfy;\bfphi_{k})$ converges monotonically to $\log p^* = \log p(\bfy;\bfphi^*)$ for some stationary point of $\log p(\bfy;\cdot)$, $\bfphi^*$.
\end{theorem}

Conditions (1--3) are from \citet{Wu:1983} and are standard assumptions needed to prove the convergence of the EM algorithm. Likewise, Conditions (4--6) are from \citet{Johnson:2013} and are standard assumptions used to prove common properties of stochastic optimization algorithms. Condition (7) is needed to prove that SVRG is a continuous mapping.

Condition (2) from \citet{Wu:1983} and condition (5) of \citet{Johnson:2013} can be restrictive and are often violated in common settings. For example, both are violated when estimating the variance of Gaussian state-dependent distributions within an HMM. This issue is well-known for maximum likelihood estimation in mixture models \citep{Chen:2009,Liu:2015b}. It can be avoided by setting lower bounds on the variance components \citep{Zucchini:2016}. 

Theorem 1 above applies only to version 2 of \texttt{EM-VRSO} when $P = \texttt{False}$ and $A = \text{SVRG}$. Unfortunately, convergence analysis for \texttt{EM-VRSO} is more complicated when $P = \texttt{True}$ than when $P = \texttt{False}$ because the E and M steps are mixed. However, Theorem 2 below shows that stationary points of the log-likelihood are fixed points of Algorithm (\ref{alg:EM-VRSO}) for all values of $P$ and $A$. We prove Theorem 2 in the online appendix. 

\begin{theorem}
    If $\nabla \log p(\bfy;\bfphi_0) = 0$, then for all $\lambda \in \bbR$, $A \in \{\text{SAGA}, \text{SVRG}\}$, $P \in \{\texttt{True},\texttt{False}\}$, $M \in \bbN$, and $K \in \bbN$, $\texttt{EM-VRSO}(\bfphi_0, \lambda, A, P, M, K) = \bfphi_0$ with probability 1, where $\texttt{EM-VRSO}$ is defined in Algorithm (\ref{alg:EM-VRSO}).
\end{theorem}

Theorem 2 is useful because it guarantees that Algorithm (\ref{alg:EM-VRSO}) does not change $\bfphi_0$ when it is a stationary point of the likelihood. However, it makes no guarantees that Algorithm (\ref{alg:EM-VRSO}) will converge if $\nabla \log p(\bfy;\bfphi_0) \neq 0$ and either $P = \texttt{True}$ or $A = \text{SAGA}$. Nonetheless, we see in our simulation and case studies that Algorithm (\ref{alg:EM-VRSO})  approaches local maxima of the log-likelihood function faster than existing full-batch baselines for all values of $P$ and $A$. For theoretical guarantees on convergence, practitioners can set $P = \texttt{True}$ or $A = \text{SAGA}$ for a predetermined number of iterations, followed by switching to $P = \texttt{False}$ and $A = \text{SVRG}$, or a full-gradient method such as BFGS \citep{Fletcher:2000}.

\section{Practical Considerations}

Algorithm (\ref{alg:EM-VRSO}) outlines a method to perform the Baum-Welch algorithm with a stochastic E and M step. This section outlines implementation details that improve its practical performance.

\subsection{Line Search for Step Size Selection}
\label{subsec:est_L}

One drawback of stochastic optimization is that the step size can greatly affect the practical performance of the algorithm. \citet{Defazio:2014} suggest a step size of $\lambda = 1/(3L)$ for SAGA, where $L$ is the Lipschitz constant defined in Theorem 1. However, the Lipschitz constants are rarely known in practice. Therefore, following \citet{Schmidt:2017}, we initialize an estimate of the Lipschitz constant, $\hat L$, and update it if the following inequality does not hold at any step $m$ of the optimization algorithm:

\begin{gather}
    F^{(m+1)}_{t_m}\Big(\bfphi^{(m)} - \frac{1}{\hat L} \nabla F_{t_m}^{(m+1)}(\bfphi^{(m)}) \Big) \leq F^{(m+1)}_{t_m}\Big(\bfphi^{(m)}\Big) - \frac{1}{2 \hat L} \Big\| \nabla F_{t_m}^{(m+1)}(\bfphi^{(m)}) \Big\| ^2, \quad \text{where} \nonumber \\
    F_{t_m}^{(m+1)} = F_{t_m}\left(\cdot ~ \Big | ~ \widehat \bfgamma^{(m+1)}_{t_m}, \widehat \bfxi^{(m+1)}_{t_m}\right)
    \label{ineq:F}
\end{gather}
The inequality above is obeyed if $\hat L = L$, so if it is violated, then we double the Lipschitz constant estimate $\hat L$. We also follow \citet{Schmidt:2017} and do not check the inequality if $\| \nabla F_{t_m}^{(m+1)}(\bfphi^{(m)})\| < 10^{-8}$ due to numerical instability. 

In addition, the Lipschitz constant $L$ is a global quantity, but the algorithm will likely remain in a neighborhood around a local maximum of $F$ later in the optimization algorithm. Within this local neighborhood, a smaller value of $L$ may apply \citep{Schmidt:2017}, so the Lipschitz constant estimate is decreased by a small amount after each parameter update:
$\hat L \leftarrow 2^{-1/T} ~ \hat L$.
Updating $\hat L$ after each parameter update allows the step size of the optimization algorithm to adapt to the smoothness of the objective function close to the optimum value.

\subsection{Multiple Step Sizes}

The optimization problem within the M step of the Baum-Welch algorithm can be written as two separate optimization problems over both $\bftheta$ and $\bfeta$ (recall that $\bfphi = \{\bftheta,\bfeta\}$). In particular, we can rewrite $F_t$ as $F_t(\bfphi \mid \bfgamma_t,  \bfxi_t) = G_t(\bftheta \mid  \bfgamma_t,  \bfxi_t) + H_t(\bfeta \mid  \bfgamma_t,  \bfxi_t)$, where

\begin{gather}
    G_t(\bftheta \mid  \bfgamma_t,  \bfxi_t) = - \sum_{i=1}^N  \gamma^{(i)}_t \log f^{(i)}(y_t;\theta^{(i)}), \quad t = 1,\ldots,T, \\
    H_1(\bfeta \mid  \bfgamma_t,  \bfxi_t) = - \sum_{i=1}^N  \gamma^{(i)}_t \log \delta^{(i)}(\bfeta), \\
    H_t(\bfeta \mid  \bfgamma_t,  \bfxi_t) =  - \sum_{i=1}^N \sum_{j=1}^N  \xi^{(i,j)}_t \log \Gamma^{(i,j)}(\bfeta), \quad t = 2,\ldots,T.
\end{gather}
To this end, let $\widehat \nabla G^{(m)}_{t}$ be equal to the components of $\widehat \nabla F^{(m)}_{t}$ that correspond to $\bftheta$, and let $\widehat \nabla H^{(m)}_{t}$ be equal to the components of $\widehat \nabla F^{(m)}_{t}$ that correspond to $\bfeta$. Likewise, let $\widehat \nabla G^{(m)}$ be equal to the components of $\widehat \nabla F^{(m)}$ that correspond to $\bftheta$, and let $\widehat \nabla H^{(m)}$ be equal to the components of $\widehat \nabla F^{(m)}$ that correspond to $\bfeta$. Then, we can rewrite the gradient step in the stochastic M step of Algorithm (\ref{alg:VRSO-PE}) as

\begin{align}
    \bftheta^{(m+1)} &= \bftheta^{(m)} - \lambda_{\bftheta} \left[\nabla G_{t_m}\left(\bftheta^{(m)} ~ \Big | ~ \widehat \bfgamma_{t_m}^{(m+1)}, \widehat \bfxi_{t_m}^{(m+1)}\right) - \widehat \nabla G_{t_m}^{(m)} + \widehat \nabla G^{(m)} \right], \\
    \bfeta^{(m+1)} &= \bfeta^{(m)} - \lambda_{\bfeta} \left[\nabla H_{t_m}\left(\bfeta^{(m)} ~ \Big | ~ \widehat \bfgamma_{t_m}^{(m+1)}, \widehat \bfxi_{t_m}^{(m+1)}\right) - \widehat \nabla H_{t_m}^{(m)} + \widehat \nabla H^{(m)} \right].
\end{align}
where $\lambda_{\bftheta} = \lambda_{\bfeta} = \lambda$. Note that $G_t$ is a function of $\bftheta$ alone and $H_t$ is a function of $\bfeta$ alone for given $\widehat \bfgamma_{t}$ and $\widehat \bfxi_{t}$ and all $t = 1,\ldots,T$. As such, we allow $\lambda_{\bftheta} \neq \lambda_{\bfeta}$ and have each depend upon different Lipschitz constant estimates: $\lambda_{\bftheta} = 1/(3\hat L_G)$ and $\lambda_{\bfeta} = 1/(3\hat L_H)$. The line search described in Section 4.1 can then be used to update the estimates $\hat L_G$ and $\hat L_H$ separately.

\subsection{Adaptive Step Size for Fixed Lipschitz Constants}
\label{subsec:L_divider}

Under certain regularity conditions, \citet{Defazio:2014} prove that SAGA converges using a step size of $1/(3L)$ for a given loss function $F$ with Lipschitz constant $L$. We therefore initialize step sizes of $\lambda_{\bftheta} = 1/(3 \hat L_G)$ and $\lambda_{\bfeta} = 1/(3 \hat L_H)$ for all experiments and algorithms. However, for Algorithm (\ref{alg:EM-VRSO}) with $P = \texttt{True}$, the objective function $F(\cdot \mid \widehat \bfgamma_{t_m}^{(m+1)} ~,~ \widehat \bfxi_{t_m}^{(m+1)}) = G(\cdot \mid \widehat \bfgamma_{t_m}^{(m+1)} ~,~ \widehat \bfxi_{t_m}^{(m+1)}) + H(\cdot \mid \widehat \bfgamma_{t_m}^{(m+1)} ~,~ \widehat \bfxi_{t_m}^{(m+1)})$ itself changes over the course of a single M step as $\widehat \bfgamma_{t_m}^{(m+1)}$ and $\widehat \bfxi_{t_m}^{(m+1)}$ are updated. As a result, more conservative (i.e. smaller) step sizes $\lambda_{\bftheta}$ and $\lambda_{\bfeta}$ may be needed, even if the Lipschitz constant estimates $\hat L_G$ and $\hat L_H$ are accurate. We therefore allow Algorithm (\ref{alg:EM-VRSO}) to change the step size after each attempt $\ell$ through the M step of the Baum-Welch algorithm. Namely, if the log-likelihood does not decrease after iteration $k$ and attempt $\ell$ through the M step of Algorithm (\ref{alg:EM-VRSO}) with $P = \texttt{True}$, we halve the step size (as a function of either $\hat L_G$ or $\hat L_H$) for attempt $\ell+1$. For example, if the step sizes are $\lambda_{\bftheta} = 1/(3 \hat L_G)$ and $\lambda_{\bfeta} = 1/(3 \hat L_H)$ for attempt $\ell$ through a given M step, and attempt $\ell$ results in an increase of the log-likelihood, we define new step-sizes $\lambda_{\bftheta} \leftarrow 1/(6 \hat L_G)$ and $\lambda_{\bfeta} \leftarrow 1/(6 \hat L_H)$ for attempt $\ell+1$. We also maintain this halved step size for the remainder of Algorithm (\ref{alg:EM-VRSO}).

\subsection{Sampling for SAGA and SVRG Without Replacement}
\label{subsec:wo_replacement}

Finally, we sample each random index $t$ \textit{without} replacement within Algorithm (\ref{alg:VRSO-PE}). If $M > T$, then we sample without replacement until all time indices are sampled, and then re-sample the data set without replacement. Sampling without replacement for SGD is often easier to implement and performs better than sampling with replacement \citep{Gurbuzbalaban:2015}. \citet{Ohad:2016} also gives several convergence results for SVRG when indices are sampled without replacement.
\label{sec:prac}

\section{Simulation Study}

\subsection{Simulation Procedure}

To test the performance of Algorithm (\ref{alg:EM-VRSO}), we ran eight simulation experiments. For a given experiment, we simulated $T \in \{10^3,10^5\}$ observations from an HMM with $N \in \{3,6\}$ hidden states and observations $y_t \in \bbR^d$, with $d \in \{3,6\}$. All possible combinations of $T$, $N$, and $d$ comprised a total of $2^3 = 8$ experiments. For each experiment, we simulated five data sets. For every experiment and data set, $Y_t \mid X_t = i$ followed a normal distribution with mean $\mu^{(i)}$ and covariance matrix $\bfSigma^{(i)}$. We defined $\mu^{(i)}$ and $\bfSigma^{(i)}$ for every data set as
$\mu^{(i)} \sim \calN(0,I)$ and $\bfSigma^{(i)} = \text{diag}[\exp(-2)]$ for $i \in \{1,\ldots,N\}$
where $I$ is the identity matrix.
We set the transition probability matrices of the generating process to depend upon $T$ to keep the expected number of total transitions at 100. This simulates sequences of observations that are sampled at either low- or high- frequencies.
Denote the true transition probability matrix from an experiment with $T$ observations and $N$ hidden states as $\bfGamma_{T,N}$. We defined $\bfGamma_{10^3,3} \in \bbR^{3 \times 3}$ to have diagonal elements of 0.9 and off-diagonal elements of 0.05, $\bfGamma_{10^3,6} \in \bbR^{6 \times 6}$ to have diagonal elements of 0.9 and off-diagonal elements of 0.02, $\bfGamma_{10^5,3} \in \bbR^{3 \times 3}$ to have diagonal elements of 0.999 and off-diagonal elements of $5 \times 10^{-4}$, and $\bfGamma_{10^3,6} \in \bbR^{6 \times 6}$ to have diagonal elements of 0.999 and off-diagonal elements of $2 \times 10^{-4}$. 
We randomly defined the initial distribution as $\bfdelta \sim \text{dir}(\onevec_N)$ for every experiment and data set.

\subsection{Optimization Procedure}

We estimated the parameters of the generating model for all five data sets and all eight experiments using six different versions of Algorithm (\ref{alg:EM-VRSO}). In particular, we used $A \in \{\text{SVRG, SAGA}\}$, and for each value of $A$, we used the combinations $\{P = \texttt{False}, ~ M = T\}$, $\{P = \texttt{True}, ~ M = T\}$, and $\{P = \texttt{True}, ~ M = 10T\}$. Recall that setting $P = \texttt{True}$ corresponds to integrating a partial E step into the variance-reduced stochastic M step. The variable $M$ corresponds to the number of iterations of SAGA or SVRG that are performed at each M step of the algorithm. It is natural to set $M=T$ to approximately balance the computational load of the E step and the M step. Nonetheless, we ran an experiment with $M=10T$ and $P = \texttt{True}$ to test the algorithm when the majority of the computational load was placed on the combined partial E / stochastic M step. We also estimated the HMM parameters using three baseline methods: BFGS \citep{Fletcher:2000}, the conjugate gradient method \citep{Fletcher:1964}, and full-batch gradient descent. 

We sampled a total of five random parameter initializations for each data set / experiment pair, and then ran all nine optimization algorithms on every parameter initialization. Each parameter initialization was re-used for each algorithm to ensure consistency. Let $\bar y$ denote the sample mean and $\bfQ$ denote the sample covariance of the observation sequence $\{y_t\}_{t=1}^T$. We initialized $\bftheta_0 = \{\mu^{(i)}_0,\bfSigma^{(i)}_0\}_{i=1}^N$ as
$\mu^{(i)}_0 \sim \calN(\bar y, \text{diag}(\bfQ))$ and $\bfSigma^{(i)}_0 = \text{diag}(\bfQ)$ for $i = 1,\ldots,N.$ We initialized $\bfeta_0$ as
$\eta^{(i)}_0 \sim \calN(0,1)$ for $i = 2,\ldots,N$ and
$\eta^{(i,j)}_0 \sim \calN(-2,2^2)$ for $i,j = 1,\ldots,N$, where $i \neq j$.
Throughout the optimization procedure, we assumed that $\bfSigma^{(i)}$ was diagonal for all $i \in \{1,\ldots,N\}$, which is in line with the generating model described above. Further, we reparameterized $\bfSigma^{(i)}$ as 
$\bfSigma^{(i)} = \text{diag}\left[\exp(\boldsymbol{\rho}^{(i)})\right]$
and performed inference on $\boldsymbol{\rho}^{(i)} = \begin{pmatrix} \rho^{(i)}_1 & \ldots & \rho^{(i)}_N \end{pmatrix}$ for $i = 1,\ldots,N$, which is unconstrained.
All six algorithms were initialized with step sizes of $\lambda_{\bftheta} = 1/3 \hat L_G$ and $\lambda_{\bfeta} = 1/3 \hat L_H$. The Lipschitz constants were initialized as $\hat L_G = \hat L_H = 100/3$ and updated during the optimization routine according to the procedure from section \ref{subsec:est_L}. 
All algorithms and baselines were run for a total of 12 hours on the Compute Canada Cedar cluster on nodes with 16GB of RAM.
All baselines were implemented using the Scipy library in Python \citep{Virtanen:2019}, and we implemented Algorithm (\ref{alg:EM-VRSO}) using a custom Python script.

We employed several measures to fairly compare the performance of each optimization algorithm. To account for differences in speed due to implementation discrepancies, we measured computational complexity in epochs rather than raw computation time. We define one epoch as either $T$ evaluations of Equations (\ref{eqn:tilde_alpha} -- \ref{eqn:tilde_xi}) in the E step of Algorithm (\ref{alg:EM-VRSO}), $T$ stochastic gradient evaluations in the M step of Algorithm (\ref{alg:EM-VRSO}), or one gradient evaluation in the full-gradient baseline algorithms. We estimated the true maximum likelihood parameters $\bfphi^*$ for each data set / experiment pair using the parameters from the best-performing optimization algorithm after 12 hours. We also recorded the epoch of convergence for each optimization algorithm, which was defined as the point when the gradient norm of the log-likelihood (divided by $T$) was less than $10^{-2}$. The tolerance was set to $10^{-2}$ because it was the lowest tolerance that all algorithms regularly converged to within 12 hours.

\subsection{Simulation Results}

The results from the simulation study are shown in Figures (2--4).

\begin{figure}
    \centering
    \includegraphics[width=5.5in]{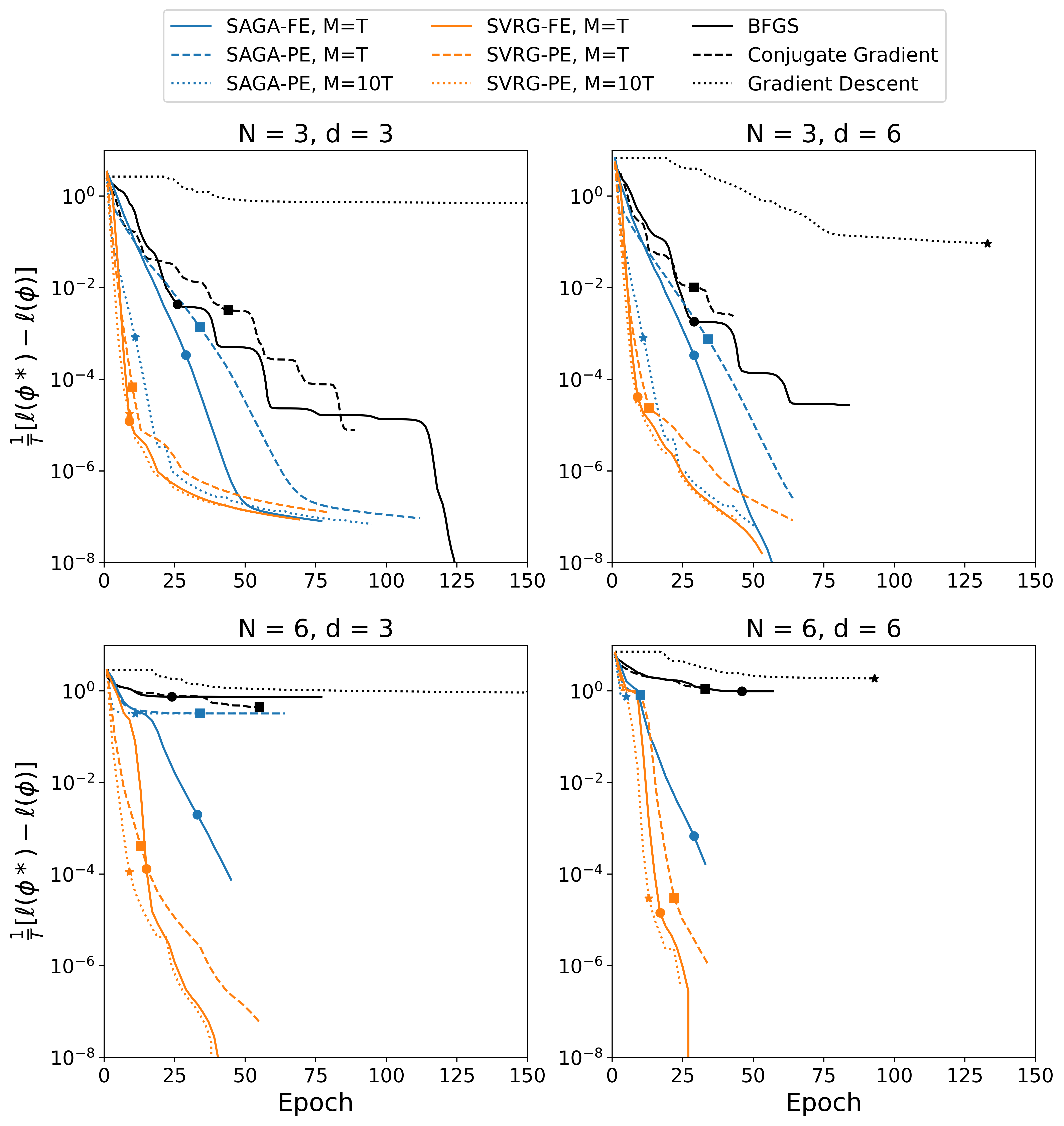}
    \caption{
    Maximum log-likelihood minus the log-likelihood (all over $T$) vs epoch for a selected run of each optimization algorithm in the simulation study. Algorithms were run for either 12 hours or 150 epochs (whichever came first) on a data set of experiments with $T=10^{5}$, $N \in \{3,6\}$, and $d \in \{3,6\}$. FE corresponds to $P = \texttt{False}$, and PE corresponds to $P = \texttt{True}$. The log-likelihood is denoted as $\ell(\bfphi)$, and the y-axis is on a log-scale. We display the random initialization of each algorithm that resulted in the highest likelihood after 12 hours. Dots correspond to the epoch and likelihood at convergence. 
    }
    \label{fig:ll_trace_sim}
\end{figure}
\begin{figure}
    \centering
    \includegraphics[width=6.5in]{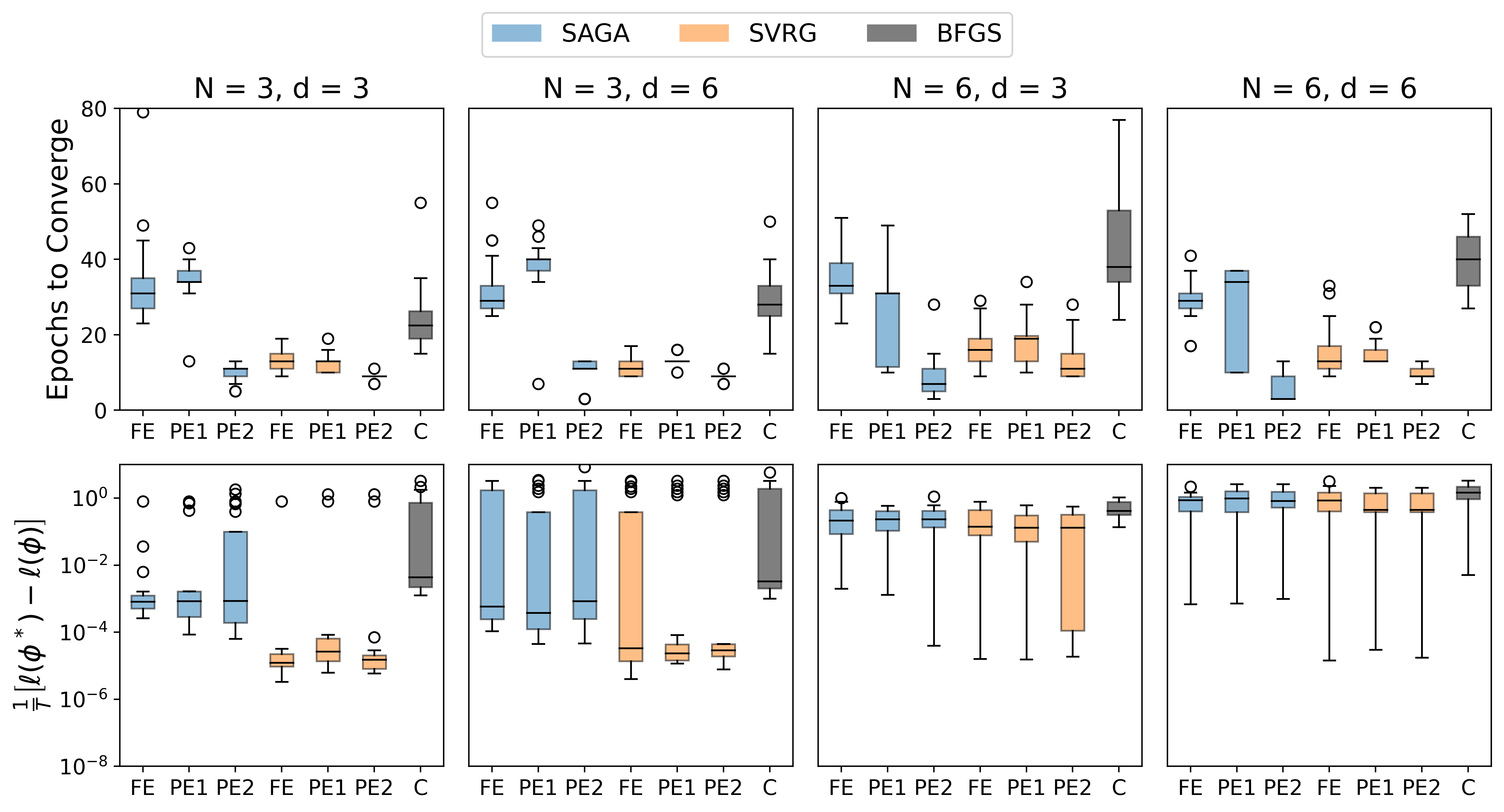}
    \caption{Box plots showing epochs to converge (top) and maximum log-likelihood minus log-likelihood at convergence (all over $T$, bottom) for each optimization algorithm in the simulation study. Unlike Figure (\ref{fig:ll_trace_sim}), results are displayed for all data sets and initial parameter values. FE corresponds to $\{P = \texttt{False}, M = T\}$, PE1 corresponds to $\{P = \texttt{True}, M = T\}$, and PE2 corresponds to $\{P = \texttt{False}, M = 10T\}$. Results are shown for all simulation studies with $T=10^{5}$. The log-likelihood is denoted as $\ell(\bfphi)$, and the y-axis of the bottom row is on a log-scale.}
    \label{fig:boxplots_sim}
\end{figure}
\begin{figure}
    \centering
    \includegraphics[width=5.5in]{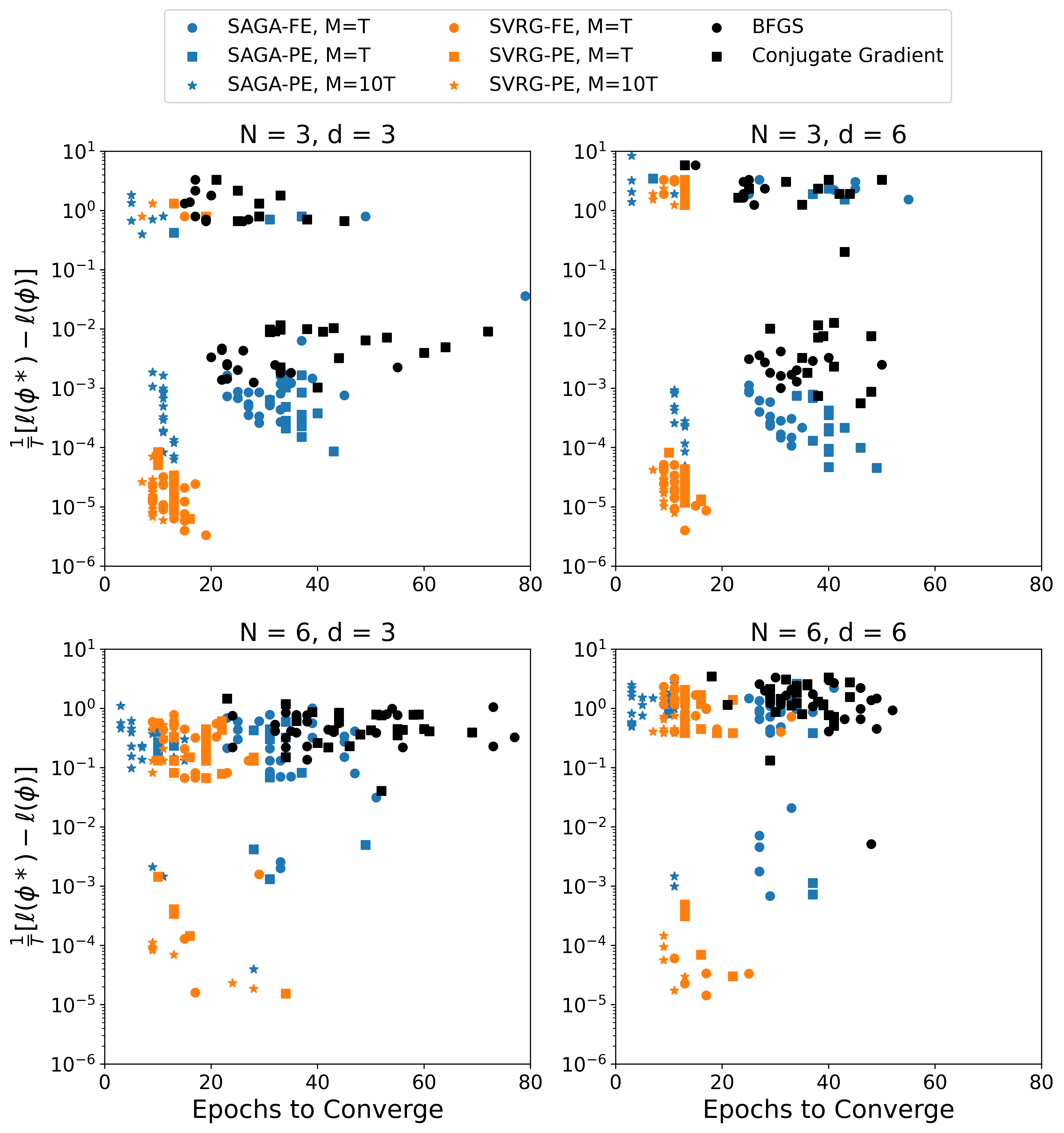}
    \caption{Maximum log-likelihood minus log-likelihood at convergence (all over $T$) versus epochs to converge for each optimization algorithm in the simulation study with $T=10^{5}$, $N \in \{3,6\}$, and $d \in \{3,6\}$. FE corresponds to $P = \texttt{False}$, and PE corresponds to $P = \texttt{True}$. The log-likelihood is denoted as $\ell(\bfphi)$ on the y-axis, which is on a log-scale.}
    \label{fig:scatter_sim}
\end{figure}
Algorithm (\ref{alg:EM-VRSO}) with $A=\text{SVRG}$ markedly sped up the optimization procedure, as it usually converged in at most half as many epochs compared to the baselines for each experiment. Algorithm (\ref{alg:EM-VRSO}) with $A=\text{SAGA}$ also tended to converge faster than the baselines for all experiments except for those with $N=3$. Algorithm (\ref{alg:EM-VRSO}) with $A=\text{SAGA}$ also converged in significantly fewer epochs when $P = \texttt{True}$ and $M=10T$. 
All methods were prone to converge to local minima, especially when $N=6$, in which case the likelihood surface was highly multi-modal. However, Algorithm (\ref{alg:EM-VRSO}) tended to converge to areas of higher likelihood than the baseline for almost all experiments. The sole exception was when $T=10^3$, $N=6$, and $d=6$, where BFGS and the conjugate gradient method were less likely to get stuck in local minima (see Figures (1--2) in Supplement A). However, the aforementioned experiment is a small-data setting, and we are primarily interested in modeling very large data sets.
We present figures analogous to Figure (\ref{fig:ll_trace_sim}) for all five data sets corresponding to all experiments in Supplement A. All figures associated experiments with $T=10^3$ can also be found in Supplement A. 

\section{Case Study}

We additionally tested the performance of our optimization algorithm by modeling the movement of eight northern resident killer whales off the coast of British Columbia, Canada with a data set of over $5000$ dives and $80000$ depth readings. Biologgers are an essential tool used to understand the behavior of marine mammals. For example, time-depth recorders allow researchers to estimate behavioral states associated with each dive \citep[e.g. foraging, resting, and traveling,][]{Tennessen:2023}. Researchers also use biologging data sets to identify and characterize dive phases, which are important for inferring behavior \citep[e.g., prey capture often occurs in the bottom phase of a foraging dive,][]{Wright:2017,Jensen:2023}. As such, we developed an HMM to identify three common dive phases (ascent, descent, and bottom) and three dive types that may indicate distinct behaviors of the animal, including resting, foraging, and traveling. We performed inference on the resulting model using our optimization algorithm in order to illustrate its computational advantages.

\subsection{Data Collection and Preprocessing}

\begin{figure}
    \centering
    \includegraphics[width=5.5in]{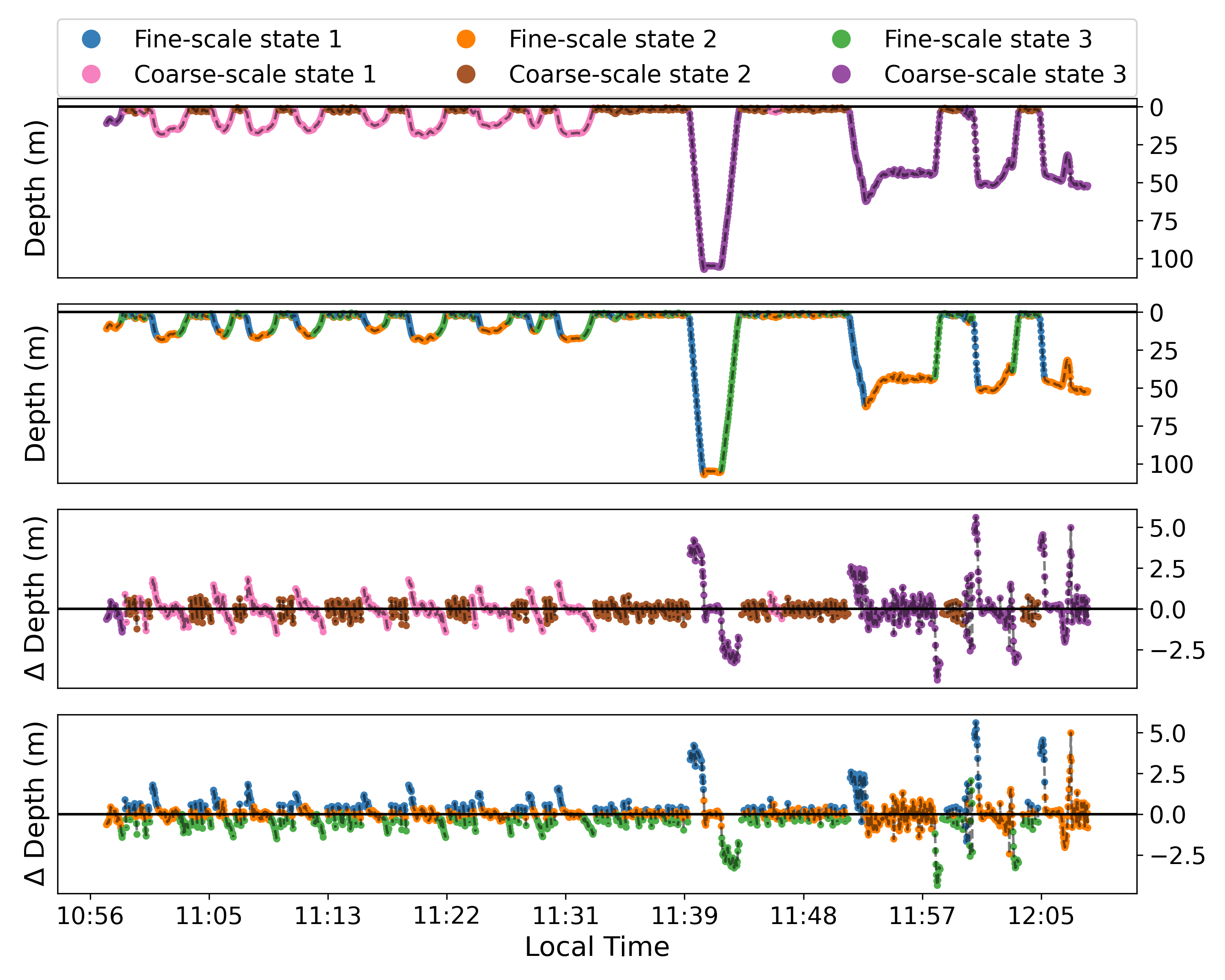}
    \caption{Depth profile and change in depth vs time of day for a selected killer whale (I107, male, born 2004) off the coast of British Columbia, Canada. The data in panels one and three are color-coded according to the most likely hidden coarse-scale state (i.e. dive type) for each dive. The data in panels two and four are color-coded according to the most likely hidden fine-scale state (i.e. dive phase) for each two-second window.}
    \label{fig:data}
\end{figure}

The data used in this case study were collected in August and September of 2020 using a CATS time-depth recorder, or TDR (Customizable Animal Tracking Solutions, {\em{www.cats.is}}). Northern resident killer whales were equipped with suction-cup attached CATs tags in Queen Charlotte Sound using an adjustable 6-8m carbon fiber pole. The tags were programmed to release within 3-24 hours of attachment. Instruments were retrieved following each deployment using a Wildlife Computers 363C SPOT tag (providing Argos satellite positions), goniometer, ultra high frequency receiver, and yagi antenna. The tags included 3D kinematic sensors (accelerometer, magnetometer, gyroscope), time-depth recorder, hydrophone and camera. All sensors were programmed to sample at 50 hertz. However, for the purposes of this study, we focus on the time-depth recorder data to discern behaviorally distinct dives. We calibrated the depth readings using a MATLAB package developed by \citet{Cade:2021}, and defined a dive as any sequence of depth readings under 0.5 meters that lasted for at least two seconds. We then down-sampled the depth readings to a frequency of 0.5 hertz. The processed data set contained a total of $5858$ dives and $89462$ depth readings. Figure (\ref{fig:data}) shows the depth and change in depth for a subset of dives for one whale in the data set. 

\subsection{Model Formulation}

Dive phases may vary depending upon the animal's behavior. For example, foraging dives tend to be deeper and longer than resting dives, so it is natural to model the phases of foraging dives differently compared to those of resting dives \citep{Tennessen:2019b}. As such, we used a hierarchical HMM to jointly model dive types and dive phases \citep{Barajas:2017}. Hierarchical HMMs are specific instances of traditional HMMs, so the machinery developed here is applicable to perform inference. 

We assumed there to be three dive types, which is consistent with other studies of cetaceans \citep[e.g. resting, foraging and traveling,][]{Barajas:2017}. We also assumed that there are three dive phases per dive type (descent, bottom, and ascent), which is also consistent with other studies of diving birds and mammals \citep[e.g.][]{Vivant:2014}. This resulted in a total of $N = 9$ hidden states, each corresponding to a different dive phase / dive type combination.
Since each dive must begin with the descent phase, we set the initial distribution $\bfdelta$ to have the form
$\bfdelta = \begin{pmatrix} \delta^{(1)} & 0 & 0 & \delta^{(2)} & 0 & 0 & \delta^{(3)} & 0 & 0 \end{pmatrix}$,
where $\delta^{(i)}$ represents the probability that a killer whale begins its dive profile with a dive of type $i$. 
For this hierarchical HMM, we allowed the dive type to change between dives, but not within a given dive. As such, the form of transition probability matrix changed over time, so we denoted the probability transition matrix as $\bfGamma_t$. We defined a coarse-scale, inter-dive transition probability matrix $\bfGamma^{(c)} \in \bbR^{3 \times 3}$. For each dive type $i$, we also defined a fine-scale, intra-dive probability transition matrix $\bfGamma^{(f,i)} \in \bbR^{3 \times 3}$. 
To ensure that the dive type did not change within a dive, we defined the probability transition matrix \textit{within} a dive to have a block-diagonal form:
\begin{equation}
    \bfGamma_t = 
    \begin{pmatrix}
        \bfGamma^{(f,1)} & \mathbf{0} & \mathbf{0} \\
        \mathbf{0} & \bfGamma^{(f,2)} & \mathbf{0} \\
        \mathbf{0} & \mathbf{0} & \bfGamma^{(f,3)} \\
    \end{pmatrix},
\end{equation}
where $\bfGamma^{(f,i)}$ is upper-triangular for $i \in \{1,2,3\}$. The transition matrices were upper-triangular because the descent and bottom phases of a dive cannot occur after ascent, and the descent phase of a dive cannot occur after the bottom phase.
However, \textit{between} dives, we allowed the coarse-scale dive type to transition, but forced each dive to begin in the descent phase. As such, $\bfGamma_t$ took the following form:
\begin{equation}
    \bfGamma_t = \bfGamma^{(c)} \otimes \begin{pmatrix} 1 & 0 & 0 \\ 1 & 0 & 0 \\ 1 & 0 & 0 \end{pmatrix},
\end{equation}
where $\otimes$ denotes the Kronecker product.

Rather than modeling raw depth every two seconds as the observation sequence, we encoded each two-second window of depth data with summary statistics. Namely, we denoted an observation as $Y_t = \{D_t,E_t\}$, where $D_t \in \bbR$ is the change in depth in meters and $E_t = 1$ if a dive ended at index $t$ and $E_t = 0$ otherwise. 
Within dive type $i$ and dive phase $j$, we assumed $D_t$ followed a normal distribution with mean $\mu^{(i,j)}$ and standard deviation $\bfSigma^{(i,j)}$, and we assumed that $E_t$ followed a Bernoulli distribution with probability $p^{(i,j)}$. We assumed that dives must end on the ascent phase, so we set $p^{(i,1)} = p^{(i,2)} = 0$ for dive types $i = 1,2,3$. Conditioned on the dive type and dive phase, $D_t$ and $E_t$ were assumed to be independent of one another.

\subsection{Optimization Procedure}

We used a procedure similar to the simulation study to initialize the case study parameters. Let $\bar D$ denote the sample mean and $s$ denote sample standard deviation of $\{D_t\}_{t=1}^T$. We initialized the initial estimates for the mean $\left(\mu^{(i,j)}_0\right)$ and standard deviation $\left(\sigma^{(i,j)}_0\right)$ of the state-dependent density of $D_t$ as
$\mu^{(i,j)}_0 \sim \calN(\bar D, s^2)$ and $\log\left(\sigma^{(i,j)}_0\right) \sim \calN(\log(s),1)$ for $i,j = 1,2,3$,
where $i$ refers to the dive type and $j$ to the dive phase. Further, let $\bar E$ represent the mean of $\{E_t\}_{t=1}^T$. We initialized the state-dependent probability of observing a dive end as
$p^{(i,1)}_0 = 0$, $p^{(i,2)}_0 = 0$ and $\logit(p^{(i,3)}_0) \sim \calN(\logit(\bar E),1)$ for $i = 1,2,3$,
where $p^{(i,j)}_0$ is the initial estimate corresponding to the Bernoulli distribution of $E_t$ during dive type $i$ and dive phase $j$. Dive phase 3 is ascent.

Let $\bfeta^{(\bfdelta)}_k \in \bbR^3$ denote the parameters associated with $\bfdelta$ at iteration $k$ of a given optimization algorithm. The reparameterization from $\bfeta^{(\bfdelta)}_k$ to $\bfdelta_k$ is given in Equation (\ref{eqn:reparam}). We initialized the first element of $\eta^{(\bfdelta)}_0$ as zero for identifiability and the second and third elements with a standard normal distribution, $\calN(0,1)$.

Let $\bfeta_k^{(c)} \in \bbR^{3 \times 3}$ denote the parameters associated with the coarse-scale probability transition matrix at iteration $k$ of a given optimization algorithm. The reparameterization from $\bfeta_k^{(c)}$ to $\bfGamma_k^{(c)}$ is given in Equation (\ref{eqn:reparam}). We initialized the diagonal elements of $\bfeta_0^{(c)}$ as zeros, and we initialized the off-diagonal elements of $\bfeta_0^{(c)}$ with a normal distribution with mean $-3$ and unit variance, $\calN(-3,1)$.

Let $\bfeta_k^{(f,i)} \in \bbR^{3 \times 3}$ denote the parameters associated with fine-scale transition probability matrix $\bfGamma_k^{(f,i)}$. The reparameterization from $\bfeta_k^{(f,i)}$ to $\bfGamma_k^{(f,i)}$ is given in Equation (\ref{eqn:reparam}). We initialized all diagonal elements of $\bfeta_0^{(f,i)}$ as zeros and all off-diagonal elements of $\bfeta_0^{(f,i)}$ with a normal distribution with mean -1 and unit variance, $\calN(-1,1)$.

Similarly to the simulation study, we estimated the parameters of the hierarchical HMM using our six inference algorithms ($A \in \{\text{SVRG, SAGA}\}$ in addition to $(P,M) \in \{(\texttt{False},T), (\texttt{True},T), (\texttt{True},10T)\}$) and three baseline algorithms (BFGS, conjugate gradient, and gradient descent). All algorithms were run using 50 random initializations for a total of 12 hours each on Compute Canada Cedar nodes with 16GB of RAM.

We employed similar measures as those in the simulation study to fairly compare the optimization algorithms. In particular, we measured computational complexity in epochs rather than raw computation time and defined an epoch using the definition from the simulation study. We also estimated the maximum likelihood parameters $\bfphi^*$ for each data set / experiment pair using the same method as the simulation study. Finally, we recorded the epoch and likelihood of convergence for each optimization algorithm using the same definition of convergence as the simulation study.

\subsection{Case Study Results}

Our model predicted dive phases and dive types that are in line with previous studies of marine mammal diving behavior. For example, dive types are separated by shallow, medium, and deep depths, which is similar to results from \citet{Barajas:2017}. Further, each dive has a well-characterized bottom phase that occurs at approximately 70\% of the maximum depth for all dive types. This finding is similar to the results from \citet{Tennessen:2019a}. See Figure (\ref{fig:data}) and supplement B for further detail.

\begin{figure}
    \centering
    \includegraphics[width=4in]{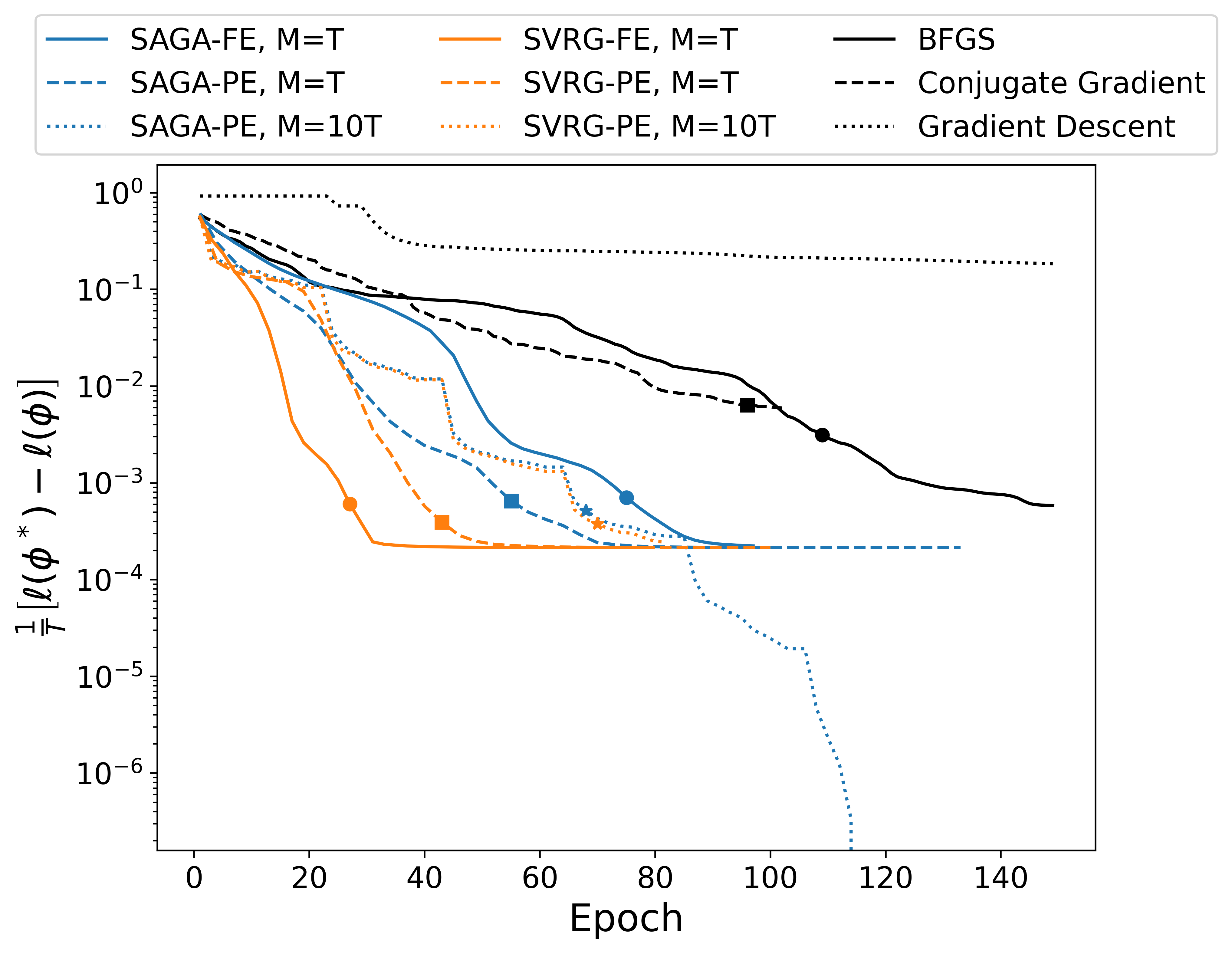}
    \caption{
    Maximum log-likelihood minus log-likelihood (all over $T$) vs epoch for 12 hours or 100 epochs (whichever came first) for the HMM from the killer whale case study. FE corresponds to $P = \texttt{False}$, and PE corresponds to $P = \texttt{True}$. The log-likelihood of $\bfphi$ is denoted as $\ell(\bfphi)$ on the y-axis, which is on a log-scale. We display the random initialization of each algorithm that resulted in the highest likelihood after 12 hours. Dots correspond to the epoch and likelihood at convergence for each algorithm.
    }
    \label{fig:ll_trace_case}
\end{figure}

\begin{figure}
    \centering
    \includegraphics[width=4in]{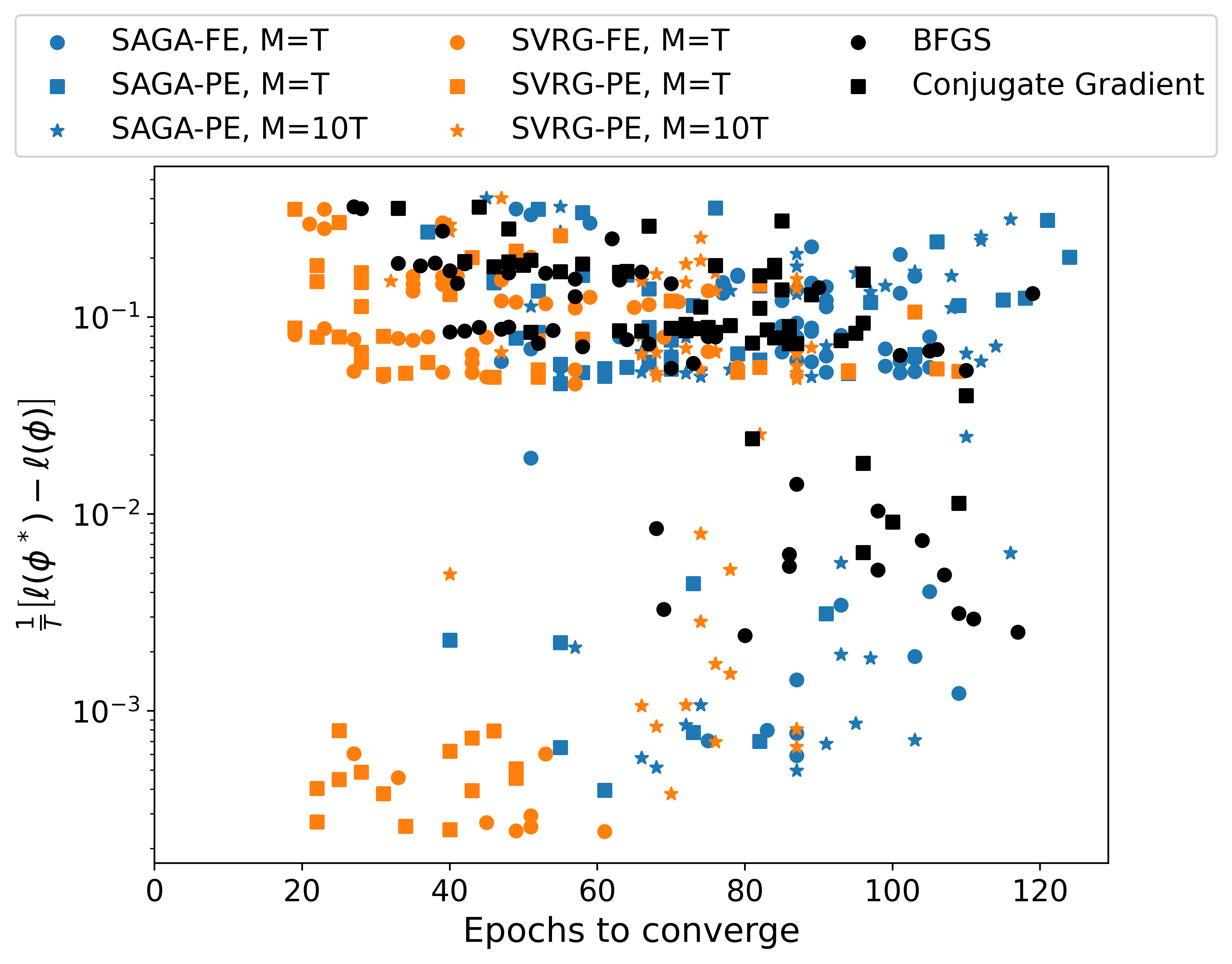}
    \caption{Maximum log-likelihood minus log-likelihood at convergence (all over $T$) versus epochs to converge for the killer whale case study. FE corresponds to $P = \texttt{False}$, and PE corresponds to $P = \texttt{True}$. The log-likelihood of $\bfphi$ is denoted as $\ell(\bfphi)$ on the y-axis, which is on a log-scale.}
    \label{fig:scatterplot_case}
\end{figure}

Most importantly, this case study demonstrates that all of our stochastic algorithms converged in fewer epochs and to regions of higher likelihood compared to the full-batch baselines; see Figures (\ref{fig:ll_trace_case}--\ref{fig:scatterplot_case}).

All algorithms occasionally converged to sub-optimal local minima, but Algorithm (\ref{alg:EM-VRSO}) with $A = \text{SVRG}$ and $P=\texttt{True}$ and $M=T$ tended to converge in the fewest epochs to regions of highest likelihood; see Figure (\ref{fig:scatterplot_case}). As with the simulation study, Algorithm (\ref{alg:EM-VRSO}) with $A = \text{SVRG}$ tended to converge in fewer epochs compared with $A=\text{SAGA}$; see Figure (\ref{fig:ll_trace_case}). Setting $P = \texttt{True}$ appears to be of particular use early in the optimization procedure (i.e the first $\approx$ 5 epochs, see Figure \ref{fig:ll_trace_case}). This behavior is intuitive because the proper weights $\left(\bfgamma(\bfphi_k^{(m)}) ~ \text{and} ~ \bfxi(\bfphi_k^{(m)})\right)$ change rapidly early in the optimization procedure. 

\section{Discussion}

The advent of high-frequency sensing technology has allowed researchers to model exceptionally long, high-frequency stochastic processes with increasingly complex HMMs \citep{Patterson:2017}. However, these complex models can be computationally expensive to fit \citep{Glennie:2023}.
We introduce an inference algorithm that speeds up maximum likelihood estimation for HMMs compared to existing batch-gradient methods. We do so without approximating the likelihood, which is required for many existing stochastic inference methods \citep{Gotoh:1998,Ye:2017}.

Our method does not require a closed-form solution for the M step, which enables quick inference for a diverse range of HMM models. This is useful in practice because many hidden Markov models lack closed-form solutions for the M step. In ecology, \citet{Pirotta:2018} used covariates in the transition probability matrix of an HMM to determine the effect of fishing boats on the behavior of Northern Fulmars (\textit{Fulmarus glacialis}). Likewise, \citet{Lawler:2019} used autoregression in the state-dependent distributions of an HMM to model the movement of grey seals (\textit{Halichoerus grypus}). In finance, \citet{Langrock:2018} modelled the relationship between the price of energy and price of the oil in Spain using non-parametric HMMs. The parameters of these models can not be estimated using a naive Baum-Welch algorithm, so our algorithm is a good candidate to speed up inference for complicated HMMs of this type. This will be especially relevant in the future as advances in biologging and tracking technology allow practitioners to collect increasingly large and high-frequency data sets \citep{Patterson:2017}.

Our new algorithm was particularly effective when performing inference over large data sets. For example, when applied to simulations with $T=10^{5}$ observations, our algorithm converged in approximately half as many epochs and tended to converge to regions of higher likelihood compared to existing baselines. Our case study of killer whale kinematic data showed similar improvements, demonstrating how our optimization procedure makes complex hierarchical HMMs less computationally expensive to fit on large biologging data sets.

The partial E step variant of our algorithm (i.e. with $P = \texttt{True}$) outperforms baselines particularly well early in the optimization procedure (in the first $\approx$ 5 epochs). As such, using a partial E step may be particularly advantageous when researchers are modeling large data sets with relatively low convergence thresholds.

One method that is particularly aligned with our algorithm is that of \citet{Zhu:2017}, who implement variance-reduced stochastic optimization to perform the M step of the EM algorithm on high-dimensional latent-variable models. Their method obtains a sub-linear computational complexity in the length of the observation sequence as well as a linear convergence rate. However, they focus primarily on mixture models rather than HMMs, and they do not combine the variance-reduced stochastic M step with a partial E step, which is an extension that we implement here. Further, their theoretical results assume independence between observations, which we do not rely on here.

While we use SVRG and SAGA in our analysis, there are other variance-reduced stochastic optimization algorithms that could be applied within our framework. For example, SARAH recursively updates the control variate in the inner loop of the optimization algorithm \citep{Nguyen:2017}, SPIDER uses a path-integrated differential estimator of the gradient \citep{Fang:2018}, and LiSSA is a second-order variance-reduced stochastic optimization scheme \citep{Agarwal:2017}. Future work can integrate these algorithms within our framework and evaluate the resulting performance. While we focus on an ecological case study here, the inference procedures we developed can unlock more complicated HMMs with larger latent spaces and bigger data sets for practitioners across a variety of disciplines.

\section*{Acknowledgements}
All killer whale data was collected under University of British Columbia Animal Care Permit no. A19-0053 and Fisheries and Oceans Canada Marine Mammal Scientific License for Whale Research no. XMMS 6 2019. Tags were deployed by Mike deRoos, the tagging boat skipper was Chris Hall, and photo-ID was done by Taryn Scarff. This research was enabled in part by support provided by WestGrid (www.westgrid.ca) and Compute Canada (www.computecanada.ca). 

\section*{Funding}
We acknowledge the support of the Natural Sciences and Engineering Research Council of Canada (NSERC) as well as the support of Fisheries and Oceans Canada (DFO). This project was supported by a financial contribution from the DFO and NSERC (Whale Science for Tomorrow). This work was also supported by the NSERC Discovery program under grant RGPIN-2020-04629; the Canadian Research Chairs program for Statistical Ecology; the BC Knowledge Development fund; the Canada Foundation for Innovation (John R. Evans Leaders Fund) under grant 37715; and the University of British Columbia via the Four-Year Doctoral Fellowship program.

\section*{Disclosure}
The authors report there are no competing interests to declare.

\bibliography{References}

\newpage
\begin{appendix}
\section{Version 2 of \texttt{EM-VRSO}}

For the purposes of readability, we abuse notation and use the shorthand $F(\bfphi \mid \bfgamma(\bfphi'), \bfxi(\bfphi')) \equiv F(\bfphi \mid \bfphi')$ and $F_{t}(\bfphi \mid \bfgamma_{t}(\bfphi'), \bfxi_{t}(\bfphi')) \equiv F_{t}(\bfphi \mid \bfphi')$ in this appendix. Further, we denote $F^*(\bfphi') \equiv \inf_{\bfphi \in \bfPhi} F(\bfphi \mid \bfphi')$.

\begin{algorithm}
\caption{\texttt{EM-VRSO}$(\bfphi_0, \lambda, A, M, K)$ (Version 2)}\label{alg:EM-VRSO-v2}
\begin{algorithmic}[1]
\Require Initial parameters ($\bfphi_{0}$), step size ($\lambda$), algorithm $A \in \{\text{SVRG, SAGA}\}$, whether to do a partial E step $P \in \{\texttt{True,False}\}$, number of iterations per update ($M$), and number of updates ($K$). Assume known Lipschitz constant $L$ and known strong convexity constant $C$.
\State $\zeta = (C \lambda(1-2L\lambda)M)^{-1} + (2L\lambda) / (1-2L\lambda) < 1$ 
\Comment{$\zeta < 1$}
\For{$k = 0,\ldots,K-1$}
\State $\{\bfalpha_{k,t},\bfbeta_{k,t},\bfgamma_{k,t},\bfxi_{k,t}\}_{t=1}^{T} = \texttt{E-step}(\bfphi_{k})$ \Comment{E step}
\State $K(\bfphi_{k}) = F^{*}(\bfphi_{k}) + \left[(1 + \zeta) / 2\right]\left(F(\bfphi_{k} \mid \bfphi_{k}) - F^{*}(\bfphi_{k})\right)$
\State $\ell \gets 0$ \Comment{M step}
\While{$\ell = 0$ or $F(\bfphi_{k,\ell} \mid \bfphi_{k}) > K(\bfphi_{k})$}
\State $\ell \gets \ell+1$
\State $\bfphi_{k,\ell} = \texttt{VRSO-PE}(\{\bfalpha_{k,t},\bfbeta_{k,t},\bfgamma_{k,t},\bfxi_{k,t}\}_{t=1}^{T},\bfphi_{k},\lambda,A,P,M)$
\EndWhile
\State $\bfphi_{k+1} = \bfphi_{k,\ell}$
\EndFor
\State \Return $\bfphi_K$
\end{algorithmic}
\end{algorithm}

One potential concern with Algorithm (\ref{alg:EM-VRSO-v2}) is that it requires quantities that are not known in practice, namely $\zeta$ and $F^{*}(\bfphi_{k})$. However, these quantities are only used in the while loop to ensure a strict increase in the likelihood for every iteration of \texttt{VRSO-PE}. In addition, version 1 of the algorithm is intuitively preferable since it will accept new parameters that increase the likelihood by any amount. We only use version 2 above to prove Theorem 1.

Another concern with Algorithm (\ref{alg:EM-VRSO-v2}) is that $F^{*}(\bfphi_{k})$ is defined using an infimum, which can be problematic in some settings. However, we prove in Lemma 1 that if the conditions of Theorem 1 hold, then $F^*(\bfphi')$ is finite, has a unique minimizer, and is uniformly continuous.

\begin{lemma}
    If the conditions of Theorem 1 hold, then for all $\bfphi_k \in \bfPhi$, $F^{*}(\bfphi_k) = \inf_{\bfphi \in \bfPhi} F(\bfphi \mid \bfphi_k)$ is finite and has a unique minimizer, so $F^*(\bfphi_k) = \min_{\bfphi \in \bfPhi} F(\bfphi \mid \bfphi_k) = F(\bfphi_{k+1}^* \mid \bfphi_k)$ for a unique $\bfphi_{k+1}^* \in \bfPhi$. Further, $F^*(\bfphi_k)$ is uniformly continuous for all $\bfphi_k \in \bfPhi$.
\end{lemma}

\begin{proof}
    For any fixed $\bfphi_k$, if the conditions of Theorem 1 hold, then there exists a unique minimizer of $F(\cdot \mid \bfphi_k)$ over $\bfPhi_{\bfphi_{k}}$ because $\bfPhi_{\bfphi_{k}}$ is compact and $F$ is strongly convex and continuous. Denote this minimizer as $\bfphi_{k+1}^* = \argmin_{\bfphi \in \bfPhi_{\bfphi_{k}}} F(\bfphi \mid \bfphi_{k})$. We show that $\bfphi_{k+1}^*$ minimizes $F(\cdot \mid \bfphi_{k})$ over the \textit{entirety} of $\bfPhi$ using two cases:
    
    \begin{enumerate}
        \item For any $\bfphi \in \bfPhi_{\bfphi_{k}}$ with $\bfphi \neq \bfphi_{k+1}^*$, we have that $F(\bfphi \mid \bfphi_{k}) > F(\bfphi_{k+1}^* \mid \bfphi_{k})$ because $\bfphi_{k+1}^*$ is the unique minimizer over $\bfPhi_{\bfphi_{k}}$. 
        \item For any $\bfphi \notin \bfPhi_{\bfphi_{k}}$, $\log p(\bfy;\bfphi) < \log p(\bfy;\bfphi_k) \leq \log p(\bfy;\bfphi^*_{k+1})$ by the definition of $\bfPhi_{\bfphi_{k}}$. This implies that $F(\bfphi \mid \bfphi_{k}) > F(\bfphi_{k} \mid \bfphi_{k}) \geq F(\bfphi^*_{k+1} \mid \bfphi_{k})$ by basic properties of the EM algorithm \citep{Dempster:1977}.
    \end{enumerate} 

    Therefore, $\bfphi_{k+1}^*$ is a global minimizer of $F(\cdot \mid \bfphi_k)$, and $F^{*}(\bfphi_{k}) = F(\bfphi_{k+1}^* \mid \bfphi_k)$.

    We now prove that $F^*(\bfphi_k)$ is uniformly continuous. Let $\epsilon > 0$ be fixed. Then, because $F(y \mid x)$ is uniformly continuous in $(x,y) \in \bfPhi \times \bfPhi$, there exists some $\delta$ such that for all $(y,x)$ and $(y,x_0)$ with $|x-x_0| < \delta$ we have that $|F(y \mid x) - F(y \mid x_0)| < \epsilon$. Now, we fix arbitrary $x$ and $x_0$ with $|x-x_0| < \delta$ and show that $|F^*(x) - F^*(x_0)| < \epsilon$. To this end, pick $y^*$ such that $F(y^* \mid x) = \min_{y \in \bfPhi} F(y \mid x)= F^*(x)$ and pick $y_0^*$ such that $F(y_0^* \mid x_0) = \min_{y \in \bfPhi} F(y \mid x_0) = F^*(x_0)$. Then: 
    
    \begin{align*}
        F^*(x_0) &= F(y_0^* \mid x_0) \leq F(y^* \mid x_0) < F(y^* \mid x) + \epsilon = F^*(x) + \epsilon, \\ 
        F^*(x_0) &= F(y_0^* \mid x_0) > F(y_0^* \mid x) - \epsilon \geq F(y^* \mid x) - \epsilon = F^*(x) - \epsilon,
    \end{align*}
    %
    so $|F^*(x) - F^*(x_0)| < \epsilon$.
\end{proof}

\section{Proof of Theorem 1}

We begin by introducing notation needed for the proof. Let $I \in \{1,\ldots,T\}^M$ be an $M$-dimensional vector of indices. Let $\texttt{VRSO-PE}^*(I,\lambda,\bfphi')$ correspond to the mapping that results from performing $M$ steps of SVRG on objective function $F(\cdot \mid \bfphi')$ starting at $\bfphi'$ with step size $\lambda$ and random realization $I$. That is, $\texttt{VRSO-PE}^*(I,\lambda,\bfphi')$ equals $\texttt{VRSO-PE}(\{\bfalpha_t(\bfphi'),\bfbeta_t(\bfphi'),\bfgamma_t(\bfphi'),\bfxi_t(\bfphi')\}_{t=1}^{T}, \bfphi', \lambda, \text{SVRG}, \texttt{False}, M)$ when the random vector of indices $\begin{pmatrix} t_0 & \cdots & t_{M-1} \end{pmatrix}$ is equal to $I$. We prove that $\texttt{VRSO-PE}^*(I,\lambda,\bfphi')$ is uniformly continuous in $\bfphi' \in \bfPhi$ for fixed $I$ and $\lambda$ in Lemma 2 below.

\begin{lemma}
    Suppose all conditions from Theorem 1 hold. Then $\texttt{VRSO-PE}^*(I,\lambda,\bfphi')$ is uniformly continuous in $\bfphi' \in \bfPhi$ for fixed $I$ and $\lambda$.
\end{lemma}

\begin{proof}
    Define $\bfphi'^{(0)} = \bfphi'$. The function $\texttt{VRSO-PE}^*(I,\lambda,\bfphi')$ involves calculating $\bfphi'^{(0)}, \bfphi'^{(1)}, \ldots, \bfphi'^{(m)}, \ldots, \bfphi'^{(M-1)}$ by iteratively applying the following mapping (with $t_m = I[m+1]$) and returning $\bfphi'^{(M)} \equiv \bfphi$:
    \begin{equation*}
        \bfphi'^{(m+1)}(\bfphi'^{(m)},\bfphi'^{(0)}) = \lambda \left[\nabla F_{t_m}(\bfphi'^{(m)} \mid \bfphi'^{(0)}) - \nabla F_{t_m}(\bfphi'^{(0)} \mid \bfphi'^{(0)}) + \frac{1}{T}\sum_{t=1}^T\nabla F_{t}(\bfphi'^{(0)} \mid \bfphi'^{(0)}) \right].
    \end{equation*}
    Note that $\nabla F_t(\bfphi \mid \bfphi')$ is uniformly continuous in $(\bfphi,\bfphi')$ for each $t$ by condition (7) of Theorem 1. Therefore, the mapping above is also uniformly continuous in $(\bfphi'^{(m)},\bfphi'^{(0)})$, and so $\texttt{VRSO-PE}^*(I,\lambda,\bfphi')$ is uniformly continuous in $\bfphi'$ for fixed $I$ and $\lambda$.
\end{proof}

If the conditions of Theorem 1 are met, iteration $k$ of Algorithm (\ref{alg:EM-VRSO-v2}) is equivalent to drawing a sequence of index vectors $I_{k,0},I_{k,1},\ldots$ and calculating $\bfphi_{k,\ell} = \texttt{VRSO-PE}^*(I_{k,\ell},\lambda,\bfphi_{k})$ until the condition to exit the while loop is satisfied, which happens in finite time by Lemma 3 below. Lemma 3 depends upon Theorem 1 of \citet{Johnson:2013}, so Table (\ref{tbl:notation}) identifies how our notation corresponds to theirs.
\begin{table}[]
\centering
\begin{tabular}{c|c}
\citet{Johnson:2013}                  & Our Notation                          \\ \hline
$\alpha$                              & $\zeta$                               \\
$\eta$                                & $\lambda$                             \\
$L$                                   & $L$                                   \\
$\gamma$                              & $C$                                   \\
$m$                                   & $M$                                   \\
$\tilde{w}_0$                         & $\bfphi_{k}$                          \\
$\tilde{w}_1$                         & $\bfphi_{k,\ell}$                     \\
$w_{*}$                               & $\bfphi^*_{k+1}$                      \\
$P$                                   & $F(\cdot \mid \bfphi_k)$              \\
$\psi_i$                              & $F_t(\cdot \mid \bfphi_k)$                         
\end{tabular}
\caption{Legend connecting this paper's notation to that of \citet{Johnson:2013}.}
\label{tbl:notation}
\end{table}

\begin{lemma}
    Suppose all conditions from Theorem 1 hold. Then, the while loop within Algorithm (\ref{alg:EM-VRSO-v2}) almost surely terminates in finite time, i.e. $\bbP\{\ell^*(k) < \infty\} = 1$ for all $k \geq 0$. Furthermore, $\bbP\{\forall k : \ell^*(k) < \infty \} = 1$.
\end{lemma}

\begin{proof}

By Lemma 1, $F^*(\bfphi_k) = \inf_{\bfphi \in \bfPhi} F(\bfphi \mid \bfphi_k)$ is finite and uniformly continuous in $\bfphi_k$. Further, Theorem 1 of \citet{Johnson:2013} applies for one iteration through $M$ steps of SVRG. In particular, for all $\ell \geq 0$:

\begin{align}
    \bbE & \left[F(\bfphi_{k,\ell} \mid \bfphi_{k}) - F^*(\bfphi_k) ~\Big\vert~ \bfphi_{k} \right] \leq \zeta \Big(F(\bfphi_{k} \mid \bfphi_{k}) - F^*(\bfphi_k) \Big), \label{eqn:SVRG_T1}
\end{align}
where $\zeta$ is defined in condition (6) of Theorem 1. Using (\ref{eqn:SVRG_T1}) in Markov's inequality gives

\begin{align}
    \bbP \Big[F(\bfphi_{k,\ell} \mid \bfphi_{k}) - F^*(\bfphi_{k}) \geq \frac{1 + \zeta}{2} \left(F(\bfphi_{k} \mid \bfphi_{k}) - F^*(\bfphi_k) \right) ~\Big\vert~ \bfphi_{k} \Big] &\leq \frac{\zeta \left( F(\bfphi_{k} \mid \bfphi_{k}) - F^*(\bfphi_k) \right)}{\frac{1+\zeta}{2} \Big( F(\bfphi_{k} \mid \bfphi_{k}) - F^*(\bfphi_k) \Big)} \\
    &= \frac{2}{1 + 1/\zeta} < 1.
\end{align}

Taking the complement of the above expression and rearranging terms gives

\begin{equation}
    \bbP \Big[F(\bfphi_{k,\ell} \mid \bfphi_{k}) \leq F^*(\bfphi_{k}) + \frac{1 + \zeta}{2} \Big(F(\bfphi_{k} \mid \bfphi_{k}) - F^*(\bfphi_{k}) \Big) ~\Big\vert~ \bfphi_{k} \Big] \geq \frac{1-\zeta}{1 + \zeta} > 0 \label{eqn:markov_ineq},
\end{equation}

so with some probability greater than or equal to $(1-\zeta)/(1+\zeta) > 0$, the condition to exit the while loop at iteration $\ell$ of Algorithm (\ref{alg:EM-VRSO-v2}) is satisfied.

Further, repeatedly calling \texttt{VRSO-PE} within the inner while loop of Algorithm (\ref{alg:EM-VRSO-v2}) corresponds to drawing \textit{independent} samples of $\bfphi_{k,\ell}$ conditioned on $\bfphi_k$. The draws are independent because $\texttt{VRSO-PE}^*$ is only random due to $I_{k,\lambda}$ after conditioning on $\bfphi_k$, and each iteration of the inner while loop of Algorithm (\ref{alg:EM-VRSO-v2}) involves drawing independent samples of $I_{k,\lambda}$. As a result, $\ell^*(k)$ follows a geometric distribution with some positive success probability $p_k \geq (1-\zeta)/(1+\zeta)$. Therefore, $\bbP\Big\{\ell^*(k) < \infty \Big\} = \sum_{\ell \geq 1} p_k(1-p_k)^{\ell-1} = 1$, and 

\begin{equation*}
    \bbP\Big\{\forall k : \ell^*(k) < \infty \Big\} = 1 - \bbP\Big\{\exists k : \ell^*(k) = \infty \Big\} = 1.
\end{equation*}
\end{proof}

Finally, we introduce $R_{M,\lambda}$ as a point-to-set map corresponding to one iteration of Algorithm (\ref{alg:EM-VRSO-v2}) with $M$ steps of step size $\lambda$. First, recall that

\begin{equation}
    K(\bfphi') = F^*(\bfphi') + [(1 + \zeta)/2] \Big(F(\bfphi' \mid \bfphi') - F^*(\bfphi') \Big). 
\end{equation}

Then, we define $R_{M,\lambda}(\bfphi')$ as

\begin{equation}
    R_{M,\lambda}(\bfphi') = \Big\{\bfphi ~ : ~ \bfphi = \texttt{VRSO-PE}^*(I,\lambda,\bfphi') ~ \text{for some} ~ I \in \{1,\ldots,T\}^M ~\text{and}~ F(\bfphi \mid \bfphi') \leq K(\bfphi')\Big\}.
\end{equation}

Iteration $k$ of Algorithm (\ref{alg:EM-VRSO-v2}) corresponds to randomly sampling $\bfphi_{k+1}$ from $R_{M,\lambda}(\bfphi_k)$. If the conditions of Theorem 1 are met, then $\bbP(\ell^*(k) < \infty) = 1$, so $R_{M,\lambda}(\bfphi_k)$ must not be the empty set. We prove two more useful Lemmas regarding $R_{M,\lambda}$ before finally proving Theorem 1.

\begin{lemma}
    Suppose all conditions from Theorem 1 hold. Then $R_{M,\lambda}$ is a closed point-to-set map for all non-stationary points $\bfphi' \in \bfPhi$. Note that $R_{M,\lambda}$ is closed at a point $\bfphi'$ if $\bfphi'_{n} \to \bfphi'$ and $\bfphi_{n} \to \bfphi$ with $\bfphi_{n} \in R_{M,\lambda}(\bfphi'_{n})$, implies that $\bfphi \in R_{M,\lambda}(\bfphi')$. 
\end{lemma}

\begin{proof}
     Given a point $\bfphi'$, suppose there exists some sequence $\bfphi'_{n} \to \bfphi'$ as well as another sequence $\bfphi_{n} \to \bfphi$ with $\bfphi_{n} \in R_{M,\lambda}(\bfphi'_{n})$. To prove the Lemma, we prove that $\bfphi \in R_{M,\lambda}(\bfphi')$, i.e. that
    \begin{enumerate}
        \item $F(\bfphi \mid \bfphi') \leq K(\bfphi')$, and
        \item $\bfphi = \texttt{VRSO-PE}^*(I,\lambda,\bfphi')$ for some $I \in \{1,\ldots,T\}^M$
    \end{enumerate}
    We start with the first condition. Since $\bfphi_{n} \to \bfphi$ and $\bfphi'_{n} \to \bfphi'$,  \ we have that $F(\bfphi_{n} \mid \bfphi'_{n}) \to F(\bfphi \mid \bfphi')$ because $F$ is continuous by condition (7) of Theorem 1. Likewise, $K(\bfphi'_{n}) \to K(\bfphi')$ because $K$ is continuous by Lemma 1. Finally, $F(\bfphi \mid \bfphi') \leq K(\bfphi')$ because for all $n$, $\bfphi_n \in R_{M,\lambda}(\bfphi_n')$, which implies that $F(\bfphi_{n} \mid \bfphi'_{n}) \leq K(\bfphi'_{n})$. Taking the limit of both sides as $n \to \infty$ gives the result.
    
    We now show the second condition. By way of contradiction, assume that $\bfphi \neq \texttt{VRSO-PE}^*(I,\lambda,\bfphi')$ for any value of $I \in \{1,\ldots,T\}^M$. Then, $\min_I ||\bfphi - \texttt{VRSO-PE}^*(I,\lambda,\bfphi')||$ is strictly positive. 
    By assumption, for each $n$, $\bfphi_{n} \in R_{M,\lambda}(\bfphi'_{n})$, so $\bfphi_{n} = \texttt{VRSO-PE}^*(I_n,\lambda,\bfphi'_{n})$ for some $I_n \in \{1,\ldots,T\}^M$. 
    Using the triangle inequality, we have:

    \begin{align}
        0 < \min_{I}||\bfphi - \texttt{VRSO-PE}^*(I,\lambda,\bfphi')|| &\leq ||\bfphi_{n} - \bfphi|| + ||\bfphi_{n} - \texttt{VRSO-PE}^*(I_n,\lambda,\bfphi')|| \\
        &= ||\bfphi_{n} - \bfphi|| + ||\texttt{VRSO-PE}^*(I_n,\lambda,\bfphi'_{n}) - \texttt{VRSO-PE}^*(I_n,\lambda,\bfphi')|| \\
        &\leq ||\bfphi_{n} - \bfphi|| + \max_{I} ||\texttt{VRSO-PE}^*(I,\lambda,\bfphi'_{n}) - \texttt{VRSO-PE}^*(I,\lambda,\bfphi')||.
    \end{align}
    
    However, the right-hand side of the equation converges to zero since $\bfphi_n \to \bfphi$, $\bfphi'_n \to \bfphi'$, and $\texttt{VRSO-PE}^*(I,\lambda,\bfphi')$ is continuous for fixed $I$ and $\lambda$. This is a contradiction, so it must be that $\bfphi = \texttt{VRSO-PE}^*(I,\lambda,\bfphi')$ for some $I$.
\end{proof}

\begin{lemma}
    Suppose all conditions from Theorem 1 hold, and consider the sequence $\{\bfphi_k\}_{k=0}^{K}$ generated from running Algorithm (\ref{alg:EM-VRSO-v2}) with $\bfphi_0 \in \bfPhi$. For $k = 0,\ldots,K-1$, we have that $\log p(\bfy;\bfphi_{k+1}) \geq \log p(\bfy;\bfphi_k)$. Further, if $\bfphi_k$ is not a stationary point of $\log p$, then $\log p(\bfy;\bfphi_{k+1}) > \log p(\bfy;\bfphi_k)$.
\end{lemma}

\begin{proof}

First, note that $\bfphi_k$ is a stationary point of $\log p$ if and only if it is a stationary point of $F(\cdot \mid \bfphi_k)$ because $\nabla \log p(\bfphi_k) = \nabla Q(\bfphi_k \mid \bfphi_k) = -\nabla F(\bfphi_k \mid \bfphi_k) / T$.

We begin with the case where $\bfphi_k$ is a stationary point of $\log p$ (and therefore a stationary point of $F(\cdot \mid \bfphi_k)$). Then, $F(\bfphi_k \mid \bfphi_k) = F^{*}(\bfphi_{k})$ since $F$ is continuously differentiable and strongly convex. To exit the while loop at iteration $k$ of Algorithm (\ref{alg:EM-VRSO-v2}), it must be the case that $F(\bfphi_{k+1} \mid \bfphi_{k}) \leq F^*(\bfphi_{k}) + [(1+\zeta)/2] \Big(F(\bfphi_{k} \mid \bfphi_{k}) - F^*(\bfphi_{k}) \Big) = F(\bfphi_k \mid \bfphi_k)$. Finally, we know from \citet{Dempster:1977} that $\log p(\bfy ; \bfphi_{k+1}) - \log p(\bfy ; \bfphi_{k}) \geq F(\bfphi_{k} \mid \bfphi_{k}) - F(\bfphi_{k+1} \mid \bfphi_{k}) \geq 0$. This implies that $\log p(\bfy ; \bfphi_{k+1}) \geq \log p(\bfy ; \bfphi_{k})$.

Next, we focus on the case where $\bfphi_k$ is \textit{not} a stationary point of $\log p$ (and therefore not a stationary point $F(\cdot \mid \bfphi_k)$). Then, $F(\bfphi_k \mid \bfphi_k) > F^{*}(\bfphi_{k})$ since $F$ is continuously differentiable and strongly convex. To exit the while loop at iteration $k$ of Algorithm (\ref{alg:EM-VRSO-v2}), it must be the case that $F(\bfphi_{k+1} \mid \bfphi_{k}) \leq F^*(\bfphi_{k}) + [(1+\zeta)/2] \Big(F(\bfphi_{k} \mid \bfphi_{k}) - F^*(\bfphi_{k}) \Big) < F(\bfphi_k \mid \bfphi_k)$. Finally, we know from \citet{Dempster:1977} that $\log p(\bfy ; \bfphi_{k+1}) - \log p(\bfy ; \bfphi_{k}) \geq F(\bfphi_{k} \mid \bfphi_{k}) - F(\bfphi_{k+1} \mid \bfphi_{k}) > 0$. This implies that $\log p(\bfy ; \bfphi_{k+1}) > \log p(\bfy ; \bfphi_{k})$.
\end{proof}

We now prove Theorem 1.

\begin{proof}

The first part of Theorem 1 states that $\bbP(\ell^*(k) < \infty) = 1$ for all $k \geq 0$ and is true by Lemma 3. The second part of Theorem 1 states that all limit points of $\{\bfphi_{k}\}_{k=0}^\infty$ are stationary points of $\log p(\bfy;\bfphi)$, and $\log p(\bfy;\bfphi_{k})$ converges monotonically to $\log p^* = \log p(\bfy;\bfphi^*)$ for some stationary point of $\log p$, $\bfphi^*$. This is a direct application of Theorem 1 of \citet{Wu:1983}, which requires the following conditions:

\begin{enumerate}[label=(\alph*)]
    \item $R_{M,\lambda}$ is a closed point-to-set map for all non-stationary points.
    \item $\log p(\bfy;\bfphi_{k+1}) \geq \log p(\bfy;\bfphi_k)$ for all $\bfphi_k  \in \bfPhi$, and $\log p(\bfy;\bfphi_{k+1}) > \log p(\bfy;\bfphi_k)$ for all $\bfphi_k \in \bfPhi$ that are not stationary points of $\log p$.
\end{enumerate}

Condition (a) is satisfied by Lemma 4, and condition (b) is satisfied by Lemma 5. 
\end{proof}

\section{Proof of Theorem 2}

We first give a basic outline of the proof. If $\nabla \log p(\bfy;\bfphi_0) = 0$, then $\nabla F(\bfphi_0 \mid \bfphi_0) = 0$, so every gradient step of SAGA and SVRG within $\texttt{VRSO-PE}$ does not change $\bfphi_0^{(m)}$, so $ = \bfphi_0^{(m+1)} =  \bfphi_0^{(m)} = \bfphi^{(0)}_0$ (we prove this in step 3 of lemma 6). As a result, the gradient approximations do not change either, so $\widehat \nabla F^{(m+1)} = \widehat \nabla F^{(m)}$ and $\widehat \nabla F^{(m+1)}_t = \widehat \nabla F^{(m)}_t$ for $t = 1,\ldots,T$ (we prove this in steps 4 and 5 of lemma 6). Finally, since the parameter estimates do not change, the conditional probability approximations $\left\{\widehat \bfalpha^{(m)}_t, \widehat \bfbeta^{(m)}_t, \widehat \bfgamma^{(m)}_t, \widehat \bfxi^{(m)}_t\right\}_{t=1}^T = \left\{\bfalpha_t(\bfphi_0), \bfbeta_t(\bfphi_0), \bfgamma_t(\bfphi_0), \bfxi_t(\bfphi_0) \right\}_{t=1}^T$ do not change either (we prove this in steps 1 and 2 of lemma 6). Because $\texttt{VRSO-PE}$ returns $\bfphi_1 = \bfphi_0^{(M)} = \bfphi_0$, each iteration of $\texttt{EM-VRSO}$ must do the same, so by induction $\bfphi_K = \bfphi_0$ (we prove this in the body of theorem 2).

To prove the Theorem more formally, we first prove a useful Lemma.

\begin{lemma}
    If:
    \begin{enumerate}
        \item $\widehat \bfalpha^{(0)}_t = \bfalpha_t(\bfphi^{(0)})$ and $\widehat \bfbeta^{(0)}_{t} = \bfbeta_t(\bfphi^{(0)})$ for all $t = 1,\ldots,T$,
        \item $\widehat \bfgamma^{(0)}_{t} = \bfgamma_t(\bfphi^{(0)})$ and $\widehat \bfxi^{(0)}_{t} = \bfxi_t(\bfphi^{(0)})$ for all $t = 1,\ldots,T$,
        \item $\nabla F(\bfphi^{(0)} \mid \bfgamma(\bfphi^{(0)}),\bfxi(\bfphi^{(0)})) = 0$,
    \end{enumerate}
    then with probability 1,
    \begin{equation}
        \texttt{VRSO-PE}\Big(\left\{\widehat \bfalpha_t^{(0)},\widehat \bfbeta_t^{(0)},\widehat \bfgamma_t^{(0)},\widehat \bfxi_t^{(0)}\right\}_{t=1}^T, \bfphi^{(0)}, \lambda, A, P, M \Big) = \bfphi^{(0)}
    \end{equation}
    for all $\lambda \in \bbR$, $A \in \{\text{SAGA}, \text{SVRG}\}$, $P \in \{\texttt{True},\texttt{False}\}$, and $M \in \bbN$.
\end{lemma}

\begin{proof}

    We use induction to show that the following conditions hold for all $m \geq 0$:

    \begin{enumerate}
        \item $\widehat \bfalpha^{(m)}_t = \bfalpha_t(\bfphi^{(0)})$ and $\widehat \bfbeta^{(m)}_{t} = \bfbeta_t(\bfphi^{(0)})$ for all $t = 1,\ldots,T$.
        \item $\widehat \bfgamma^{(m)}_{t} = \bfgamma_t(\bfphi^{(0)})$ and $\widehat \bfxi^{(m)}_{t} = \bfxi_t(\bfphi^{(0)})$ for all $t = 1,\ldots,T$.
        \item $\bfphi^{(m)} = \bfphi^{(0)}$
        \item $\widehat \nabla F^{(m)}_{t} = \nabla F_t\left(\bfphi^{(0)} \mid \bfgamma_t(\bfphi^{(0)}),\bfxi_t(\bfphi^{(0)})\right)$ for all $t = 1,\ldots,T$.
        \item $\widehat \nabla F^{(m)} = 0$
    \end{enumerate}

    Algorithm (\ref{alg:VRSO-PE}) returns $\bfphi^{(M)}$ for some finite $M$, so proving the above will prove the lemma. 
    The base case ($m=0$) is trivial either from the assumptions of the Lemma or from lines 1--4 of Algorithm (\ref{alg:VRSO-PE}), so we focus on the inductive case. We assume that all of the conditions (1--5) above hold for $m$, and prove each condition holds for $m+1$ sequentially.

    \begin{enumerate}
        \item First note that if $P = \texttt{False}$ or if $P = \texttt{True}$ and $t \neq t_m$, then the conditional probability approximations are not updated, so $\widehat \bfalpha^{(m+1)}_t = \widehat \bfalpha^{(m)}_t  = \bfalpha_t(\bfphi^{(0)})$ and $\widehat \bfbeta^{(m+1)}_t = \widehat \bfbeta^{(m)}_t = \bfbeta_t(\bfphi^{(0)})$ by inductive hypothesis (1).
        
        If $P = \texttt{True}$ and $t = t_m$, then both $\widehat \bfalpha^{(m+1)}_{t_m}$ and $\widehat \bfbeta^{(m+1)}_{t_m}$ are updated in line 9 of Algorithm (\ref{alg:EM-VRSO}). The approximation $\widehat \bfalpha^{(m+1)}_{t_m}$ is updated using the mapping 
        \begin{align}
            \widehat \bfalpha^{(m+1)}_{t_m} &= \widetilde \bfalpha_{t_m}(\widehat \bfalpha^{(m)}_{t_m-1},\bfphi^{(m)}) \nonumber \\
            &= \widetilde \bfalpha_{t_m}(\bfalpha_{t_m-1}(\bfphi^{(0)}),\bfphi^{(0)}) \nonumber \\
            &= 
            \begin{cases}
                \bfdelta(\bfeta^{(0)}) ~ P(y_1;\bftheta^{(0)}), & \text{for } t_m = 1 \\
                \bfalpha_{t_m-1}(\bfphi^{(0)}) ~ \bfGamma(\bfeta^{(0)}) ~P(y_{t_m};\bftheta^{(0)}), & \text{for } t_m = 2,\ldots,T
            \end{cases} \nonumber \\
            &= \bfalpha_{t_m}(\bfphi^{(0)}) \label{eqn:a_inductive_step}
        \end{align}
        The second line is true by inductive hypotheses (1) and (3), the third line is the definition of $\widetilde \bfalpha_{t_m}$, and the final line is the definition of $\bfalpha_{t_m}(\bfphi^{(0)})$. Similar logic can be used to show that 
        
        \begin{equation}
            \widehat \bfbeta^{(m+1)}_{t_m} = \bfbeta_{t_m}(\bfphi^{(0)}). \label{eqn:b_inductive_step}
        \end{equation}
        \item First note that if $P = \texttt{False}$ or if $P = \texttt{True}$ and $t \neq t_m$, then the conditional probability approximations are not updated, so $\widehat \bfgamma^{(m+1)}_t = \widehat \bfgamma^{(m)}_t = \bfgamma_t(\bfphi^{(0)})$ and $\widehat \bfxi^{(m+1)}_t = \widehat \bfxi^{(m)}_t = \bfxi_t(\bfphi^{(0)})$ by inductive hypothesis (2).
        
        If $P = \texttt{True}$ and $t = t_m$, both $\widehat \bfgamma^{(m+1)}_{t_m}$ and $\widehat \bfxi^{(m+1)}_{t_m}$ are calculated in line 10 of Algorithm (\ref{alg:EM-VRSO}). $\widehat \bfxi^{(m+1)}_{t_m}$ is updated using the mapping 
        
        \begin{align}
            \widehat \bfxi^{(m+1)}_{t_m} &= \widetilde \bfxi_{t_m}(\widehat \bfalpha^{(m+1)}_{t_m-1},\widehat \bfbeta^{(m+1)}_{t_m}, \bfphi^{(m)}) \nonumber \\
            &= \widetilde \bfxi_{t_m}(\bfalpha_{t_m-1}(\bfphi^{(0)}),\bfbeta_{t_m}(\bfphi^{(0)}), \bfphi^{(0)}) \nonumber  \\
            &= \frac{\text{diag}(\bfalpha_{t-1}(\bfphi^{(0)})) ~~ \bfGamma(\bfeta^{(0)}) ~~ P(y_t;\bftheta^{(0)}) ~~ \text{diag}(\bfbeta_t(\bfphi^{(0)}))}{\bfalpha_{t-1}(\bfphi^{(0)}) ~~ \bfGamma(\bfeta^{(0)}) ~~ P(y_{t};\bftheta^{(0)}) ~~ \bfbeta_{t}(\bfphi^{(0)})^\top} \nonumber  \\
            &= \bfxi_{t_m}(\bfphi^{(0)}). \label{eqn:g_inductive_step}
        \end{align}
        The second line is true by Equations (\ref{eqn:a_inductive_step} -- \ref{eqn:b_inductive_step}) and inductive hypothesis (3), the third line is the definition of $\widetilde \bfxi_{t_m}$, and the final line is the definition of $\bfxi_{t_m}(\bfphi^{(0)})$. Similar logic can be used to show that 
        
        \begin{equation}
            \widehat \bfgamma^{(m+1)}_{t_m} = \bfgamma_{t_m}(\bfphi^{(0)}). \label{eqn:x_inductive_step}
        \end{equation}
        \item The parameter $\bfphi^{(m+1)}$ is calculated in line 12 of \texttt{VRSO-PE} as follows: 
        
        \begin{align*}
            \bfphi^{(m+1)} &= \bfphi^{(m)} - \lambda \left[\nabla F_{t_m}\left(\bfphi^{(m)} \mid \widehat \bfgamma^{(m+1)}_{t_m}, \widehat \bfxi^{(m+1)}_{t_m}\right) - \widehat \nabla F^{(m)}_{t_m} + \widehat \nabla F^{(m)} \right] \\
            &= \bfphi^{(0)} - \lambda \left[\nabla F_{t_m}\left(\bfphi^{(0)} \mid \widehat \bfgamma^{(m+1)}_{t_m}, \widehat \bfxi^{(m+1)}_{t_m}\right) - \nabla F_{t_m}\left(\bfphi^{(0)} \mid \bfgamma_{t_m}(\bfphi^{(0)}), \bfxi_{t_m}(\bfphi^{(0)})\right) + 0 \right] \\
            &= \bfphi^{(0)} - \lambda \left[ \nabla F_{t_m}\left(\bfphi^{(0)} \mid \bfgamma_{t_m}(\bfphi^{(0)}), \bfxi_{t_m}(\bfphi^{(0)})\right) - \nabla F_{t_m}\left(\bfphi^{(0)} \mid \bfgamma_{t_m}(\bfphi^{(0)}), \bfxi_{t_m}(\bfphi^{(0)})\right) \right] \\
            &= \bfphi^{(0)}.
        \end{align*}
        The second line is true by the inductive hypotheses (3), (4), and (5), and the third line is true by Equations (\ref{eqn:g_inductive_step} -- \ref{eqn:x_inductive_step}).
        \item Note that if $A = \text{SVRG}$ or if $A = \text{SAGA}$ and $t \neq t_m$, then the gradient approximation at index $t$ is not updated, so $\widehat \nabla F^{(m+1)}_t = \widehat \nabla F^{(m)}_t = \nabla F_t\left(\bfphi^{(0)} \mid \bfgamma_t(\bfphi^{(0)}),\bfxi_t(\bfphi^{(0)})\right)$ by inductive hypothesis (4).
        
        If $A = \text{SAGA}$ and $t = t_m$, then $\widehat \nabla F^{(m+1)}_{t_m}$ is calculated in line 14 of \texttt{VRSO-PE} as follows:
        
        \begin{align}
            \widehat \nabla F^{(m+1)}_{t_m} &= \nabla F_{t_m}\left(\bfphi^{(m)} \mid \widehat \bfgamma^{(m+1)}_{t_m}, \widehat \bfxi^{(m+1)}_{t_m}\right) \nonumber \\
            &= \nabla F_{t_m}\left(\bfphi^{(0)} \mid \widehat \bfgamma^{(m+1)}_{t_m}, \widehat \bfxi^{(m+1)}_{t_m}\right) \nonumber \\
            &= \nabla F_{t_m}\left(\bfphi^{(0)} \mid \bfgamma_{t_m}(\bfphi^{(0)}), \bfxi_{t_m}(\bfphi^{(0)})\right). \label{eqn:nabla_hat_inductive_step}
        \end{align}
        The second line is true by inductive hypothesis (3), and the third line is true by Equations (\ref{eqn:g_inductive_step} -- \ref{eqn:x_inductive_step}).
        \item If $A = \text{SVRG}$, then the full gradient approximation is not updated, so $\widehat \nabla F^{(m+1)} = \widehat \nabla F^{(m)} = 0$ by inductive hypothesis (5).
        
        If $A = \text{SAGA}$, then $\widehat \nabla F^{(m+1)}$ is calculated in line 14 of \texttt{VRSO-PE} as follows:
        
        \begin{align*}
            \widehat \nabla F^{(m+1)} &= \widehat \nabla F^{(m)} + \frac{1}{T} \left(\widehat \nabla F^{(m+1)}_{t_m} - \widehat \nabla F^{(m)}_{t_m}\right) \\
            &= 0 + \frac{1}{T} \left[\widehat \nabla F^{(m+1)}_{t_m} - \nabla F_{t_m}\left(\bfphi^{(0)} \mid \bfgamma_{t_m}(\bfphi^{(0)}), \bfxi_{t_m}(\bfphi^{(0)})\right)\right] \\
            &= \frac{1}{T} \left[\nabla F_{t_m}\left(\bfphi^{(0)} \mid \bfgamma_{t_m}(\bfphi^{(0)}), \bfxi_{t_m}(\bfphi^{(0)})\right) - \nabla F_{t_m}\left(\bfphi^{(0)} \mid \bfgamma_{t_m}(\bfphi^{(0)}), \bfxi_{t_m}(\bfphi^{(0)})\right)\right] \\
            &= 0.
        \end{align*}
        The second line is true by inductive hypotheses (4) and (5), and the third line is true by Equation (\ref{eqn:nabla_hat_inductive_step}).
    \end{enumerate}
    This completes the inductive step, and proves the lemma.
\end{proof}

We now prove Theorem 2.

\begin{proof}

We use induction and show that if $\nabla \log p(\bfy;\bfphi_0) = 0$, then $\bfphi_{k} = \bfphi_0$ for all iterations $k$ of Algorithm (\ref{alg:EM-VRSO}). The base case, $k = 0$, is trivial. For the inductive case, we assume that $\bfphi_k = \bfphi_0$ and prove that $\bfphi_{k+1} = \bfphi_k$. First, note that 

\begin{align}
    \nabla F(\bfphi_{0} \mid \bfgamma(\bfphi_{0}), \bfxi(\bfphi_{0})) &= \frac{1}{T} \sum_{t=1}^T \nabla F_t(\bfphi_0 \mid \bfgamma_t(\bfphi_0), \bfxi_t(\bfphi_0)) \nonumber \\
    &= -\frac{1}{T} \nabla Q(\bfphi_0 \mid \bfphi_0) \nonumber \\
    &= -\frac{1}{T} \nabla \log p(\bfy ~;~ \bfphi_0) \nonumber \\
    &= 0 \label{eqn:nabla_zero}.
\end{align}
Equation (\ref{eqn:nabla_zero}) shows that the condition for Lemma 6 are satisfied with $\bfphi^{(0)} = \bfphi_0$. Further, step 2 of Algorithm (\ref{alg:EM-VRSO}) performs the E step of the EM algorithm so that $\{\bfalpha_{k,t}, \bfbeta_{k,t}, \bfgamma_{k,t}, \bfxi_{k,t}\}_{t=1}^T = \left\{\bfalpha_t(\bfphi_k), \bfbeta_t(\bfphi_k), \bfgamma_t(\bfphi_k), \bfxi_t(\bfphi_k) \right\}_{t=1}^T$. So:

\begin{align*}
    \bfphi_{k,1} &= \texttt{VRSO-PE}\Big(\left\{\bfalpha_{k,t}, \bfbeta_{k,t}, \bfgamma_{k,t}, \bfxi_{k,t}\right\}_{t=1}^T, \bfphi_k, \lambda, A, P, M \Big) \\
    &= \texttt{VRSO-PE}\Big(\left\{\bfalpha_t(\bfphi_k),\bfbeta_t(\bfphi_k),\bfgamma_t(\bfphi_k),\bfxi_t(\bfphi_k)\right\}_{t=1}^T, \bfphi_k, \lambda, A, P, M \Big)\\
    &= \texttt{VRSO-PE}\Big(\left\{\bfalpha_t(\bfphi_0),\bfbeta_t(\bfphi_0),\bfgamma_t(\bfphi_0),\bfxi_t(\bfphi_0)\right\}_{t=1}^T, \bfphi_0, \lambda, A, P, M \Big)\\
    &= \bfphi_0.
\end{align*}

The second line is true by the definition of the E step in the EM algorithm, the third line is true by the inductive hypothesis, and the final line is true by Lemma 6. Since $\log p(\bfy; \bfphi_{k,1}) = \log p(\bfy;\bfphi_{k}) = \log p(\bfy; \bfphi_{0})$, $\bfphi_{k,1}$ satisfies the condition to exit the while loop of Algorithm (\ref{alg:EM-VRSO}), so $\bfphi_{k+1} = \bfphi_{k,1} = \bfphi_{0}$. 

This completes the inductive step, so $\bfphi_k = \bfphi_0$ for all $k \geq 0$. To complete the proof, note that Algorithm (\ref{alg:EM-VRSO}) returns $\bfphi_K$, so $\texttt{EM-VRSO}(\bfphi_0, \lambda, A, M, K) = \bfphi_K = \bfphi_0$.
\end{proof}
\end{appendix}

\end{document}


\title{Variance-reduced stochastic optimization for efficient inference of hidden Markov models: Supplement A: Simulation Study}

\author{
  \textbf{Evan Sidrow} \\
  Department of Statistics \\
  University of British Columbia\\
  Vancouver, Canada \\
\texttt{evan.sidrow@stat.ubc.ca} \\
  \and
  \textbf{Nancy Heckman} \\
  Department of Statistics \\
  University of British Columbia \\
  Vancouver, Canada \\
  \and
  \textbf{Alexandre Bouchard-C\^ot\'e} \\
  Department of Statistics \\
  University of British Columbia \\
  Vancouver, Canada \\
  \and
  \textbf{Sarah M. E. Fortune} \\
  Department of Oceanography \\
  Dalhousie University \\
  Halifax, Canada \\
  \and
  \textbf{Andrew W. Trites} \\
  Department of Zoology \\
  Institute for the Oceans and Fisheries \\
  University of British Columbia \\
  Vancouver, Canada \\
  \and
  \textbf{Marie Auger-M\'eth\'e} \\
  Department of Statistics \\
  Institute for the Oceans and Fisheries \\
  University of British Columbia \\
  Vancouver, Canada \\
}
\maketitle

\newpage

\section{Simulation study results, $T = 10^{3}$}

Below are results for the experiments with $T=10^{3}$. They are analogous to Figure (3) and Figure (4) in the main text.

\begin{figure}[H]
    \centering
    \includegraphics[width=6.5in]{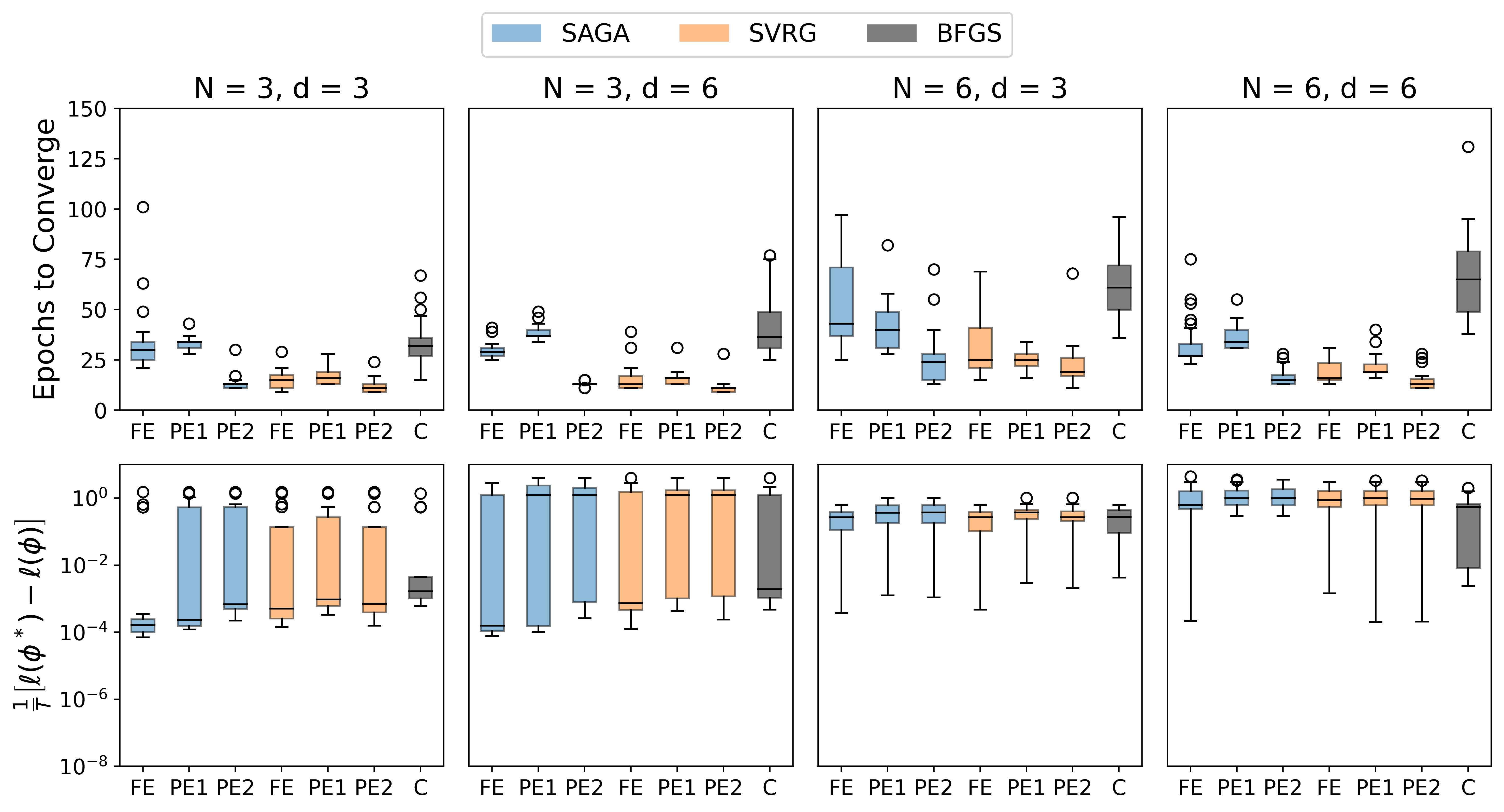}
    \caption{Box plots showing epochs to converge (top) and the maximum log-likelihood minus the log-likelihood at convergence (all divided by $T$, bottom) for each optimization algorithm. FE corresponds to $\{P = \texttt{False}, M = T\}$, PE1 corresponds to $\{P = \texttt{True}, M = T\}$, and PE2 corresponds to $\{P = \texttt{False}, M = 10T\}$. Blue corresponds to SAGA and orange corresponds to SVRG. Results are shown for all simulation studies with $T=10^{3}$, $N=3$ and $d=3$ (left), $N=3$ and $d=6$ (left-middle), $N=6$ and $d=3$ (right-middle), and $N=6$ and $d=6$ (right). One epoch represents either one full E step, $T$ iterations with the M step, or one gradient step for full-gradient algorithms. The y-axis of the bottom row is on a log-scale.}
    \label{fig:boxplots_sim}
\end{figure}
\begin{figure}[H]
    \centering
    \includegraphics[width=6.5in]{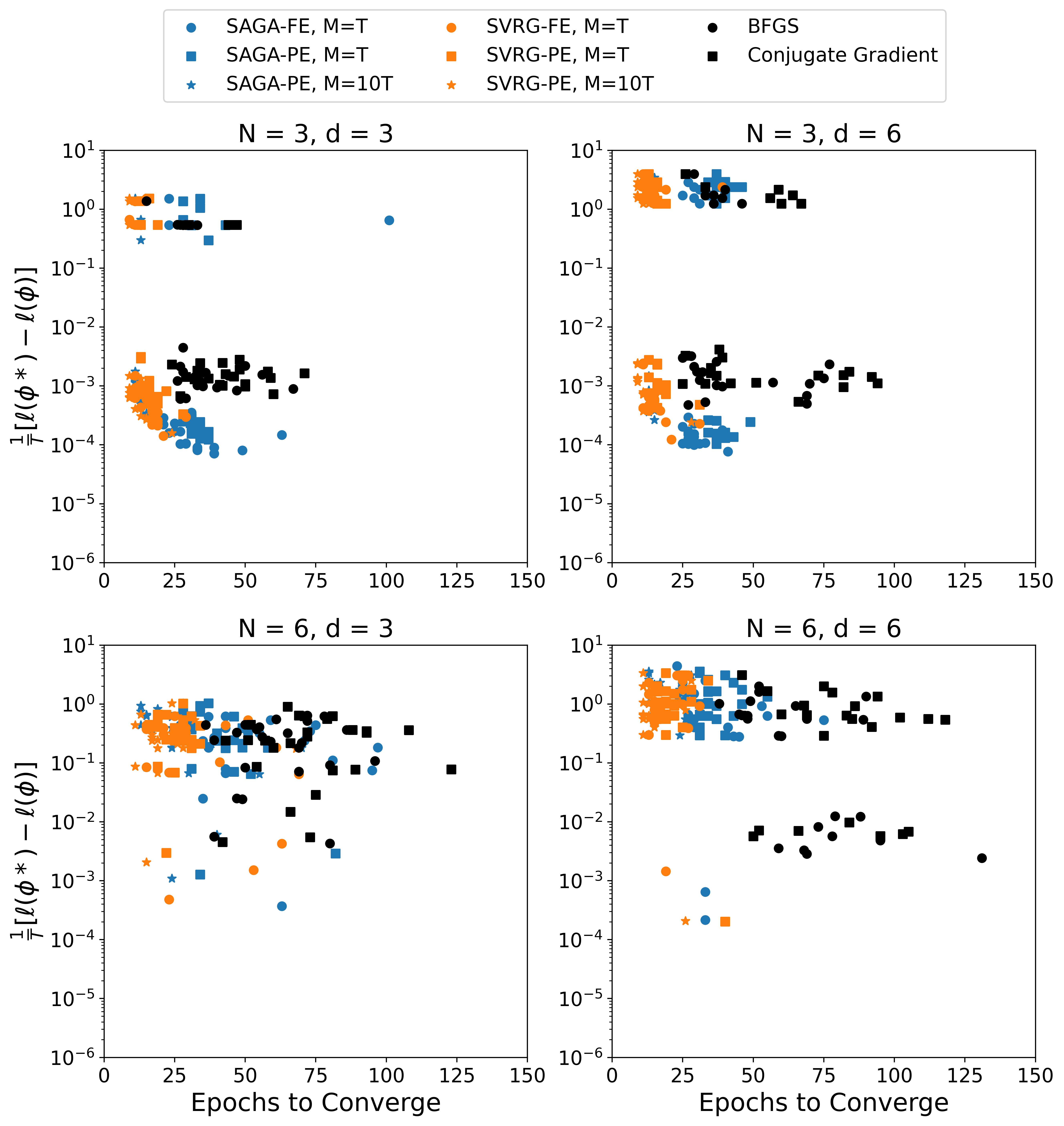}
    \caption{Maximum log-likelihood minus the log-likelihood at convergence (all over $T$) versus epochs to converge for all simulation studies with $T=10^{3}$, $N=3$ and $d=3$ (top left), $N=3$ and $d=6$ (top right), $N=6$ and $d=3$ (bottom left), and $N=6$ and $d=6$ (bottom left). One epoch represents either one full E step, $T$ iterations with the M step, or one gradient step for full-gradient algorithms. The log-likelihood of $\bfphi$ is denoted as $\ell(\bfphi)$ on the y-axis, which is on a log-scale.}
    \label{fig:scatter_sim}
\end{figure}


\newpage 

The figures below show the optimality gap vs epoch for each data set of every experiment with $T=10^{3}$ for all nine tested algorithms. They are similar to Figure (2) from the main text. $\texttt{EM-VRSO}$ tended to converge more quickly than BFGS for experiments with $N=3$. BFGS often out-performed $\texttt{EM-VRSO}$ in the long-run (after approximately 100 epochs), but BFGS is a second-order method and this behaviour is expected in the long run on small data sets. Our method is particularly well-suited to fitting HMMs on large and complex data sets. 

\begin{figure}[H]
    \centering
    \includegraphics[width=6.5in]{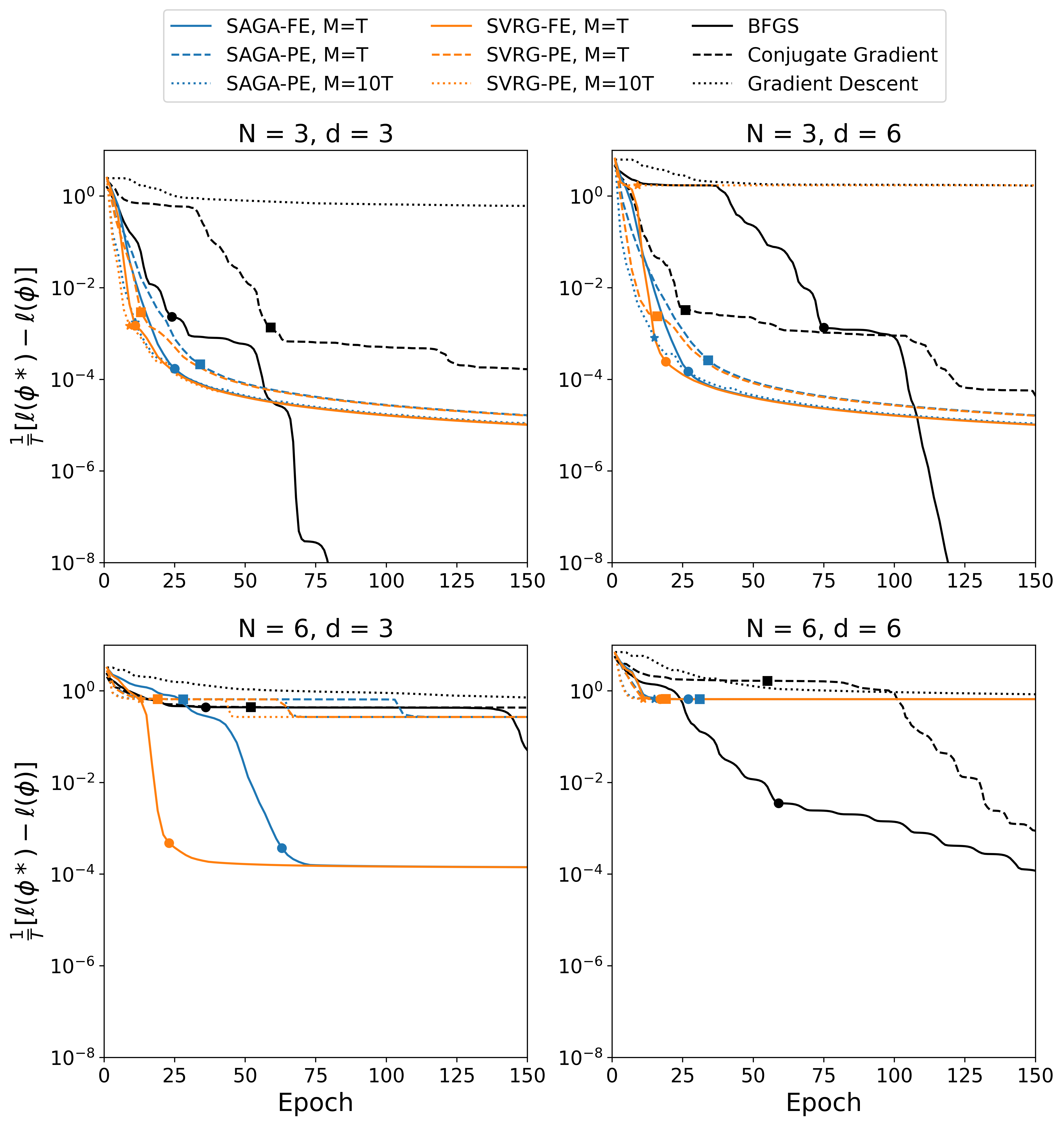}
    \caption{Optimally gap between the current log-likelihood and optimal log-likelihood for the first simulated data sets associated with experiments $T=10^{3}$, $N \in \{3,6\}$ and $d \in \{3,6\}$. One epoch represents either one full E step, $T$ iterations with the M step, or one gradient step for full-gradient algorithms. The log-likelihood of $\bfphi$ is denoted as $\ell(\bfphi)$ on the y-axis, which is on a log-scale.}
\end{figure}
\begin{figure}[H]
    \centering
    \includegraphics[width=6.5in]{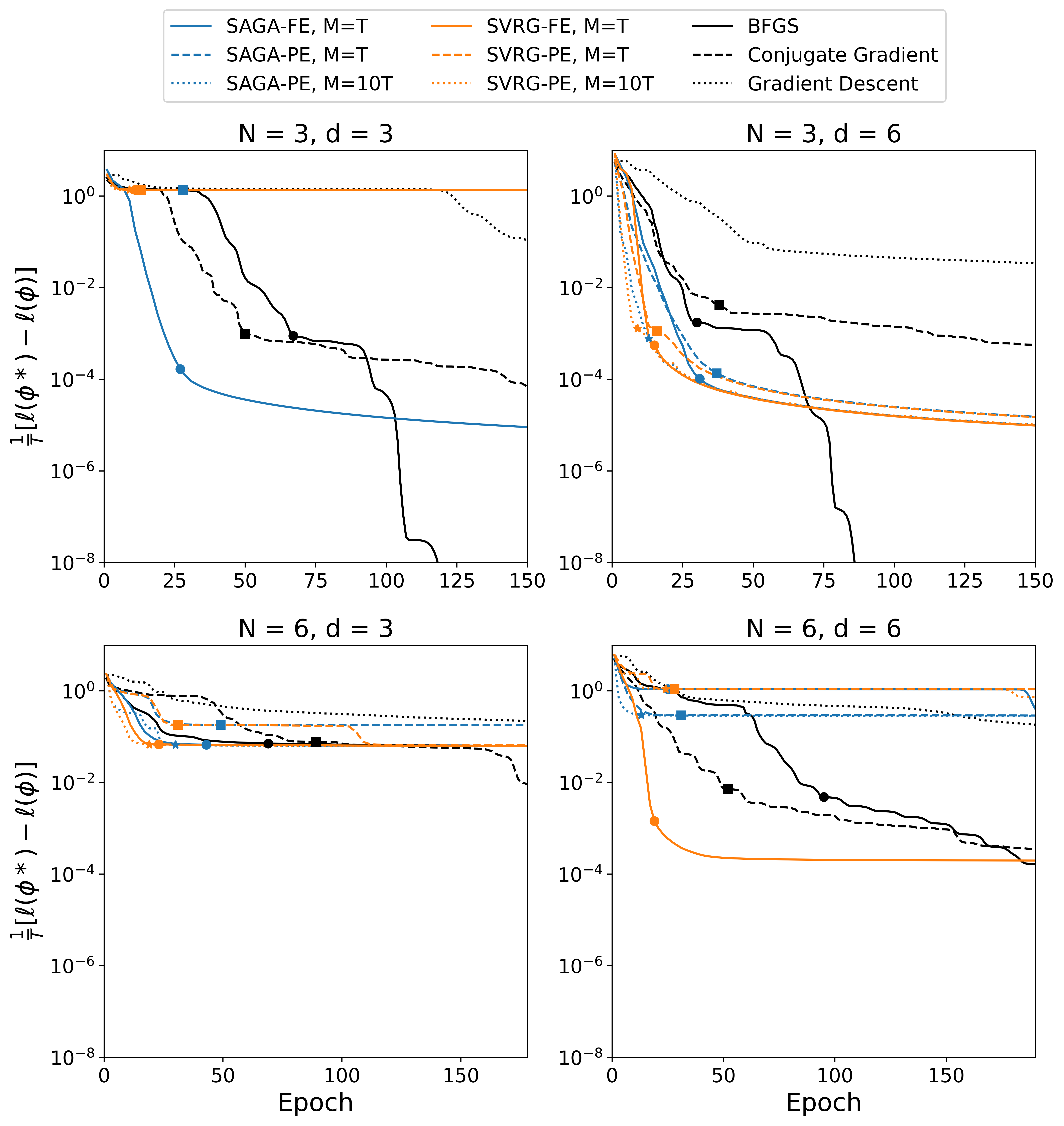}
    \caption{Optimally gap between the current log-likelihood and optimal log-likelihood for the second simulated data sets associated with experiments $T=10^{3}$, $N \in \{3,6\}$ and $d \in \{3,6\}$. One epoch represents either one full E step, $T$ iterations with the M step, or one gradient step for full-gradient algorithms. The log-likelihood of $\bfphi$ is denoted as $\ell(\bfphi)$ on the y-axis, which is on a log-scale.}
\end{figure}
\begin{figure}[H]
    \centering
    \includegraphics[width=6.5in]{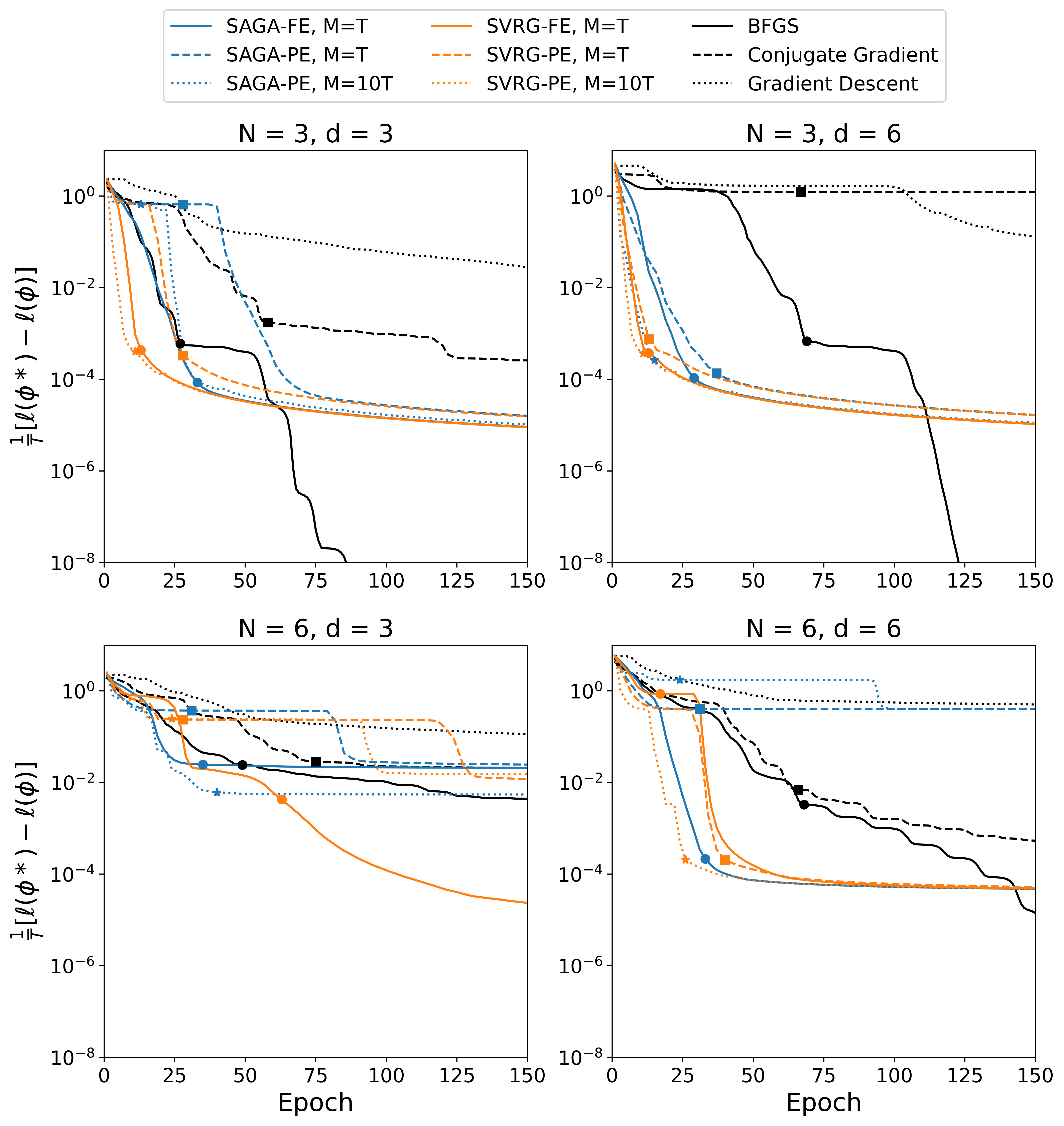}
    \caption{Optimally gap between the current log-likelihood and optimal log-likelihood for the third simulated data sets associated with experiments $T=10^{3}$, $N \in \{3,6\}$ and $d \in \{3,6\}$. One epoch represents either one full E step, $T$ iterations with the M step, or one gradient step for full-gradient algorithms. The log-likelihood of $\bfphi$ is denoted as $\ell(\bfphi)$ on the y-axis, which is on a log-scale.}
\end{figure}
\begin{figure}[H]
    \centering
    \includegraphics[width=6.5in]{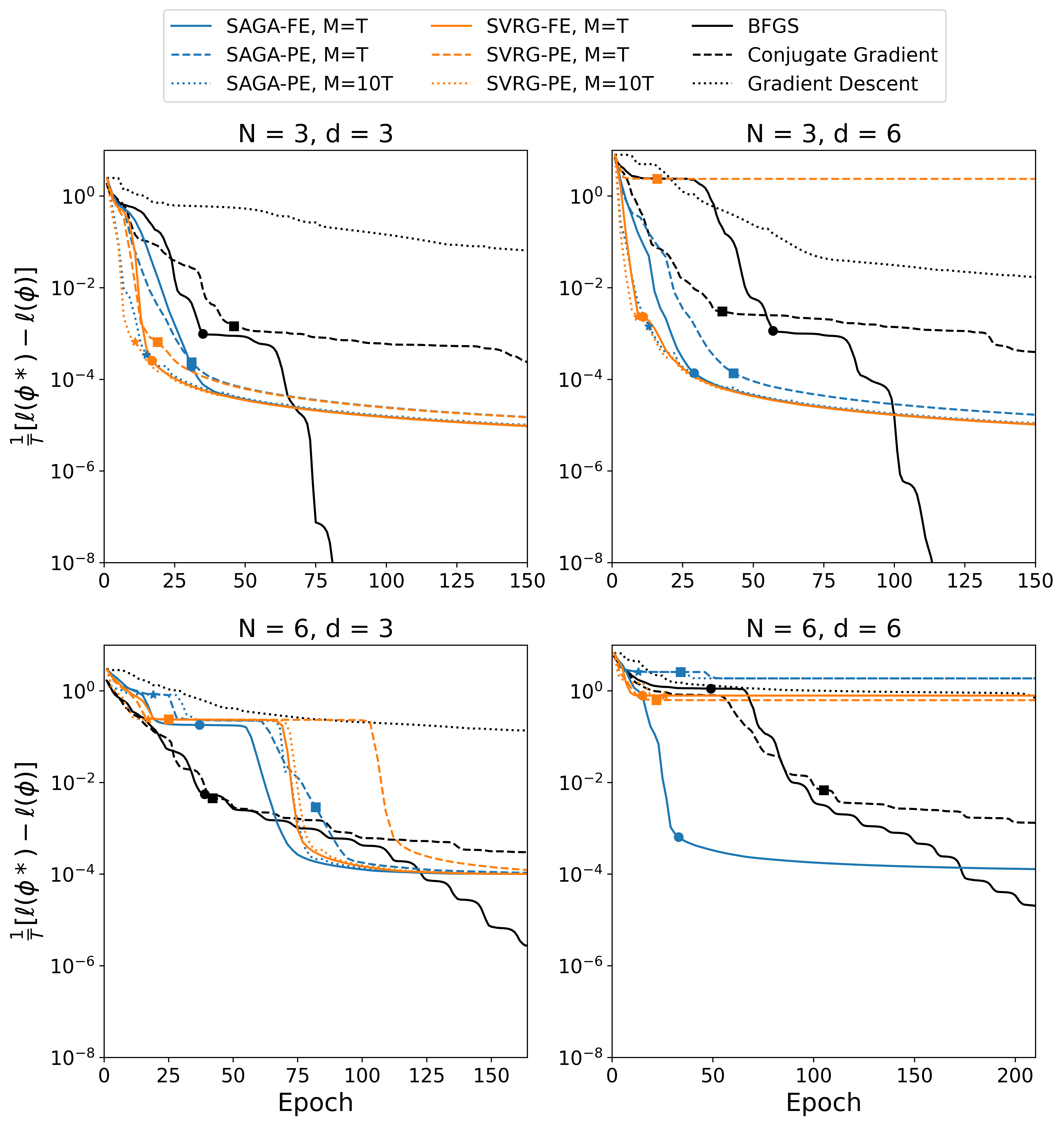}
    \caption{Optimally gap between the current log-likelihood and optimal log-likelihood for the fourth simulated data sets associated with experiments $T=10^{3}$, $N \in \{3,6\}$ and $d \in \{3,6\}$. One epoch represents either one full E step, $T$ iterations with the M step, or one gradient step for full-gradient algorithms. The log-likelihood of $\bfphi$ is denoted as $\ell(\bfphi)$ on the y-axis, which is on a log-scale.}
\end{figure}
\begin{figure}[H]
    \centering
    \includegraphics[width=6.5in]{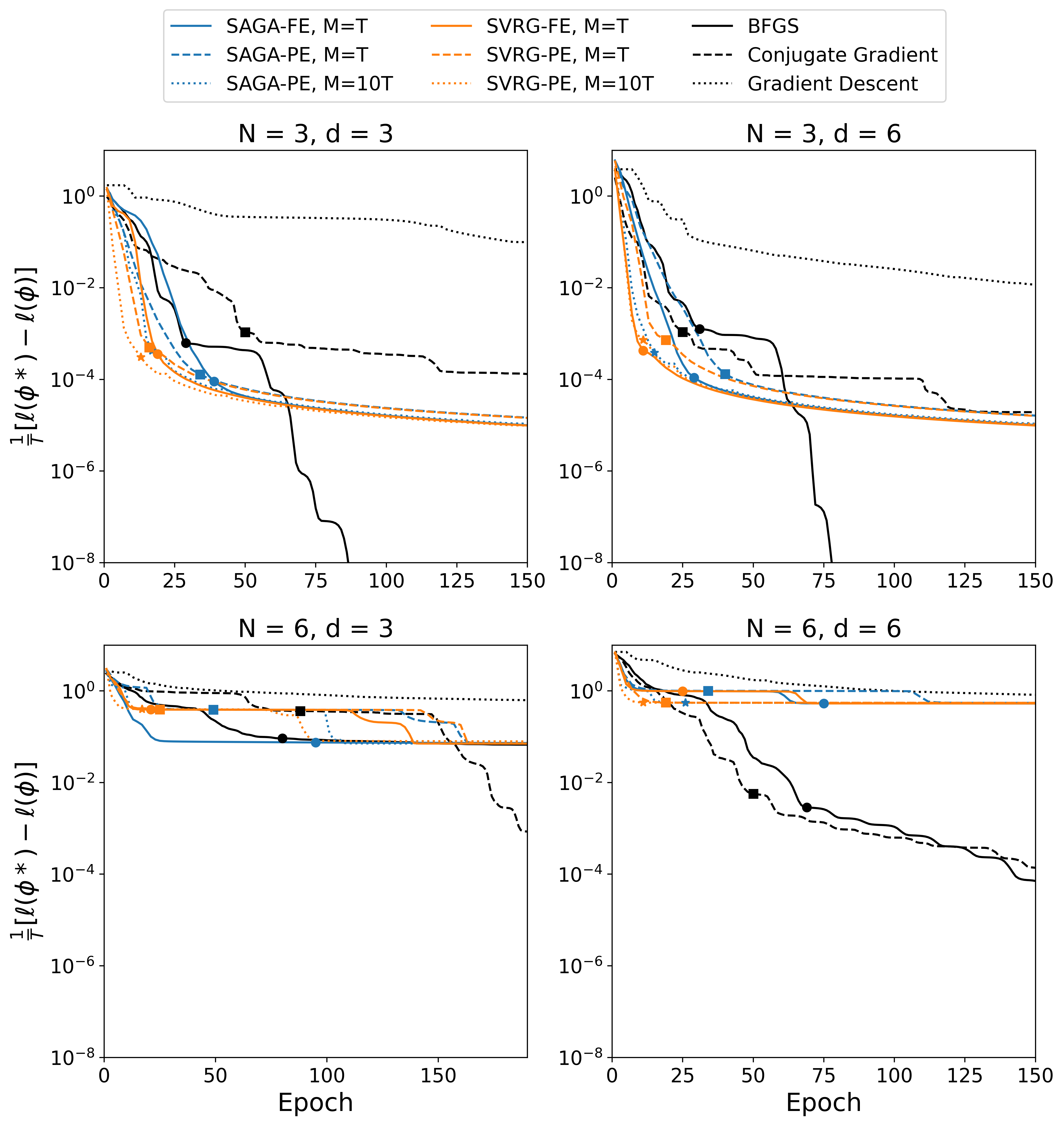}
    \caption{Optimally gap between the current log-likelihood and optimal log-likelihood for the fifth simulated data sets associated with experiments $T=10^{3}$, $N \in \{3,6\}$ and $d \in \{3,6\}$. One epoch represents either one full E step, $T$ iterations with the M step, or one gradient step for full-gradient algorithms. The log-likelihood of $\bfphi$ is denoted as $\ell(\bfphi)$ on the y-axis, which is on a log-scale.}
\end{figure}

\newpage

\section{Simulation study results, $T = 10^{5}$}

The figures below are similar to Figure (2) from the simulation study in the main text. They show the optimality gap vs epoch for each experiment and each of the nine tested algorithms. However, we include all simulated data sets here.

\begin{figure}[H]
    \centering
    \includegraphics[width=6.5in]{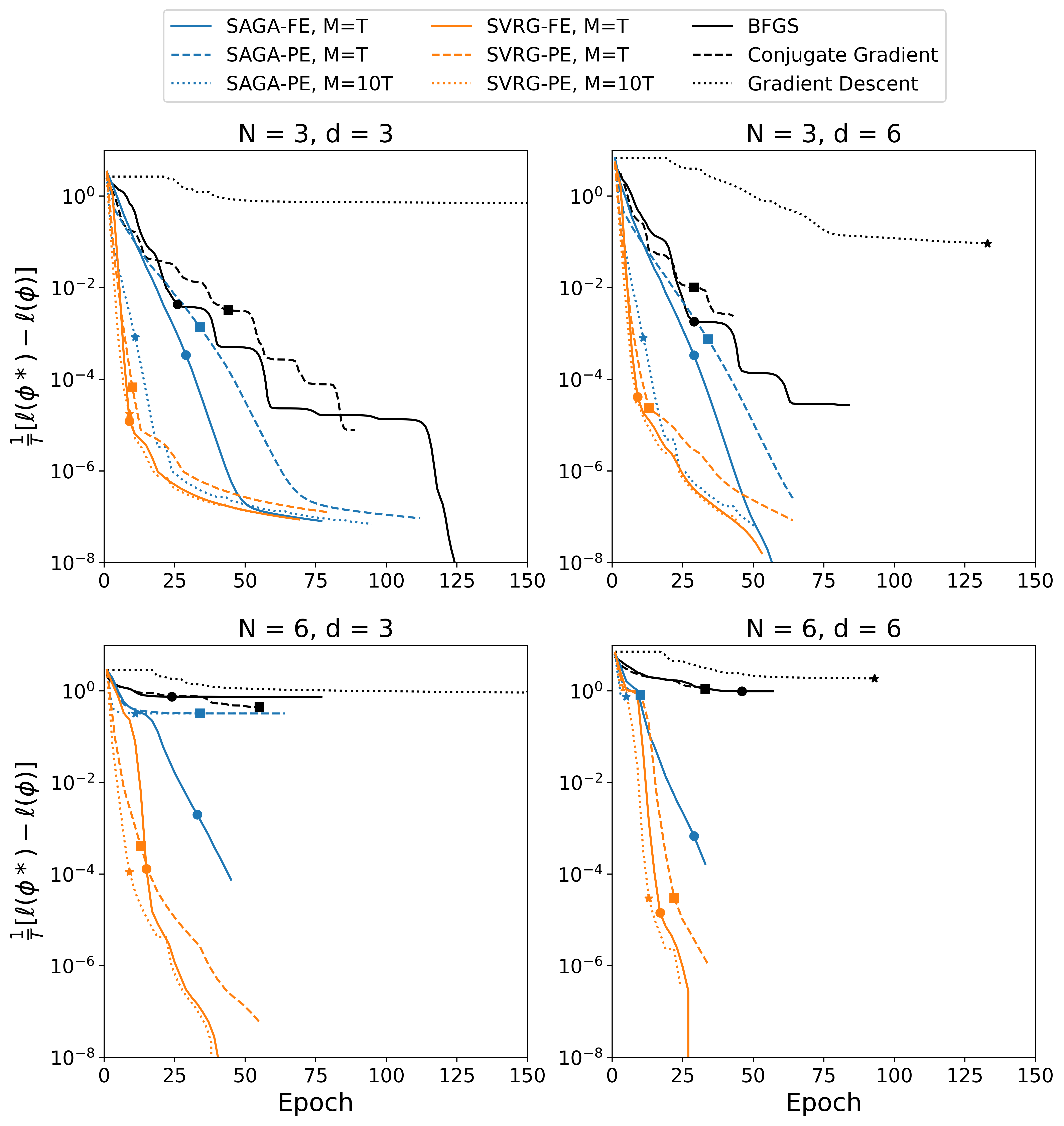}
    \caption{Optimally gap between the current log-likelihood and optimal log-likelihood for the first simulated data sets associated with experiments $T=10^{5}$, $N \in \{3,6\}$ and $d \in \{3,6\}$. One epoch represents either one full E step, $T$ iterations with the M step, or one gradient step for full-gradient algorithms. The log-likelihood of $\bfphi$ is denoted as $\ell(\bfphi)$ on the y-axis, which is on a log-scale.}
\end{figure}
\begin{figure}[H]
    \centering
    \includegraphics[width=6.5in]{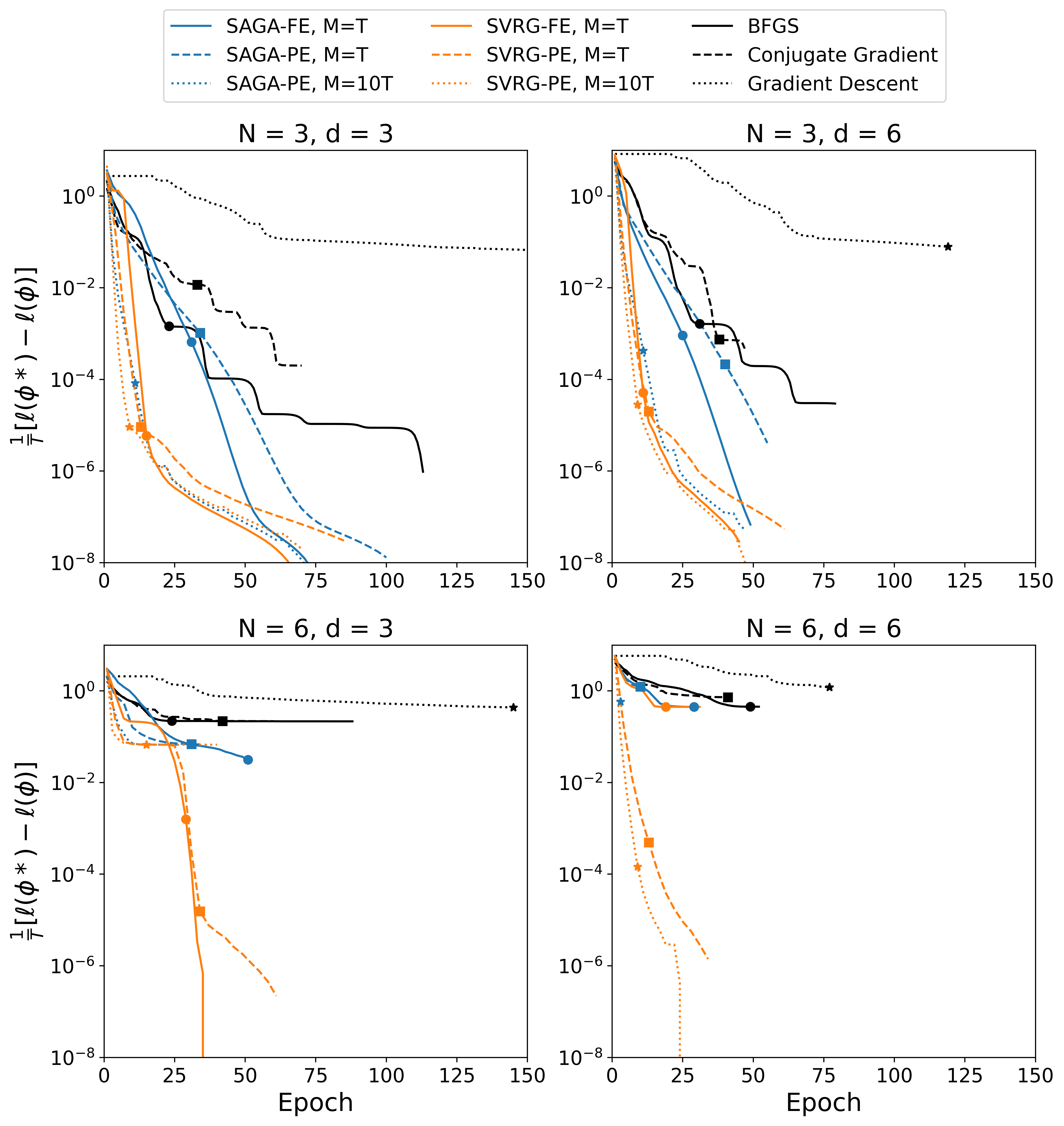}
    \caption{Optimally gap between the current log-likelihood and optimal log-likelihood for the second simulated data sets associated with experiments $T=10^{5}$, $N \in \{3,6\}$ and $d \in \{3,6\}$. One epoch represents either one full E step, $T$ iterations with the M step, or one gradient step for full-gradient algorithms. The log-likelihood of $\bfphi$ is denoted as $\ell(\bfphi)$ on the y-axis, which is on a log-scale.}
\end{figure}
\begin{figure}[H]
    \centering
    \includegraphics[width=6.5in]{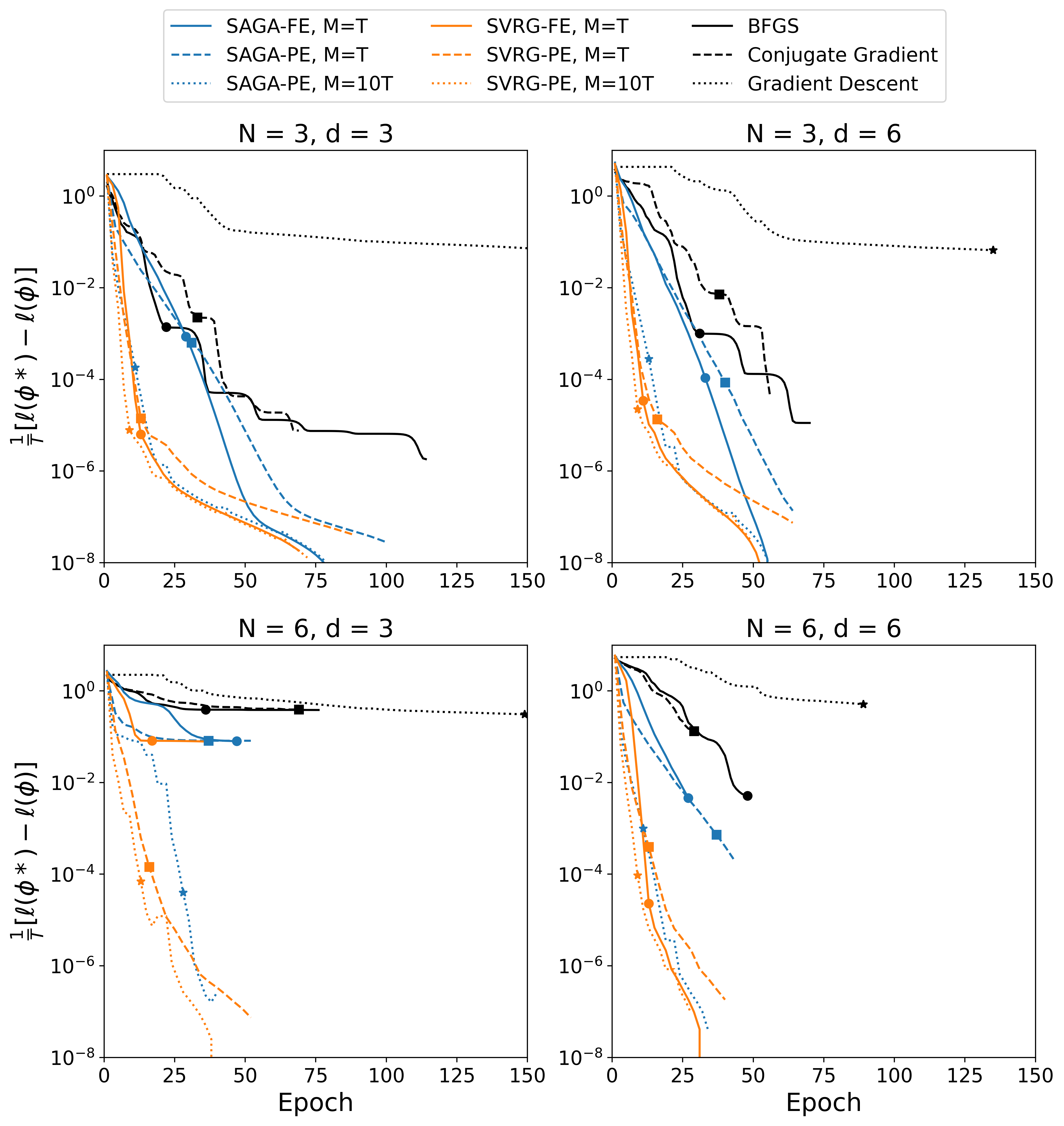}
    \caption{Optimally gap between the current log-likelihood and optimal log-likelihood for the third simulated data sets associated with experiments $T=10^{5}$, $N \in \{3,6\}$ and $d \in \{3,6\}$. One epoch represents either one full E step, $T$ iterations with the M step, or one gradient step for full-gradient algorithms. The log-likelihood of $\bfphi$ is denoted as $\ell(\bfphi)$ on the y-axis, which is on a log-scale.}
\end{figure}
\begin{figure}[H]
    \centering
    \includegraphics[width=6.5in]{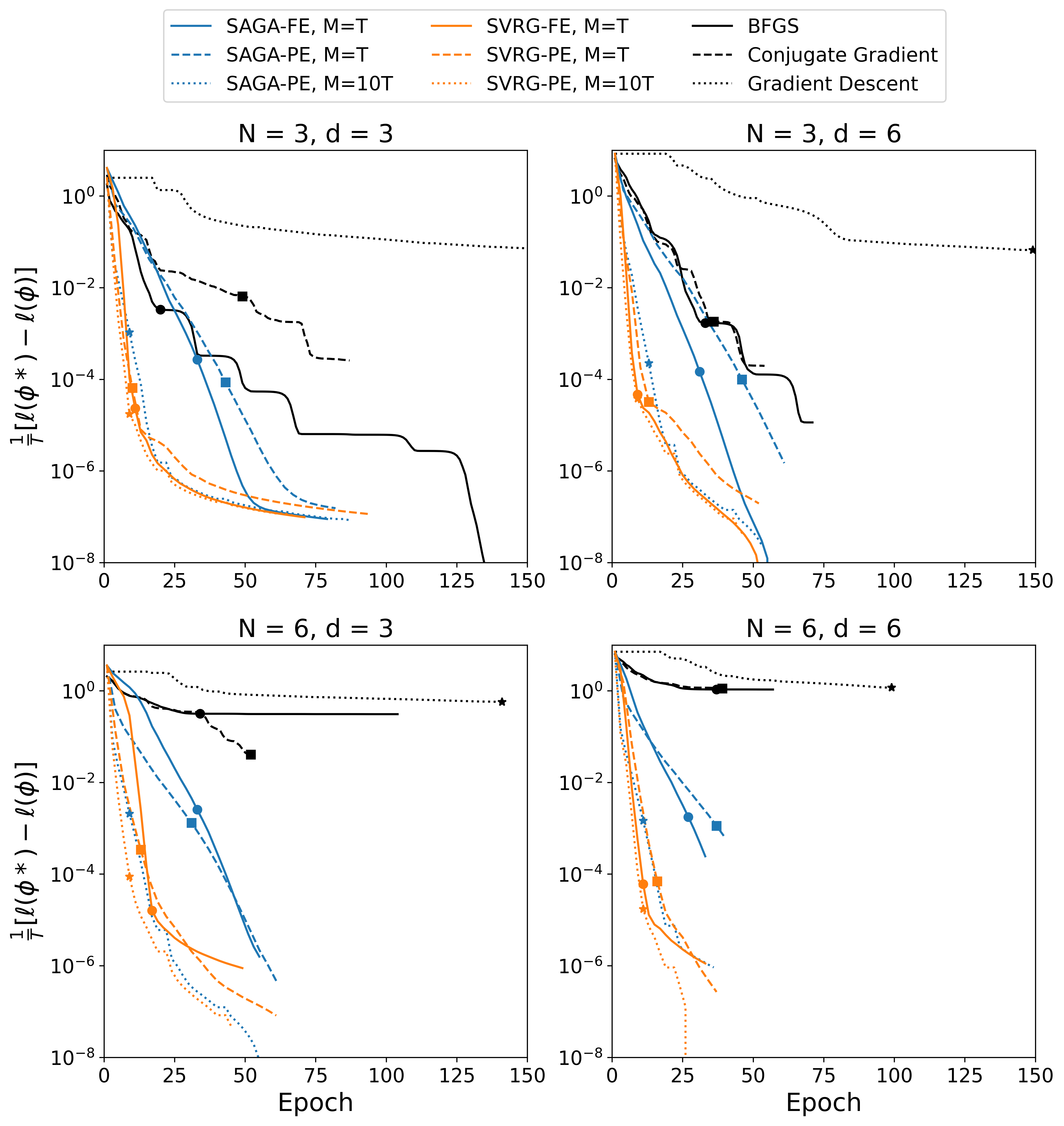}
    \caption{Optimally gap between the current log-likelihood and optimal log-likelihood for the fourth simulated data sets associated with experiments $T=10^{5}$, $N \in \{3,6\}$ and $d \in \{3,6\}$. One epoch represents either one full E step, $T$ iterations with the M step, or one gradient step for full-gradient algorithms. The log-likelihood of $\bfphi$ is denoted as $\ell(\bfphi)$ on the y-axis, which is on a log-scale.}
\end{figure}
\begin{figure}[H]
    \centering
    \includegraphics[width=6.5in]{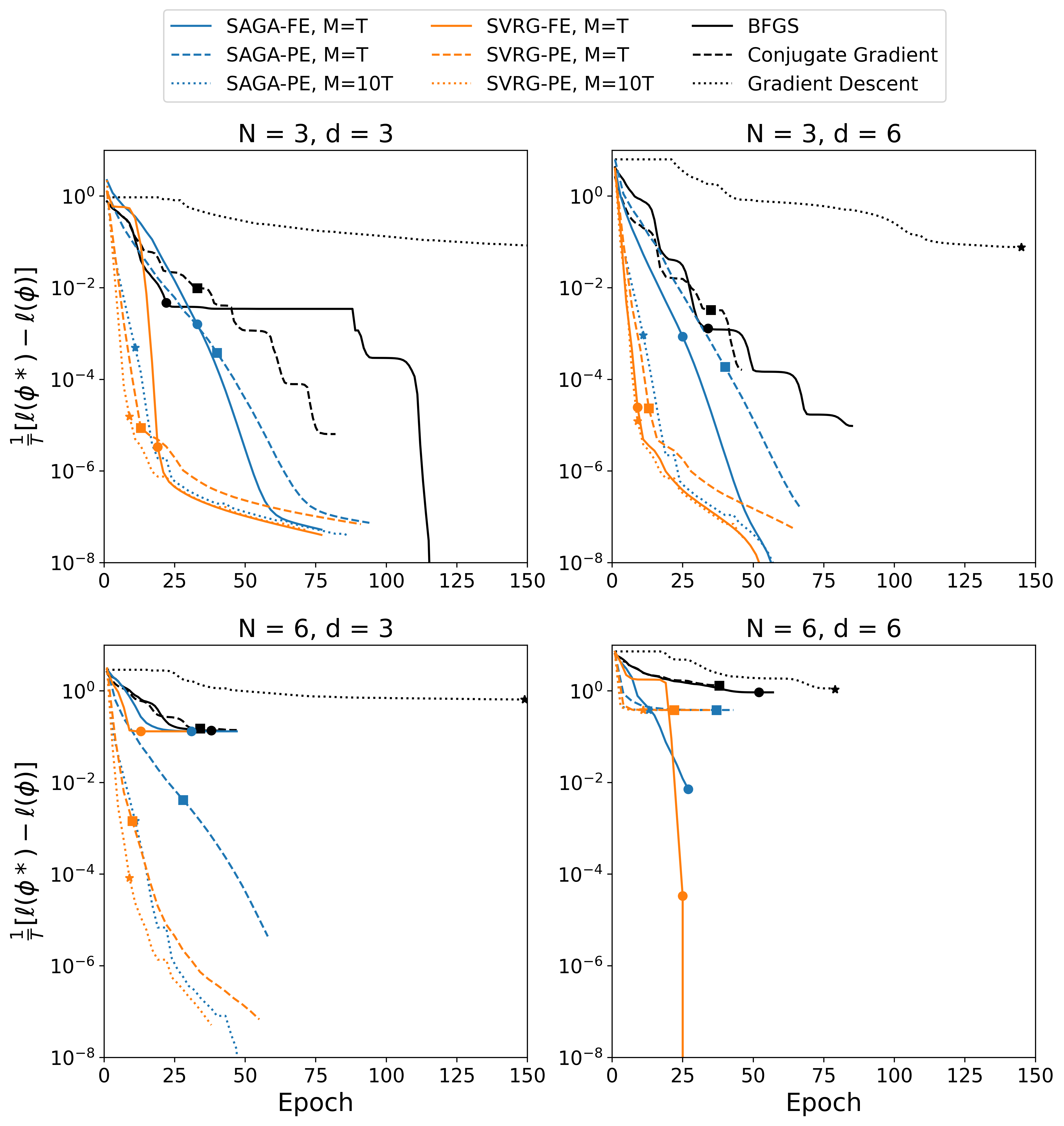}
    \caption{Optimally gap between the current log-likelihood and optimal log-likelihood for the fifth simulated data sets associated with experiments $T=10^{5}$, $N \in \{3,6\}$ and $d \in \{3,6\}$. One epoch represents either one full E step, $T$ iterations with the M step, or one gradient step for full-gradient algorithms. The log-likelihood of $\bfphi$ is denoted as $\ell(\bfphi)$ on the y-axis, which is on a log-scale.}
\end{figure}


\title{Variance-Reduced Stochastic Optimization for Efficient Inference of Hidden Markov Models: Supplement B: Case Study}

\author{
  \textbf{Evan Sidrow} \\
  Department of Statistics \\
  University of British Columbia\\
  Vancouver, Canada \\
  \texttt{evan.sidrow@stat.ubc.ca} \\
  \and
  \textbf{Nancy Heckman} \\
  Department of Statistics \\
  University of British Columbia \\
  Vancouver, Canada \\
  \and
  \textbf{Alexandre Bouchard-C\^ot\'e} \\
  Department of Statistics \\
  University of British Columbia \\
  Vancouver, Canada \\
  \and
  \textbf{Sarah M. E. Fortune} \\
  Department of Oceanography \\
  Dalhousie University \\
  Halifax, Canada \\
  \and
  \textbf{Andrew W. Trites} \\
  Department of Zoology \\
  Institute for the Oceans and Fisheries \\
  University of British Columbia \\
  Vancouver, Canada \\
  \and
  \textbf{Marie Auger-M\'eth\'e} \\
  Department of Statistics \\
  Institute for the Oceans and Fisheries \\
  University of British Columbia \\
  Vancouver, Canada \\
}
\maketitle

\newpage

\section{Case study parameter results}

The parameters presented here come from a run of \texttt{EM-VRSO} with $A = \text{SAGA}$, $P = \texttt{True}$, and $M=10T$. They represent the parameters with the highest log-likelihood among all random initializations and optimization algorithms in the case study from the main text. 

\begin{table}[H]
\centering
\begin{tabular}{c|c|lll}
\multicolumn{1}{l|}{Dive Type} & Parameter Estimate & Descent & Bottom & Ascent \\ \hline
\multirow{3}{*}{1}            & $\hat \mu$     & 0.70    & 0.02   & -0.67  \\
                              & $\hat \sigma$  & 0.45    & 0.23   & 0.40   \\
                              & $\hat p$       & \bf{0}  & \bf{0} & 0.12   \\ \hline 
\multirow{3}{*}{2}            & $\hat \mu$     & 0.33    & 0.00   & -0.44  \\
                              & $\hat \sigma$  & 0.15    & 0.15   & 0.21   \\
                              & $\hat p$       & \bf{0}  & \bf{0} & 0.47   \\ \hline
\multirow{3}{*}{3}            & $\hat \mu$     & 2.71    & 0.01   & -2.50  \\
                              & $\hat \sigma$  & 1.85    & 0.64   & 1.58   \\
                              & $\hat p$       & \bf{0}  & \bf{0} & 0.04  
\end{tabular}
\caption{Maximum likelihood parameter estimates from the Killer Whale case study described in the main text. Recall that $D_t$ represents the change in depth in meters at time index $t$ and $E_t \in \{0,1\}$ encodes whether a dive ends at time index $t$. The parameter $\mu$ corresponds to the state-dependent mean of $D_t$, $\sigma$ corresponds to the state-dependent standard deviation of $D_t$, and $p$ corresponds to the state-dependent probability that $E_t=1$.}
\end{table}

Figure (\ref{fig:emission_dist}) below shows the estimated densities of change in depth $D_t$ for all dive types and dive durations.

\begin{figure}[H]
    \centering
    \includegraphics[width=6.5in]{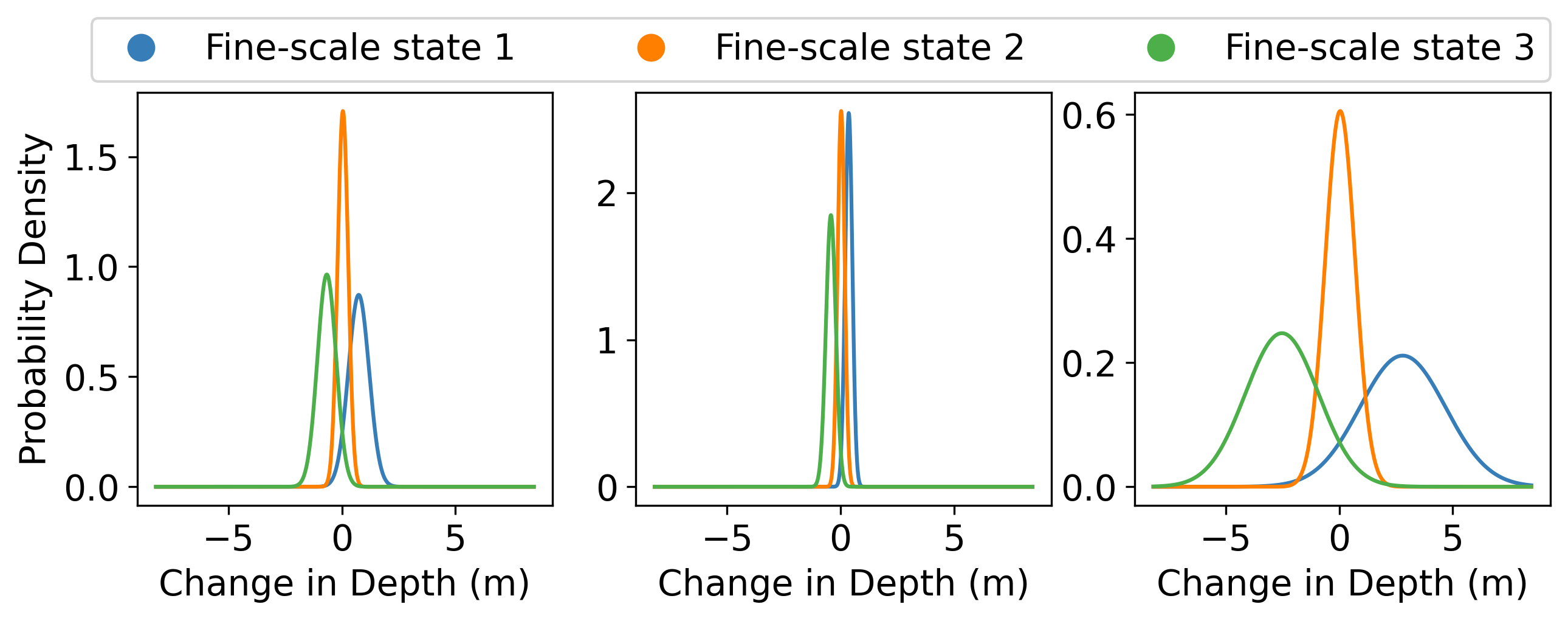}
    \caption{State-dependent density of change in depth for dive types 1, 2, and 3 (left to right). Fine-scale state 1 corresponds to descent, fine-scale state 2 corresponds to bottom, and fine-scale state 3 corresponds to ascent. Dive type 1 corresponds to medium-depth dives, dive type 2 corresponds to shallow dives, and dive type 3 corresponds to deep dives.}
    \label{fig:emission_dist}
\end{figure}

The coarse-scale, inter-dive transition probability matrix and initial distribution estimates were 

\begin{gather}
    \hat \bfGamma^{(c)} = 
    \begin{pmatrix} 
    0.223 & 0.758 & 0.017 \\
    0.123 & 0.853 & 0.027 \\
    0.074 & 0.857 & 0.068
    \end{pmatrix}, \\ \nonumber \\
    \hat \bfdelta^{(c)} = \begin{pmatrix} 0.258 & 0.002 & 0.740 \end{pmatrix}.
\end{gather}

The long-run proportion of each dive type is found by solving for the stationary distribution of $\hat \bfGamma^{(c)}$, i.e. solving the equation $\hat \bfpi^{(c)} \hat \bfGamma^{(c)} = \hat \bfpi^{(c)}$. Doing so gives the following:
\begin{equation}
    \hat \bfpi^{(c)} = 
    \begin{pmatrix}
        0.136 & 0.840 & 0.024
    \end{pmatrix}.
\end{equation}
In words, approximately 13.6\% of all dives are of type 1 (medium-depth), 84.0\% of all dives are of type 2 (shallow), and 2.4\% of all dives are of type 3 (deep).

The fine-scale, intra-dive transition probability matrices and initial distributions were estimated as

\begin{gather}
    \hat \bfGamma^{(f,1)} = 
    \begin{pmatrix} 
    0.871 & 0.123 & 0.006 \\
    \bf{0} & 0.962 & 0.038 \\
    \bf{0} & \bf{0} & \bf{1}
    \end{pmatrix}, \quad
    \hat \bfGamma^{(f,2)} = 
    \begin{pmatrix} 
    0.668 & 0.304 & 0.028 \\
    \bf{0} & 0.788 & 0.212 \\
    \bf{0} & \bf{0} & \bf{1}
    \end{pmatrix}, ~
    \\ \nonumber \\
    \hat \bfGamma^{(f,3)} = 
    \begin{pmatrix} 
    0.958 & 0.040 & 0.002 \\
    \bf{0} & 0.982 & 0.018 \\
    \bf{0} & \bf{0} & \bf{1}
    \end{pmatrix}, \\ \nonumber \\
    \hat \bfdelta^{(f,1)} = \begin{pmatrix} \bf{1} & \bf{0} & \bf{0} \end{pmatrix}, \quad ~
    \hat \bfdelta^{(f,2)} = \begin{pmatrix} \bf{1} & \bf{0} & \bf{0} \end{pmatrix}, \quad ~
    \hat \bfdelta^{(f,3)} = \begin{pmatrix} \bf{1} & \bf{0} & \bf{0} \end{pmatrix}
\end{gather}

Recall that each dive must begin with descent, so all initial distributions are identical. In addition, note that the Markov chains defined by $\hat \bfGamma^{(f,i)}$ above are not ergodic, so stationary distributions do not exist for the fine-scale, intra-dive Markov chains.

\section{Case study decoded dive profiles}

The figures below show approximately one hour of dive profiles associated with each of the eight whales used to train the hierarchical HMM in the case study.

\begin{figure}[H]
    \centering
    \includegraphics[width=6.5in]{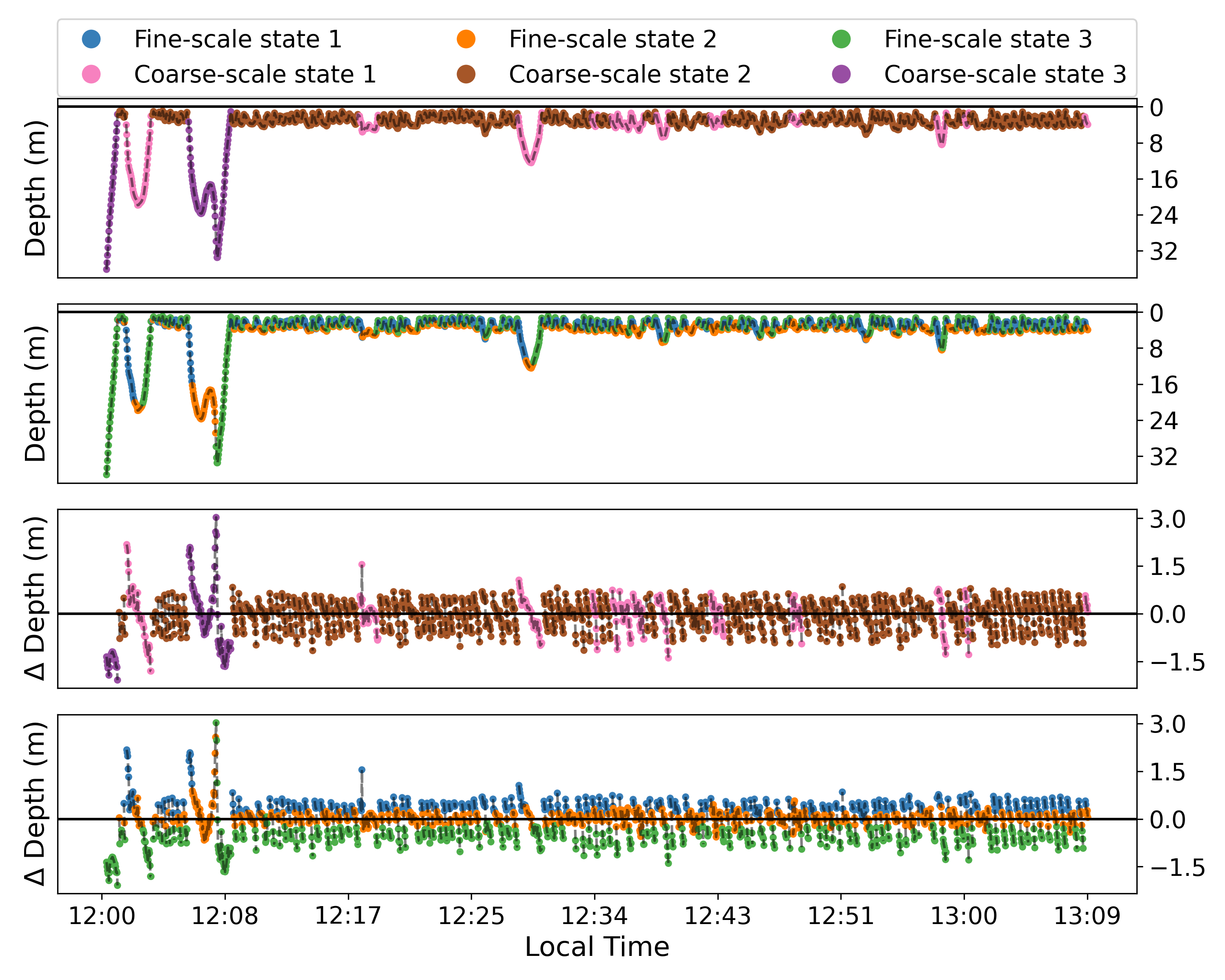}
    \caption{Dive profiles and change in depth vs time for a selected hour of dive time for killer whale A100. All data is color-coded by most likely dive type (rows one and three) as well as most likely dive phase (rows two and four).}
    \label{fig:A100}
\end{figure}

\begin{figure}[H]
    \centering
    \includegraphics[width=6.5in]{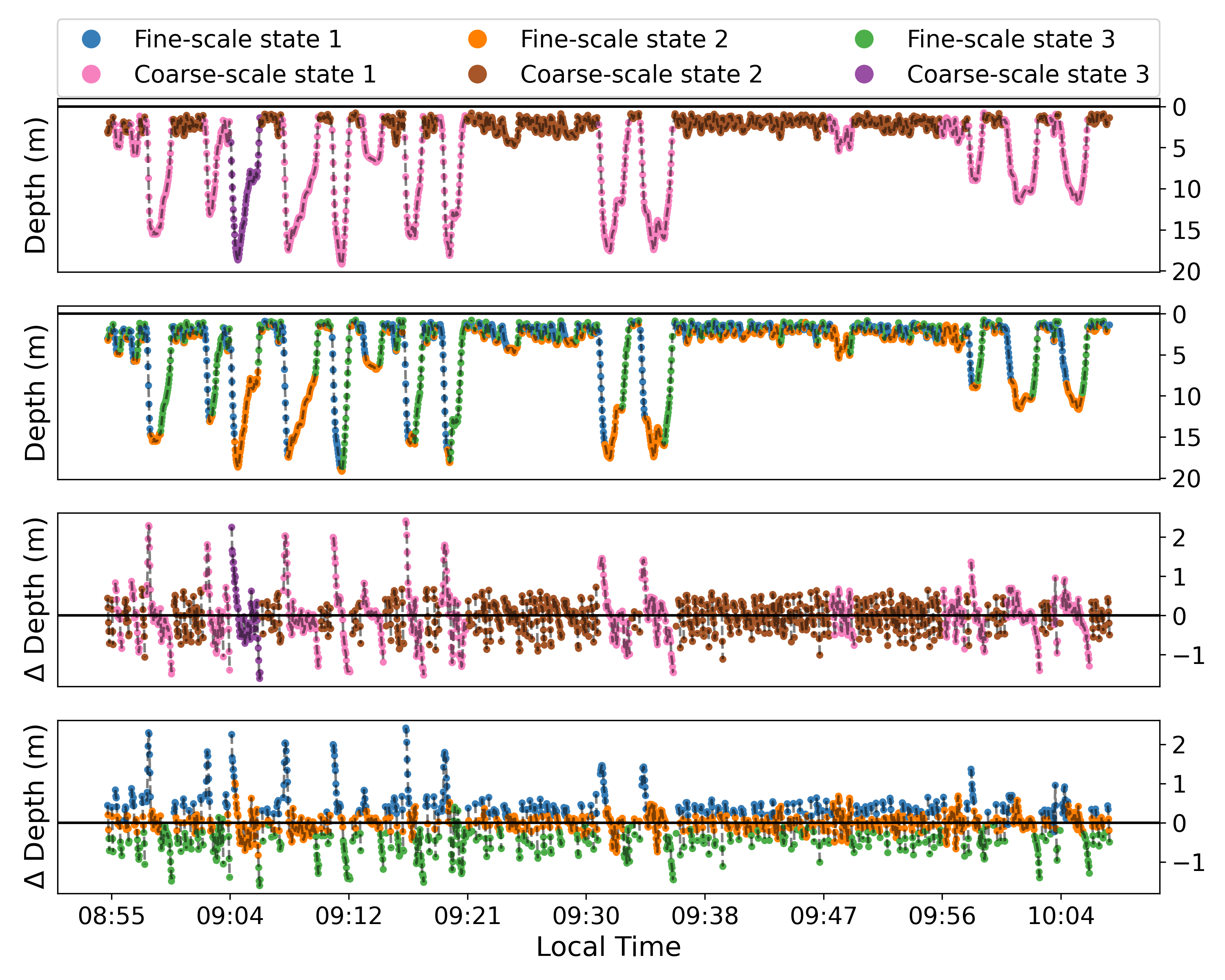}
    \caption{Dive profiles and change in depth vs time for a selected hour of dive time for killer whale A113. All data is color-coded by most likely dive type (rows one and three) as well as most likely dive phase (rows two and four).}
    \label{fig:A113}
\end{figure}

\begin{figure}[H]
    \centering
    \includegraphics[width=6.5in]{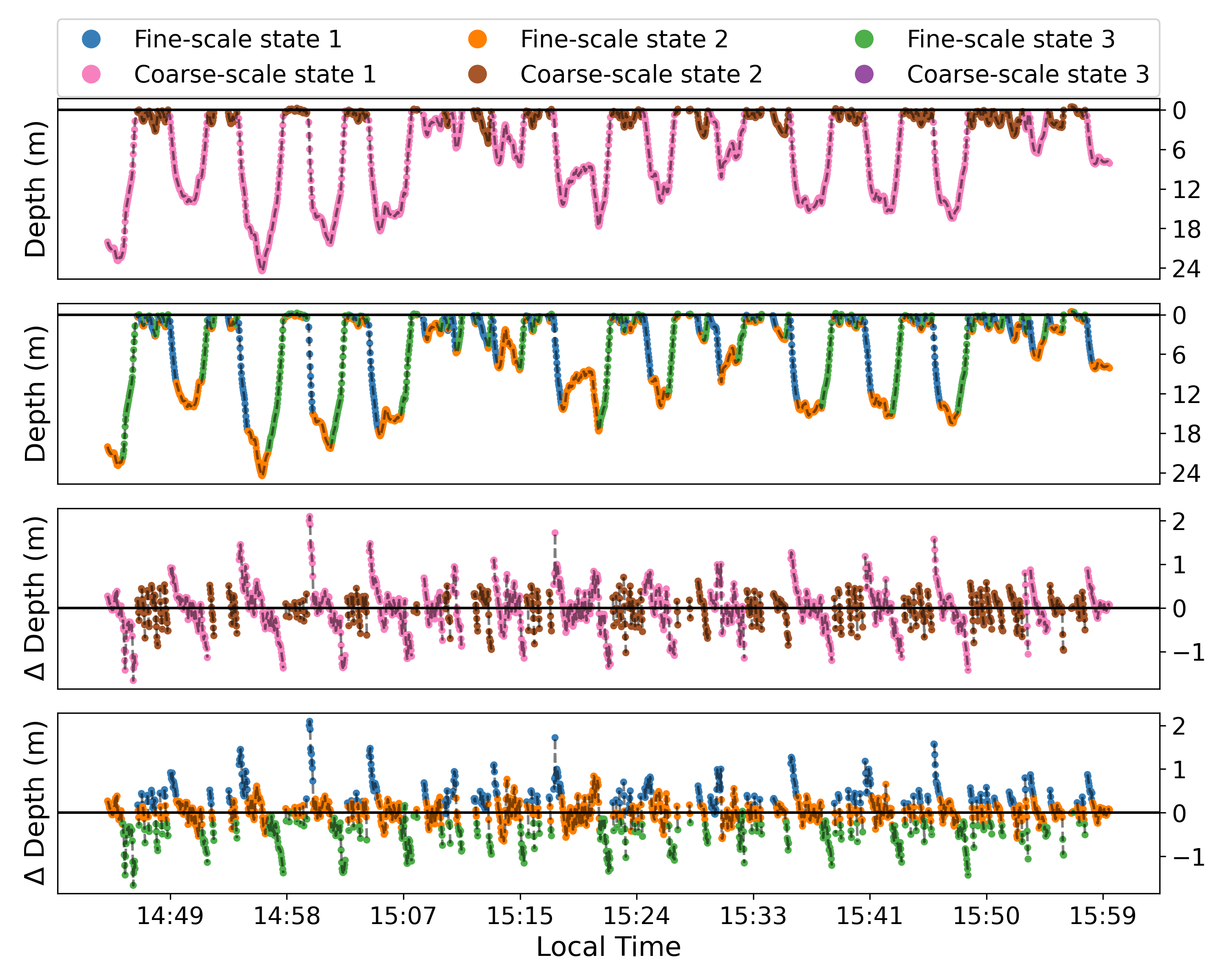}
    \caption{Dive profiles and change in depth vs time for a selected hour of dive time for killer whale D21. All data is color-coded by most likely dive type (rows one and three) as well as most likely dive phase (rows two and four).}
    \label{fig:D21}
\end{figure}

\begin{figure}[H]
    \centering
    \includegraphics[width=6.5in]{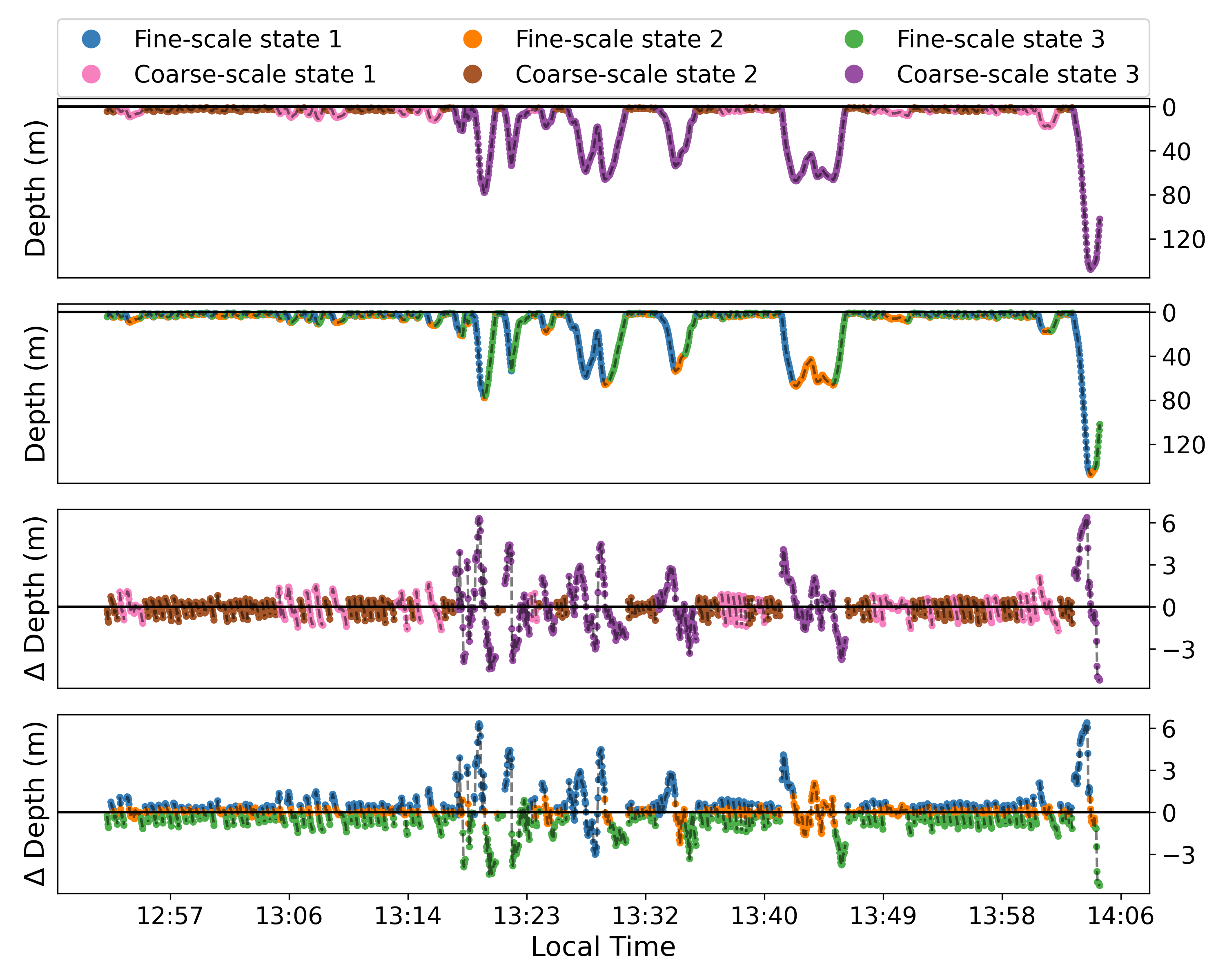}
    \caption{Dive profiles and change in depth vs time for a selected hour of dive time for killer whale D26. All data is color-coded by most likely dive type (rows one and three) as well as most likely dive phase (rows two and four).}
    \label{fig:D26}
\end{figure}

\begin{figure}[H]
    \centering
    \includegraphics[width=6.5in]{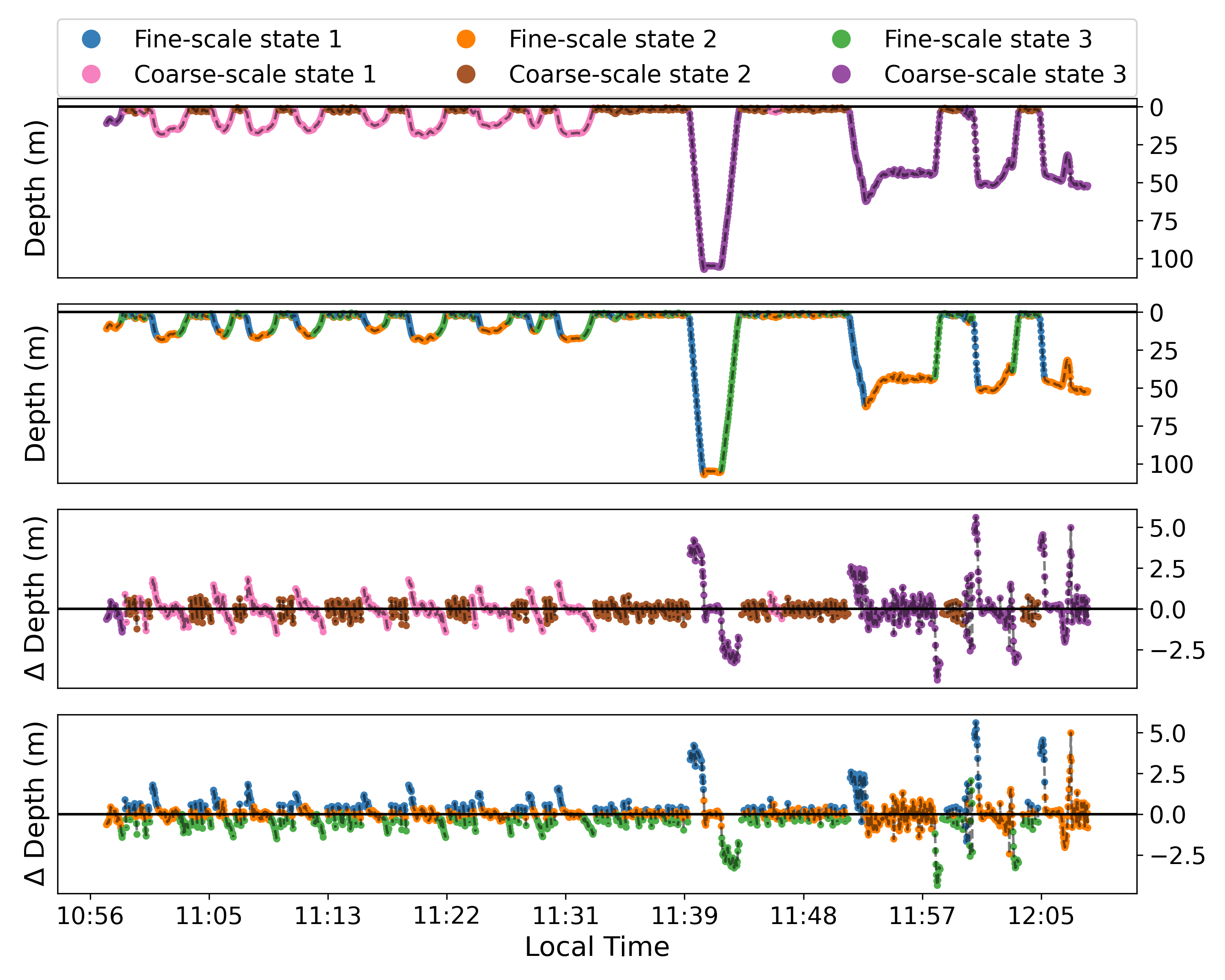}
    \caption{Dive profiles and change in depth vs time for a selected hour of dive time for killer whale I107. All data is color-coded by most likely dive type (rows one and three) as well as most likely dive phase (rows two and four).}
    \label{fig:I107}
\end{figure}

\begin{figure}[H]
    \centering
    \includegraphics[width=6.5in]{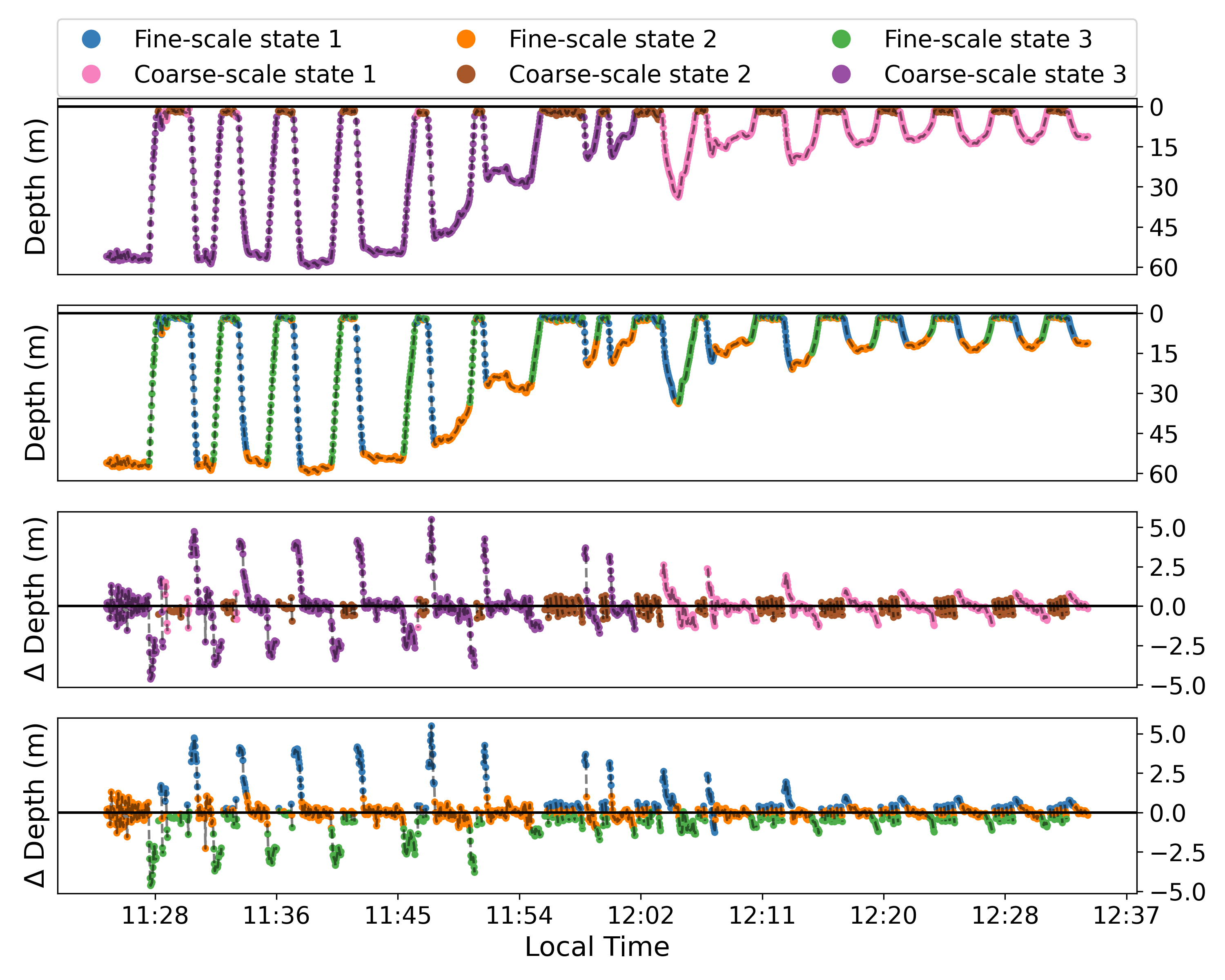}
    \caption{Dive profiles and change in depth vs time for a selected hour of dive time for killer whale I145. All data is color-coded by most likely dive type (rows one and three) as well as most likely dive phase (rows two and four).}
    \label{fig:I145}
\end{figure}

\begin{figure}[H]
    \centering
    \includegraphics[width=6.5in]{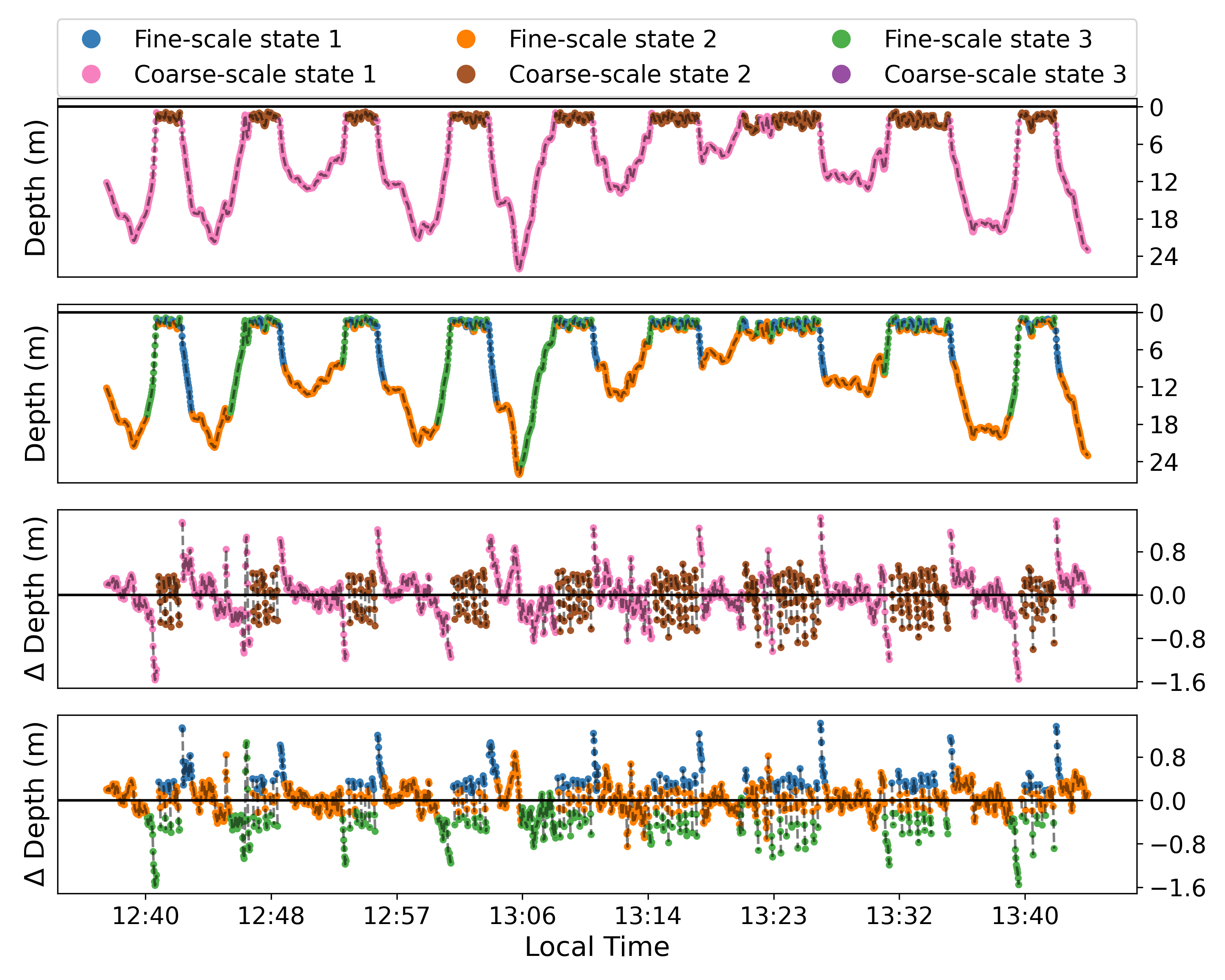}
    \caption{Dive profiles and change in depth vs time for a selected hour of dive time for killer whale R48. All data is color-coded by most likely dive type (rows one and three) as well as most likely dive phase (rows two and four).}
    \label{fig:R48}
\end{figure}

\begin{figure}[H]
    \centering
    \includegraphics[width=6.5in]{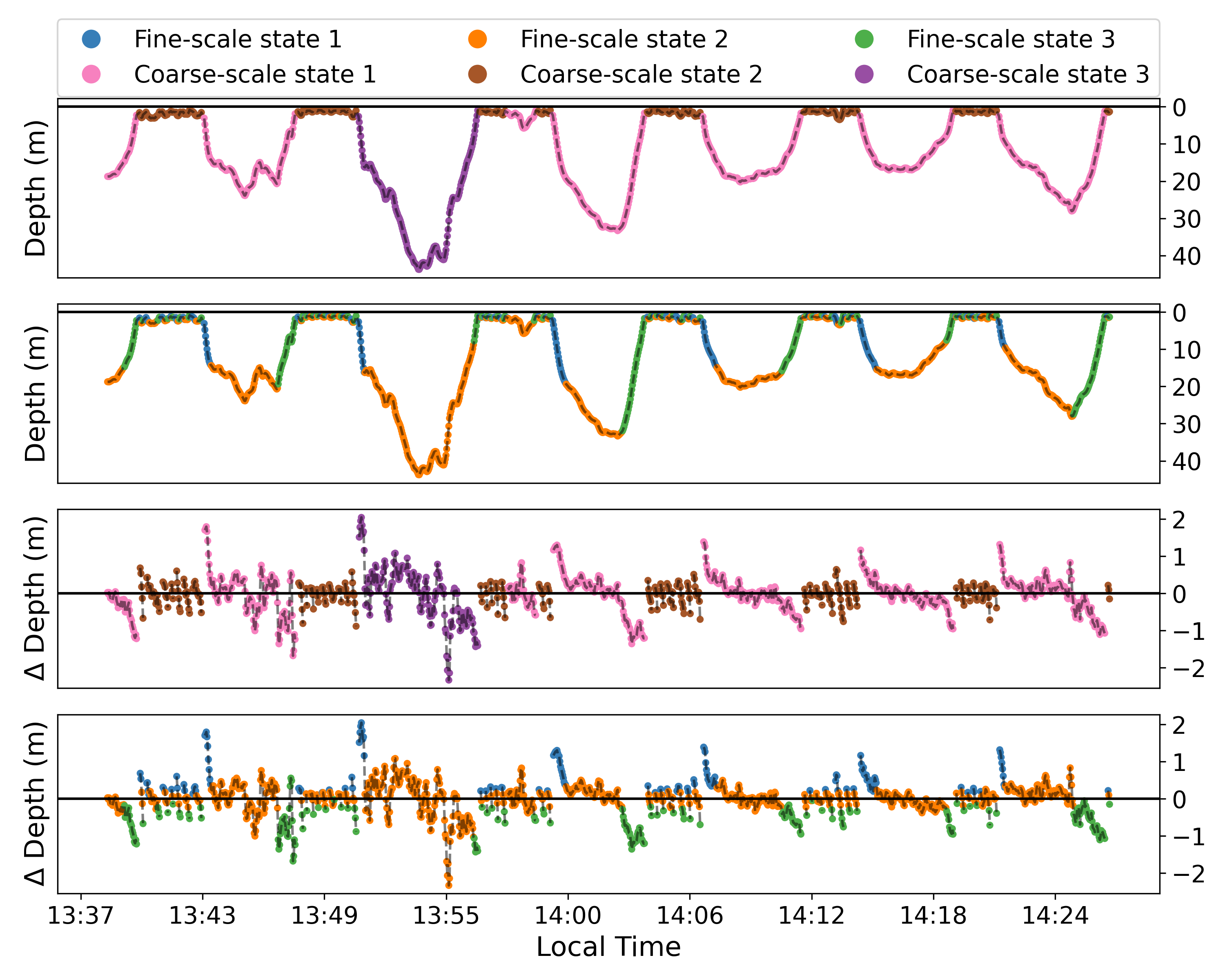}
    \caption{Dive profiles and change in depth vs time for a selected hour of dive time for killer whale R58. All data is color-coded by most likely dive type (rows one and three) as well as most likely dive phase (rows two and four).}
    \label{fig:R58}
\end{figure}